\documentclass[10pt]{article}

\usepackage[dvips]{epsfig} 
\usepackage{amssymb,amsmath,amsthm,mathrsfs} 
\usepackage{graphicx}
\usepackage{lineno}

\usepackage[authoryear,sort&compress]{natbib}
\usepackage{url}
\usepackage{enumerate}
\usepackage{colordvi,color,pspicture}
\usepackage{psfrag}
\usepackage{rotating}

\newcommand{\X}{\mathcal{X}}
\newcommand{\U}{\mathcal{U}}
\newcommand{\I}{\mathbf{I}}

\newcommand{\E}{\mathbf{E}}
\newcommand{\Var}{\mathbf{Var}}
\newcommand{\Cov}{\mathbf{Cov}}

\newcommand{\ah}{\widehat{\alpha}}
\newcommand{\bh}{\widehat{\beta}}
\newcommand{\pasy}{p_{\text{asy}}}
\newcommand{\pmc}{p_{\text{mc}}}
\newcommand{\prand}{p_{\text{rand}}}

\theoremstyle{plain}
\newtheorem{theorem}{Theorem}[section]

\theoremstyle{definition}

\theoremstyle{remark}

\newtheorem{remark}[theorem]{Remark}

\begin{document}


\title{Technical Report \# KU-EC-08-1:\\
Overall and Pairwise Segregation Tests Based on Nearest Neighbor Contingency Tables}
\author{
Elvan Ceyhan
\thanks{Department of Mathematics, Ko\c{c} University, Sar{\i}yer, 34450, Istanbul, Turkey.}
}
\date{\today}
\maketitle


\begin{abstract}
\noindent
Multivariate interaction between two or more classes (or species)
has important consequences in many fields
and causes multivariate clustering patterns
such as segregation or association.
The spatial \emph{segregation} occurs when members of a class
tend to be found near members of the same class (i.e., near conspecifics)
while spatial \emph{association} occurs when members of a class tend to be found
near members of the other class or classes.
These patterns can be studied using a nearest neighbor contingency table (NNCT).
The null hypothesis is randomness in the nearest neighbor (NN) structure,
which may result from --- among other patterns ---
random labeling (RL) or complete spatial randomness (CSR)
of points from two or more classes (which is called the \emph{CSR independence}, henceforth).
In this article, we introduce new versions of overall and cell-specific tests based on NNCTs
(i.e., NNCT-tests) and compare them with Dixon's overall and cell-specific tests.
These NNCT-tests provide information
on the spatial interaction between the classes at small scales
(i.e., around the average NN distances between the points).
Overall tests are used to detect any deviation from the null case,
while the cell-specific tests are post hoc pairwise spatial interaction tests
that are applied when the overall test yields a significant result.
We analyze the distributional properties of these tests;
assess the finite sample performance of the tests
by an extensive Monte Carlo simulation study.
Furthermore, we show
that the new NNCT-tests have
better performance in terms of Type I error and power.
We also illustrate these NNCT-tests on two real life data sets.
\end{abstract}

\noindent
{\small {\it Keywords:} Association; clustering; completely mapped data;
complete spatial randomness; random labeling; spatial pattern

\vspace{.25 in}

$^*$corresponding author.\\
\indent {\it e-mail:} elceyhan@ku.edu.tr (E.~Ceyhan) }


\newpage


\section{Introduction}
\label{sec:intro}
Multivariate clustering patterns such as segregation or association.
result from multivariate interaction between two or more classes (or species).
Such patterns are of interest in ecological sciences and other application areas.
See, for example, \cite{pielou:1961}, \cite{whipple:1980}, and \cite{dixon:1994, dixon:NNCTEco2002}.
For convenience and generality,
we refer to the different types of points as ``classes",
but the class can stand for any characteristic of an observation at
a particular location.
For example, the spatial segregation pattern
has been investigated for \emph{species} (\cite{diggle:2003}),
\emph{age classes} of plants (\cite{hamill:1986}),
\emph{fish species} (\cite{herler:2005}),
and \emph{sexes} of dioecious plants (\cite{nanami:1999}).
Many of the epidemiological applications are for a two-class system of
case and control labels (\cite{waller:2004}).
These methods can also be applied to social and ethnic segregation of residential areas.
For simplicity,
we discuss the spatial interaction between two and three classes only;
the extension to the case with more classes is straightforward.
The null pattern is usually one of the two (random) pattern
types: \emph{random labeling} (RL) or \emph{complete spatial randomness} (CSR).
We consider two major types of spatial clustering patterns as alternatives:
\emph{segregation} and \emph{association}.
Segregation (association) occurs when objects of a given class have NNs
that are more (less) frequently of the same (other) class than
would be expected if there were randomness in the NN structure.


In statistical and other literature,
many univariate and multivariate
spatial clustering tests have been proposed (\cite{kulldorff:2006}).
These include comparison of Ripley's $K(t)$ and $L(t)$ functions (\cite{ripley:2004}),
comparison of nearest neighbor (NN) distances (\cite{cuzick:1990}, \cite{diggle:2003}),
and analysis of nearest neighbor contingency tables (NNCTs)
which are constructed using the NN frequencies of classes (\cite{pielou:1961}, \cite{meagher:1980}).
\cite{pielou:1961} proposed various tests and \cite{dixon:1994} introduced an
overall test of segregation, cell-specific and class-specific tests based on NNCTs
in a two-class setting and extended his tests to
multi-class case in (\cite{dixon:NNCTEco2002}).

In this article, we introduce new overall and cell-specific tests of segregation
based on NNCTs for testing spatial clustering patterns in a multi-class setting.
We compare these tests with Dixon's NNCT-tests
which are introduced for testing against the RL of points (\cite{dixon:1994}).
We extend the use of these tests for the CSR independence pattern also.
We also compare the NNCT-tests with Ripley's $K$ or $L$-functions
and pair correlation function $g(t)$ (\cite{stoyan:1994}),
which are methods for second-order analysis of the point pattern.
We only consider \emph{completely mapped data};
i.e., for our data sets, the locations of all events in a defined area are observed.
We show through simulation that Dixon's cell-specific test can
have undesirable properties in some situations.
The newly proposed cell-specific tests perform
better (in terms of empirical size and power) than Dixon's cell-specific tests.
Likewise the new overall test tends to have
higher power compared to Dixon's overall test under segregation of the classes.
Furthermore,
we demonstrate that NNCT-tests and Ripley's $L$-function (and related methods)
answer different questions about the pattern of interest.

We provide the null and alternative patterns in Section \ref{sec:null-alt},
describe the NNCTs in Section \ref{sec:NNCT},
provide the cell-specific tests in Section \ref{sec:cell-spec},
overall tests in Section \ref{sec:overall},
empirical significance levels in the two- and three-class cases in Sections \ref{sec:CSR-emp-sign-2Cl}
and \ref{sec:CSR-emp-sign-3Cl}, respectively,
rejection rates of the tests under various Poisson
processes in Section \ref{sec:other-process},
empirical power comparisons under the segregation and association alternatives
in the two-class case in Section \ref{sec:emp-power-2Cl},
in the three-class case in Section \ref{sec:emp-power-3Cl},
examples in Section \ref{sec:examples},
and our conclusions and guidelines for using the tests in Section \ref{sec:disc-conc}.

\section{Null and Alternative Patterns}
\label{sec:null-alt}
In this section, for simplicity,
we describe the spatial point patterns for two classes only;
the extension to multi-class case is straightforward.

In the univariate spatial point pattern analysis,
the null hypothesis is usually \emph{complete spatial randomness (CSR)} (\cite{diggle:2003}).
Given a spatial point pattern $\mathcal P=\{X_i(D), i=1,\ldots,n: D \subset \mathbb{R}^2 \}$
where $X_i(D)$ is the Bernoulli random variable denoting the event that
point $i$ is in region $D$.
The pattern $\mathcal P$ exhibits CSR if
given $n$ events in domain $D$, the events are an independent
random sample from a uniform distribution on $D$.
This implies there is no spatial interaction, i.e., the locations of these points have
no influence on one another.
Furthermore, when the reference region $D$ is large,
the number of points in any planar region with area
$A(D)$ follows (approximately) a Poisson distribution with
intensity $\lambda$ and mean $\lambda \cdot A(D)$.

To investigate the spatial interaction between two or more classes
in a multivariate process, usually there are two benchmark hypotheses:
(i) \emph{independence}, which implies two classes of points are generated by a pair of independent
univariate processes and
(ii) \emph{random labeling} (RL), which implies that the class labels
are randomly assigned to a given set of locations in the region of interest (\cite{diggle:2003}).
In this article, we will consider two random pattern types as our null hypotheses:
CSR of points from two classes (this pattern will be called the \emph{CSR independence},
henceforth) or RL.
In the CSR independence pattern,
points from each of the two classes independently satisfy the CSR pattern
in the region of interest.
On the other hand, {\em random labeling} (RL) is the pattern in which,
given a fixed set of points in a region,
class labels are assigned to these fixed
points randomly so that the labels are independent of the locations.
So RL is less restrictive than CSR independence.
CSR independence is a process defining the spatial distribution
of event locations, while RL is a process, conditioned on locations,
defining the distribution of labels on these locations.

Our null hypothesis is
$$H_o: \text{randomness in the NN structure}.$$
Although RL and CSR independence are not same,
they lead to the same null model in tests using NNCT,
which does not require spatially-explicit information.
That is, when the points from two classes are assumed to be independently uniformly
distributed over the region of interest, i.e., under the CSR independence pattern,
or
when only the labeling (or marking) of a set of fixed points,
where the allocation of the points might be regular, aggregated, or clustered,
or of lattice type, is considered, i.e., under the RL pattern,
there is randomness in the NN structure.
The distinction between RL and CSR independence is very important
when defining the appropriate null model in practice,
i.e., the null model depends on the particular ecological context.
\cite{goreaud:2003} state that CSR independence implies that
the two classes are \emph{a priori}
the result of different processes (e.g., individuals of different species or age cohorts),
whereas RL implies that some processes affect \emph{a posteriori}
the individuals of a single population
(e.g., diseased vs. non-diseased individuals of a single species).
We provide the differences in the proposed tests
under these two patterns.

We treat CSR independence or RL as the main null model of interest,
since this is the logical point of departure (\cite{diggle:2003}).
However, in the ecological and epidemiological settings,
CSR independence is the exception rather than rule.
Furthermore, it is conceivable for other models to
imply randomness in the NN structure also.
We also consider patterns that deviate from
stationarity or homogeneity in the point process.
In particular, we consider various types of Poisson cluster processes (\cite{diggle:2003})
and other inhomogeneous Poisson processes (\cite{baddeley:2005}).
Randomness in the NN structure will hold if both classes independently
follow the same process with points having the same support.
For example, in a Poisson cluster process,
NN structure will be random if parents are the same for each class.
If classes have different parent sets,
then the Poisson cluster process will  imply segregation of the classes.
If parent and offspring sets are treated as two different classes,
then Poisson cluster process will imply association of the two classes.
Further, if the two classes are from the same inhomogeneous Poisson pattern,
again randomness in the NN structure will follow.
But when the two classes follow different inhomogeneous Poisson patterns
whose point intensities differ in space,
it might imply the segregation or association of the classes.

As clustering alternatives, we consider two major types of spatial patterns:
\emph{segregation} and \emph{association}.
{\em Segregation} occurs if the
NN of an individual is more likely to be of the same
class as the individual than to be from a different class;
i.e., the members of the same class tend to be clumped or clustered
(see, e.g., \cite{pielou:1961}).
For instance, one type of plant might not grow well
around another type of plant and vice versa.
In plant biology, one class of points might represent
the coordinates of trees from a species with large canopy,
so that other plants (whose coordinates are the other class of points)
that need light cannot grow (well or at all) around these trees.
See, for instance, (\cite{dixon:1994}; \cite{coomes:1999}).

{\em Association} occurs if the NN of an individual is more
likely to be from another class than to be of the same class as the individual.
For example, in plant biology, the two classes of points
might represent the coordinates of mutualistic plant species,
so the species depend on each other to survive.
As another example, one class of points
might be the geometric coordinates of
parasitic plants exploiting  the other plant
whose coordinates are of the other class.
In epidemiology, one class of points might
be the geographical coordinates of residences of cases
and the other class of points might be the coordinates of the residences of controls.

Each of the two patterns of segregation and association are not symmetric in the sense that,
when two classes are segregated (or associated), they do not necessarily
exhibit the same degree of segregation (or association).
For example, when points from each of two classes labeled as $X$ and $Y$
are clustered at different locations,
but class $X$ is loosely clustered (i.e., its point intensity in the clusters is smaller)
compared to class $Y$.
So classes $X$ and $Y$ are segregated but class $Y$
is more segregated than class $X$.
Similarly, when class $Y$ points are clustered around class $X$ points
but not vice versa,
classes $Y$ and $X$ are associated, but class $Y$ is more associated with class $X$
compared to the other way around.
Many different forms of segregation (and association) are possible.
Although it is not possible to list all
segregation types, its existence can be tested by an analysis of the
NN relationships between the classes (\cite{pielou:1961}).

\section{Nearest Neighbor Contingency Tables}
\label{sec:NNCT}
NNCTs are constructed using the NN frequencies of classes.
We describe the construction of NNCTs for two classes; extension to multi-class
case is straightforward.
Consider two classes with labels $\{1,2\}$.
Let $N_i$ be the number of points from class $i$ for $i \in \{1,2\}$ and
$n$ be the total sample size, so $n=N_1+N_2$.
If we record the class
of each point and the class of its NN, the NN
relationships fall into four distinct categories:
$(1,1),\,(1,2);\,(2,1),\,(2,2)$ where in cell $(i,j)$, class $i$ is
the \emph{base class}, while class $j$ is the class of its
\emph{NN}.
That is, the $n$ points constitute $n$
(base, NN) pairs.
Then each pair can be categorized with respect to
the base label (row categories) and NN label (column categories).
Denoting $N_{ij}$ as the frequency of cell $(i,j)$ for $i,j \in
\{1,2\}$, we obtain the NNCT in Table \ref{tab:NNCT-2x2} where $C_j$
is the sum of column $j$; i.e., number of times class $j$ points
serve as NNs for $j \in \{1,2\}$.
Furthermore, $N_{ij}$ is the cell count for
cell $(i,j)$ that is the count of all (base, NN) pairs each of which
has label $(i,j)$.
Note also that
$n=\sum_{i,j}N_{ij}$; $n_i=\sum_{j=1}^2\, N_{ij}$; and
$C_j=\sum_{i=1}^2\, N_{ij}$.
By construction,
if $N_{ij}$ is larger than expected,
then class $j$ serves as NN more frequently to class $i$ than expected,
which implies segregation if $i=j$ and
association of class $j$ with class $i$ if $i\not=j$.
On the other hand, if $N_{ij}$ is smaller than expected,
then class $j$ serves as NN less frequently to class $i$ than expected,
which implies lack of segregation if $i=j$ and
lack of association of class $j$ with class $i$ if $i\not=j$.
Furthermore, we adopt the convention that variables denoted by upper case letters
are random quantities, while variables denoted by lower case letters
fixed quantities.
Hence, column sums and cell counts are random, while row sums and the overall sum are fixed
quantities in a NNCT.

\begin{table}[ht]
\centering
\begin{tabular}{cc|cc|c}
\multicolumn{2}{c}{}& \multicolumn{2}{c}{NN class}& \\
\multicolumn{2}{c|}{}& class 1 &  class 2 & sum  \\
\hline
&class 1 &    $N_{11}$  &   $N_{12}$  &   $n_1$  \\
\raisebox{1.5ex}[0pt]{base class}
&class 2 &    $N_{21}$ &  $N_{22}$    &   $n_2$  \\
\hline
& sum    &    $C_1$   & $C_2$         &   $n$  \\
\end{tabular}
\caption{
\label{tab:NNCT-2x2}
The NNCT for two classes.}
\end{table}


\cite{pielou:1961} used Pearson's $\chi^2$ test of
independence for testing segregation.
Due to the ease in computation and interpretation, Pielou's test of
segregation is used frequently (\cite{meagher:1980}) for
both completely mapped and sparsely sampled data.
For example, Pielou's test is used for the segregation
between males and females in dioecious species (see, e.g., \cite{herrera:1988}
and \cite{armstrong:1989}) and between different species (\cite{good:1982}).
However Pielou's test is not appropriate for completely mapped data
(\cite{meagher:1980}, \cite{dixon:1994}),
since the $\chi^2$ test of independence requires independence between cell-counts
(and rows and columns also), which is violated under RL or CSR independence.
In fact, this independence between cell-counts is
violated for spatial data in general
and in particular it is violated under the null patterns,
so Pielou's test is not of the desired size.
This problem was first noted by \cite{meagher:1980} who identify the main source
of it to be reflexivity of (base, NN) pairs.
A (base, NN) pair $(X,Y)$ is \emph{reflexive} if $(Y,X)$ is also a (base, NN) pair.
As an alternative,
they suggest using Monte Carlo simulations for Pielou's test.
\cite{dixon:1994} derived the appropriate asymptotic sampling
distribution of cell counts using Moran join count statistics (\cite{moran:1948}) and
hence the appropriate test which also has a $\chi^2$-distribution asymptotically.
\cite{dixon:1994} also states that
although Pielou's test is not appropriate for completely mapped data,
it may be appropriate for sparsely sampled data.

\section{Cell-Specific Tests of Segregation}
\label{sec:cell-spec}

In this section, we describe Dixon's cell-specific test of segregation
and introduce a new type of cell-specific test based on NNCTs.

\subsection{Dixon's Cell-Specific Tests of Segregation}
\label{sec:dix-cell-spec}
The level of segregation can be estimated by
comparing the observed cell counts to the expected cell counts
under RL of points whose locations are fixed or a realization of points from CSR independence.
Dixon demonstrates that under RL,
one can write down the cell frequencies as Moran join count statistics (\cite{moran:1948}).
He then derives the means, variances, and covariances of
the cell counts (i.e., frequencies) (see, \cite{dixon:1994} and \cite{dixon:NNCTEco2002}).

Under RL, the expected cell count for cell $(i,j)$ is
\begin{equation}
\label{eqn:Exp[Nij]}
\E[N_{ij}]=
\begin{cases}
n_i(n_i-1)/(n-1) & \text{if $i=j$,}\\
n_i\,n_j/(n-1)     & \text{if $i \not= j$,}
\end{cases}
\end{equation}
where $n_i$ is a realization of $N_i$, i.e., is the fixed sample size for
class $i$ for $i=1,2,\ldots,q$.
Observe that the expected cell counts depend only on the size of
each class (i.e., row sums), but not on column sums.

The test statistic suggested by Dixon is given by
\begin{equation}
\label{eqn:dixon-Zij}
Z^D_{ij}=\frac{N_{ij}-\E[N_{ij}]}{\sqrt{\Var[N_{ij}]}},
\end{equation}
where
{\small
\begin{equation}
\label{eqn:VarNij}
\Var[N_{ij}]=
\begin{cases}
(n+R)\,p_{ii}+(2\,n-2\,R+Q)\,p_{iii}+(n^2-3\,n-Q+R)\,p_{iiii}-(n\,p_{ii})^2 & \text{if $i=j$,}\\
n\,p_{ij}+Q\,p_{iij}+(n^2-3\,n-Q+R)\,p_{iijj} -(n\,p_{ij})^2             & \text{if $i \not= j$,}
\end{cases}
\end{equation}
}
\noindent
with $p_{xx}$, $p_{xxx}$, and $p_{xxxx}$
are the probabilities that a randomly picked pair,
triplet, or quartet of points, respectively, are the indicated classes and
are given by
\begin{align}
\label{eqn:probs}
p_{ii}&=\frac{n_i\,(n_i-1)}{n\,(n-1)},  & p_{ij}&=\frac{n_i\,n_j}{n\,(n-1)},\nonumber\\
p_{iii}&=\frac{n_i\,(n_i-1)\,(n_i-2)}{n\,(n-1)\,(n-2)}, &
p_{iij}&=\frac{n_i\,(n_i-1)\,n_j}{n\,(n-1)\,(n-2)},\\
p_{iiii}&=\frac{n_i\,(n_i-1)\,(n_i-2)\,(n_i-3)}{n\,(n-1)\,(n-2)\,(n-3)}, &
p_{iijj}&=\frac{n_i\,(n_i-1)\,n_j\,(n_j-1)}{n\,(n-1)\,(n-2)\,(n-3)}.\nonumber
\end{align}
Furthermore, $R$ is twice the number of reflexive pairs
and $Q$ is the number of points with shared  NNs,
which occurs when two or more
points share a NN.
Then $Q=2\,(Q_2+3\,Q_3+6\,Q_4+10\,Q_5+15\,Q_6)$
where $Q_k$ is the number of points that serve
as a NN to other points $k$ times.
Furthermore, under RL $Q$ and $R$ are fixed quantities,
as they depend only on the location of the points, not the types of NNs.
So the sampling distribution is appropriate under RL (see also Remark \ref{rem:QandR})
and $Z^D_{ii}$ asymptotically has $N(0,1)$ distribution.
But unfortunately, for $q>2$ the asymptotic normality of the off-diagonal
cells in NNCTs is not rigorously established yet,
although extensive Monte Carlo simulations indicate approximate
normality for large samples (\cite{dixon:NNCTEco2002}).
One-sided and two-sided tests are possible for each cell $(i,j)$
using the asymptotic normal approximation of $Z^D_{ij}$ given in
Equation (\ref{eqn:dixon-Zij}) (\cite{dixon:1994}).

Under CSR independence, the quantities $Q$ and $R$ are random,
hence the sampling distributions of the cell counts are conditional on
these quantities.
Hence the expected cell counts in \eqref{eqn:Exp[Nij]} and
the cell-specific test in \eqref{eqn:dixon-Zij}
and the relevant discussion are similar to the RL case.
The only difference is that under RL, the quantities $Q$ and $R$
are fixed, while under CSR independence they are random.
That is, under CSR independence, $Z^D_{ij}$ asymptotically has $N(0,1)$
distribution conditional on $Q$ and $R$.

\subsection{A New Cell-Specific Test of Segregation}
\label{sec:new-cell-spec}
In standard cases like multinomial sampling with fixed
row totals and conditioning on the column totals,
the expected cell count for cell $(i,j)$ in contingency tables
is $\E[N_{ij}]=\frac{N_i\,C_j}{n}$.
We first consider the difference $\Delta_{ij}:=N_{ij}-\frac{N_i\,C_j}{n}$ for cell $(i,j)$.
Notice that under RL, $N_i=n_i$ are fixed,
but $C_j$ are random quantities and $C_j=\sum_{i=1}^q N_{ij}$,
hence
$$\Delta_{ij}=N_{ij}-\frac{n_i\,C_j}{n}.$$
Then under RL,
\begin{equation}
\E[\Delta_{ij}]=
\begin{cases}
\frac{n_i(n_i-1)}{(n-1)}-\frac{n_i}{n}\,\E[C_j] & \text{if $i=j$,}\\
\frac{n_i\,n_j}{(n-1)}-\frac{n_i}{n}\,\E[C_j]  & \text{if $i \not= j$.}
\end{cases}
\end{equation}

\noindent
For all $j$, $\E[C_j]=n_j$, since
\begin{multline*}
\E[C_j]=\sum_{i=1}^q \E[N_{ij}]=
\frac{n_j(n_j-1)}{(n-1)}+\sum_{i\neq j} \frac{n_i n_j}{(n-1)}
=\frac{n_j(n_j-1)}{(n-1)}+\frac{n_j}{(n-1)}\sum_{i\neq j} n_i\\
=\frac{n_j(n_j-1)}{(n-1)}+\frac{n_j}{(n-1)}(n-n_j)=n_j.
\end{multline*}

\noindent
Therefore,
\begin{equation}
\label{eqn:Exp[Deltaij]}
\E[\Delta_{ij}]=
\begin{cases}
\frac{n_i(n_i-1)}{(n-1)}-\frac{n_i^2}{n} & \text{if $i=j$,}\\
\frac{n_i\,n_j}{(n-1)}-\frac{n_i\,n_j}{n}  & \text{if $i \not= j$.}
\end{cases}
\end{equation}

\noindent
Notice that the expected value of $\Delta_{ij}$ is not zero under RL.
Hence, instead of $\Delta_{ij}$,
in order to obtain 0 expected value for our test statistic,
we suggest the following:
\begin{equation}
\label{eqn:Tij}
T_{ij}=
\begin{cases}
N_{ij}-\frac{(n_i-1)}{(n-1)}C_j & \text{if $i=j$,}\\
N_{ij}-\frac{n_i}{(n-1)}C_j     & \text{if $i \not= j$.}
\end{cases}
\end{equation}
Then $\E[T_{ij}]=0$, since,
for $i=j$,
$$
\E[T_{ii}]=\E[N_{ii}]-\frac{(n_i-1)}{(n-1)}\E[C_i]=
\frac{n_i(n_i-1)}{(n-1)}-\frac{(n_i-1)}{(n-1)}n_i=0,
$$
and
for $i \neq j$,
$$
\E[T_{ij}]=\E[N_{ij}]-\frac{(n_i-1)}{(n-1)}\E[C_j]=
\frac{n_i\,n_j}{(n-1)}-\frac{(n_i-1)}{(n-1)}n_j=0.
$$

For the variance of $T_{ij}$,
we have
\begin{equation}
\label{eqn:Var[Tij]}
\Var[T_{ij}]=
\begin{cases}
\Var[N_{ij}]+\frac{(n_i-1)^2}{(n-1)^2}\Var[C_j]-2\frac{(n_i-1)}{(n-1)}\Cov[N_{ij},C_j]
& \text{if $i=j$,} \vspace{.05 in}\\
\Var[N_{ij}]+\frac{n_i^2}{(n-1)^2}\Var[C_j]-2\frac{n_i}{(n-1)}\Cov[N_{ij},C_j] & \text{if $i \not= j$,}
\end{cases}
\end{equation}
where
$\Var[N_{ij}]$ are as in Equation \eqref{eqn:VarNij},
$\Var[C_j]=\sum_{i=1}^q \Var[N_{ij}]+\sum_{k \neq i}\sum_i \Cov[N_{ij},N_{kj}]$
and
$\Cov[N_{ij},C_j]=\sum_{k=1}^q \Cov[N_{ij},N_{kj}]$
with $\Cov[N_{ij},N_{kl}]$ are as in Equations (4)-(12) of \cite{dixon:NNCTEco2002}.

As a new cell-specific test, we propose
\begin{equation}
\label{eqn:new-Zij}
Z_{ij}^N=\frac{T_{ij}}{\sqrt{\Var[T_{ij}]}}.
\end{equation}

Recall that in the two-class case,
each cell count $N_{ij}$ has asymptotic normal distribution (\cite{cuzick:1990}).
Hence, $Z_{ij}^N$ also converges in law to $N(0,1)$ as $n\rightarrow \infty$.
Moreover, one and two-sided versions of this test are also possible.

Under CSR independence, the distribution of the test statistics above
is similar to the RL case.
The only difference is that $Z_{ij}^N$ asymptotically has
$N(0,1)$ distribution conditional on $Q$ and $R$.

\section{Overall Tests of Segregation}
\label{sec:overall}
In this section, we describe Dixon's overall test of segregation
and introduce a new overall test based on NNCTs.

\subsection{Dixon's Overall Test of Segregation}
\label{sec:dix-overall}
In the multi-class case with $q$ classes,
combining the $q^2$ cell-specific tests in Section \ref{sec:dix-cell-spec},
\cite{dixon:NNCTEco2002} suggests the quadratic
form to obtain the overall segregation test as follows.
\begin{equation}
\label{eqn:dix-chisq-qxq}
C_D=(\mathbf{N}-\E[\mathbf{N}])'\Sigma_D^-(\mathbf{N}-\E[\mathbf{N}])
\end{equation}
where $\mathbf{N}$ is the $q^2\times1$ vector of $q$ rows
of NNCT concatenated row-wise,
$\E[\mathbf{N}]$ is the vector of $\E[N_{ij}]$ which are as in Equation (\ref{eqn:Exp[Nij]}),
$\Sigma_D$ is the $q^2 \times q^2$ variance-covariance matrix for
the cell count vector $\mathbf{N}$ with diagonal entries equal to
$\Var[N_{ii}]$ and off-diagonal entries being $\Cov[N_{ij},\,N_{kl}]$
for $(i,j) \neq (k,l)$.
The explicit forms of the variance and covariance terms are provided in (\cite{dixon:NNCTEco2002}).
Also, $\Sigma_D^-$ is a generalized inverse of $\Sigma_D$ (\cite{searle:2006})
and $'$ stands for the transpose of a vector or matrix.
Then under RL $C_D$ has a $\chi^2_{q(q-1)}$ distribution asymptotically.
Furthermore, the test statistics $Z_{ij}^D$ are dependent,
hence their squares do not sum to $C_N$.
Under CSR independence, the distribution of $C_D$
is conditional on $Q$ and $R$.

\subsection{A New Overall Test of Segregation}
\label{sec:new-overall}
Instead of combining the cell-specific tests in Section \ref{sec:dix-cell-spec},
we can also combine the new cell-specific tests in Section \ref{sec:new-cell-spec}.
Let $\mathbf T$ be the vector of $q^2$ $T_{ij}$ values, i.e.,
$$\mathbf T=[T_{11},T_{12},\ldots,T_{1q},T_{21},T_{22},\ldots,T_{2q},\ldots,T_{qq}]',$$
and let $\E[\mathbf T]$ be the vector of $T_{ij}$ values.
Note that $\E[\mathbf T]=\mathbf 0$.
Hence to obtain a new overall segregation test, we use the following quadratic form:
\begin{equation}
\label{eqn:new-overall]}
C_N=\mathbf T'\Sigma_N^-\mathbf T
\end{equation}
where $\Sigma_N$ is the $q^2 \times q^2$ variance-covariance matrix of $\mathbf T$.
Under RL $C_N$ has a $\chi^2_{(q-1)^2}$ distribution asymptotically,
since rank of $\Sigma_N$ is $(q-1)^2$.
Furthermore, the test statistics $Z_{ij}^N$ are dependent,
hence their squares do not sum to $C_N$.

Under RL, the diagonal entries in the variance-covariance matrix $\Sigma_N$ are
$\Var[T_{ij}]$ which are provided in Equation \eqref{eqn:Var[Tij]}.
For the off-diagonal entries in $\Sigma_N$, i.e., $\Cov[T_{ij},T_{kl}]$ with $i \not= k$ and $j \neq l$,
there are four cases to consider:\\
\noindent
\textbf{case 1:} $i=j$ and $k=l$,
then
\begin{multline}
\Cov[T_{ii},T_{kk}]=\Cov \left[ N_{ii}-\frac{(n_i-1)}{(n-1)}C_i,N_{kk}-\frac{(n_k-1)}{(n-1)}C_k \right]=\\
\Cov[N_{ii},N_{kk}]-\frac{(n_k-1)}{(n-1)}\Cov[N_{ii},C_k]-\frac{(n_i-1)}{(n-1)}\Cov[N_{kk},C_i]
+\frac{(n_i-1)(n_k-1)}{(n-1)^2}\Cov[C_i,C_k].
\end{multline}

\noindent
\textbf{case 2:} $i=j$ and $k \neq l$,
then
\begin{multline}
\Cov[T_{ii},T_{kl}]=\Cov \left[N_{ii}-\frac{(n_i-1)}{(n-1)}C_i,N_{kl}-\frac{n_k}{(n-1)}C_l \right]=\\
\Cov[N_{ii},N_{kl}]-\frac{n_k}{(n-1)}\Cov[N_{ii},C_l]-\frac{(n_i-1)}{(n-1)}\Cov[N_{kl},C_i]
+\frac{(n_i-1)n_k}{(n-1)^2}\Cov[C_i,C_l].
\end{multline}

\noindent
\textbf{case 3:} $i \neq j$ and $k = l$,
then
$
\Cov[T_{ij},T_{kk}]=\Cov[T_{kk},T_{ij}],
$
which is essentially \textbf{case 2} above.

\noindent
\textbf{case 4:} $i \neq j$ and $k \neq l$,
then
\begin{multline}
\Cov[T_{ij},T_{kl}]=\Cov \left[ N_{ij}-\frac{n_i}{(n-1)}C_j,N_{kl}-\frac{n_k}{(n-1)}C_l \right]=\\
\Cov[N_{ij},N_{kl}]-\frac{n_k}{(n-1)}\Cov[N_{ij},C_l]-\frac{n_i}{(n-1)}\Cov[N_{kl},C_j]
+\frac{n_i n_k}{(n-1)^2}\Cov[C_j,C_l].
\end{multline}

In all the above cases, $\Cov[N_{ij},N_{kl}]$ are as in \cite{dixon:NNCTEco2002},
$\Cov[N_{ij},C_l]=\sum_{k=1}^q \Cov[N_{ij},N_{kl}]$ and
$\Cov[C_i,C_j]=\sum_{k=1}^q \sum_{l=1}^q \Cov[N_{ki},N_{lj}]$.

Under CSR independence, the distribution of $C_N$
is as in the RL case, except that it is conditional on $Q$ and $R$.

\begin{remark}
\label{rem:dixon-vs-new}
\textbf{Comparison of Dixon's and New NNCT-Tests:}
Dixon's cell-specific test in \eqref{eqn:dixon-Zij} depends
on the frequencies of (base, NN) pairs (i.e., cell counts),
and measures deviations from expected cell counts.
On the other hand, the new cell-specific test in \eqref{eqn:new-Zij}
can be seen as a difference of two statistics and has expected
value is 0 for each cell.
For the cell-specific tests,
the $z$-score for cell $(i,j)$ indicates the level and direction
of the interaction of spatial patterns of base class $i$ and NN class $j$.
If $Z^D_{ii}>0$ then class $i$ exhibits segregation from other classes,
and if $Z^D_{ii}<0$ then class $i$ exhibits lack of segregation from other classes.
The same holds for the new cell-specific tests.
Furthermore, cell-specific test for cell $(i,j)$
measures the interaction of class $j$ with class $i$.
When $i=j$ this interaction is the segregation for class $i$,
but if $i \neq j$, it is the association of class $j$ with class $i$.
Hence for $i \neq j$ cell-specific test for cell $(i,j)$ is not symmetric,
as interaction of class $j$ with class $i$ could be different from
the interaction of class $i$ with class $j$.
However, new cell-specific tests use more of the information in the NNCT
compared to Dixon's tests,
hence they potentially will have better performance
in terms of size and power.

Dixon's overall test combines Dixon's cell-specific tests
in one compound summary statistic, while new overall test combines
the new cell-specific tests.
Hence the new overall test might have better performance in terms of size and power,
as it depends on the new cell-specific tests.
$\square$
\end{remark}

\begin{remark}
\label{rem:QandR}
\textbf{The Status of $Q$ and $R$ under RL and CSR Independence:}
Under RL, $Q$ and $R$ are fixed quantities,
but under CSR independence they are random.
The variances and covariances $\Var[N_{ij}]$ and $\Cov[N_{ij},N_{kl}]$
and all the quantities depending on these
quantities also depend on $Q$ and $R$.
Hence under CSR independence,
they are variances and covariances conditional on $Q$ and $R$.
The unconditional variances and covariances can be obtained
by replacing $Q$ and $R$ by their expectations.

Unfortunately, given the difficulty of calculating the
expectations of $Q$ and $R$ under CSR independence,
it is reasonable and convenient to use test statistics employing the
conditional variances and covariances even when assessing their
behavior under CSR independence.
Alternatively, one can estimate the expected values of $Q$ and $R$ empirically
and substitute these estimates in the expressions.
For example,
for the homogeneous planar Poisson process,
we have $\E[Q/n] \approx .632786$ and $\E[R/n] \approx 0.621120$.
(estimated empirically by 1000000 Monte Carlo simulations
for various values of $n$ on unit square).
$\square$
\end{remark}

\subsection{The Two-Class Case}
\label{sec:two-class}
In the two-class case, \cite{dixon:1994} calculates
$Z_{ii}=(N_{ii}-\E[N_{ii}])\big /\sqrt{\Var[N_{ii}]}$ for both $i \in
\{1,2\}$ and then combines these test statistics into a statistic
that is equivalent to $C_D$ in Equation \eqref{eqn:dix-chisq-qxq}
and asymptotically distributed as $\chi^2_2$.
The suggested test statistic is
{\small
\begin{equation}
\label{eqn:dix-chisq-2x2}
C_D=\mathbf{Y}'\Sigma^{-1}\mathbf{Y}=
\left[
\begin{array}{c}
N_{11}-\E[N_{11}] \\
N_{22}-\E[N_{22}]
\end{array}
\right]' \left[
\begin{array}{cc}
\Var[N_{11}] & \Cov[N_{11},N_{22}] \\
\Cov[N_{11},N_{22}] & \Var[N_{22}] \\
\end{array}
\right]^{-1} \left[
\begin{array}{c}
N_{11}-\E[N_{11}] \\
N_{22}-\E[N_{22}]
\end{array}
\right]
\end{equation}

Notice that this is also equivalent to $C=\frac{Z_{AA}^2+Z_{BB}^2-2\,r\,Z_{AA}Z_{BB}}{1-r^2}$
where $Z_{AA}=\frac{N_{11}-\E[N_{11}]}{\sqrt{\Var[N_{11}]}}$,
$Z_{BB}=\frac{N_{22}-\E[N_{22}]}{\sqrt{\Var[N_{22}]}}$,
and $r=\Cov[N_{11},N_{22}]\big /\sqrt{\Var[N_{11}]\Var[N_{22}]}$.
Notice that $Z_{AA}=Z^D_{11}$ and $Z_{BB}=Z^D_{22}$.
Furthermore, $C_D$ has a $\chi^2_2$ distribution and
$C_N$ has a $\chi^2_1$ distribution asymptotically.

In the two-class case, segregation of class $i$ from class $j$ implies lack of association
between classes $i$ and $j$ ($i \not= j$)
and lack of segregation of class $i$ from class $j$ implies association between
classes $i$ and $j$ ($i \not= j$),
since $Z^D_{i1}=-Z^D_{i2}$ for $i=1,2$.
Likewise for the new cell-specific tests,
since $Z^N_{1j}=-Z^N_{2j}$ for $j=1,2$.
In the multi-class case,
a positive $z$-score, $Z^D_{ii}$, for the diagonal cell $(i,i)$ indicates segregation,
but it does not necessarily mean lack of association between class $i$
and class $j$ ($i\not=j$), since it could be the case that class $i$
could be associated with one class, yet not associated with another one.
Likewise for the new cell-specific tests.

\begin{remark}
\textbf{Asymptotic Structure for the NNCT-Tests:}
\label{remsec:asymptotics}
There are two major types of asymptotic structures for spatial data (\cite{lahiri:1996}).
In the first, any two observations are required to be at least a fixed distance apart,
hence as the number of observations increase, the region on which the process
is observed eventually becomes unbounded.
This type of sampling structure is called ``increasing domain asymptotics".
In the second type, the region of interest is a fixed
bounded region and more and more points are observed in this region.
Hence the minimum distance between data points tends to zero
as the sample size tends to infinity.
This type of structure is called ``infill asymptotics" due to \cite{cressie:1993}.
The sampling structure in our asymptotic sampling distribution
could be either one of increasing domain or infill asymptotics,
as we only consider the class sizes and hence the total sample size
tending to infinity regardless of the size of the study region.
$\square$
\end{remark}

\section{Empirical Significance Levels in the Two-Class Case}
\label{sec:monte-carlo-2Cl}
In this section, we provide the empirical significance levels
for Dixon's and the new overall and the cell-specific tests
in the two-class case under RL and CSR independence patterns.

\subsection{Empirical Significance Levels under CSR Independence of Two Classes}
\label{sec:CSR-emp-sign-2Cl}
First, we consider the two-class case with classes $X$ and $Y$.
We generate $n_1$ points from class $X$ and $n_2$ points from class $Y$
both of which are independently uniformly distributed on the unit square, $(0,1) \times (0,1)$.
Hence, all $X$ points are independent of each other and so are $Y$ points;
and $X$ and $Y$ are independent data sets.
Thus, we simulate the CSR independence pattern for the performance of the tests
under the null case.
Notice that this will imply randomness in the NN structure,
which is the null hypothesis for our NNCT-tests.
We generate $X$ and $Y$ points for some combinations of $n_1,n_2 \in \{10,30,50,100\}$
and repeat the sample generation $N_{mc}=10000$ times for each sample size combination
in order to obtain sufficient precision of the results in reasonable time.
At each Monte Carlo replication, we construct the NNCT,
then compute the overall and cell-specific tests.
Out of these 10000 samples the number of significant outcomes by each test is recorded.
The nominal significance level used in all these tests is $\alpha=.05$.
The empirical sizes are calculated as
the ratio of number of significant results to the number of Monte
Carlo replications, $N_{mc}$.
For example empirical size for Dixon's overall test, denoted by $\ah_{D}$,
is calculated as $\ah_{D}:=\sum_{i=1}^{N_{mc}} \I(\X^2_{D,i} \ge \chi^2_2(.05))$
where $\X^2_{D,i}$ is the value of Dixon's overall test statistic for iteration $i$,
$\chi^2_2(.05)$ is the $95^{th}$ percentile of $\chi^2_2$ distribution,
and $\I(\cdot)$ is the indicator function.
The empirical sizes for other tests are calculated similarly.

We present the empirical significance
levels for the NNCT-tests in Table \ref{tab:cell-spec-null},
where $\ah_{i,j}^D$ and $\ah_{i,j}^N$ are the empirical significance
levels of Dixon's and the new cell-specific
tests, respectively,
$\ah_D$ is for Dixon's and $\ah_N$ is for the new overall tests of segregation.
The empirical sizes significantly smaller (larger) than .05 are marked with $^c$ ($^{\ell}$),
which indicate that the corresponding test is conservative (liberal).
The asymptotic normal approximation to proportions are used in determining the significance of
the deviations of the empirical sizes from the nominal level of .05.
For these proportion  tests,
we also use $\alpha=.05$ to test against empirical size being equal to .05.
With $N_{mc}=10000$, empirical sizes less than .0464 are deemed conservative,
greater than .0536 are deemed liberal at $\alpha=.05$ level.
Notice that in the two-class case
$\ah_{1,1}^D=\ah_{1,2}^D$ and $\ah_{2,1}^D=\ah_{2,2}^D$,
since $N_{12}=n_1-N_{11}$ and $N_{21}=n_2-N_{22}$.
Notice also that $\ah_{1,1}^N=\ah_{2,1}^N$ and $\ah_{1,2}^N=\ah_{2,2}^N$,
since $T_{11}=-T_{21}$ and $T_{12}=-T_{22}$.
So only $\ah_{1,1}^D$, $\ah_{2,2}^D$, $\ah_{1,1}^N$, $\ah_{2,2}^N$,
$\ah_D$, and $\ah_N$ are presented in Table \ref{tab:cell-spec-null}.
The empirical sizes are also plotted in Figure \ref{fig:emp-size-CSR-2cl}
where the horizontal lines are the nominal level of .05
and upper and lower limits for the empirical size (i.e., .0464 and .0536).

Observe that Dixon's cell-specific test for cell $(1,1)$
(i.e., the diagonal entry with base and NN classes are from the smaller class)
is about the desired level for equal and large samples (i.e., $n_1=n_2 \ge 30$),
is conservative when at least one sample is small (i.e., $n_i \leq 10$),
liberal when sample sizes are large but different (i.e., $30 \le n_1 < n_2$).
It is most conservative for $(n_1,n_2)=(10,50)$.
On the other hand, Dixon's cell-specific test for cell $(2,2)$
(i.e., the diagonal entry with base and NN classes are from the larger class)
is about the desired level for almost all sample size combinations.
For Dixon's cell-specific tests, if at least one sample size is small,
the normal approximation is not appropriate.
\cite{dixon:1994} recommends Monte Carlo randomization
instead of the asymptotic approximation for the corresponding cell-specific tests
when cell counts are less than or equal to 10;
and when some cell counts are less than 5, he recommends
Monte Carlo randomization for the overall test.
When sample sizes are small, $n_i \leq 10$
or large but different $(30 \leq n_1 < n_2)$
it is more likely to have cell count for cell $(1,1)$ to be $< 10$,
however for cell $(2,2)$ cell counts
are usually much larger than 10, hence normal approximation is
more appropriate for cell $(2,2)$.

The new cell-specific tests yield very similar empirical sizes for both
cells $(1,1)$ and $(2,2)$ and are both conservative
when $n_1 \leq 30$ and about the desired level otherwise.
However, new cell-specific test for cell $(1,1)$ is less conservative than
that of Dixon's, since $T_{11}$ is less likely to be small
because it also depends on the column sum.

Dixon's overall test is about the desired level
for equal and large samples (i.e., $n_1=n_2 \geq 30$),
is conservative when at least one sample is small (i.e., $n_i \leq 10$),
liberal when sample sizes are large but different (i.e., $30 \le n_1 < n_2$).
It is most conservative for $(n_1,n_2)=(10,50)$.
The new overall segregation test is conservative for small samples
and has the desired level for moderate to large samples.

Moreover, we not only vary samples size but also the
relative abundance of the classes in our simulation study.
The differences in the relative abundance of classes seem to affect
Dixon's tests more than the new tests.
See for example cell-specific tests for cell $(1,1)$ for sample sizes
$(30,50)$ and $(50,100)$,
where Dixon's test suggests that class $X$ (i.e., class with the smaller size)
is more segregated which is only an artifact of the difference in the relative abundance.
Likewise, Dixon's overall test seems to be affected more by the
differences in the relative abundance.
On the other hand, the new tests are more robust to differences
in the relative abundance, since they depend on both row and column sums.

Thus we conclude that Type I error rates of the new overall and cell-specific tests are
more robust to the differences in sample sizes.
Furthermore, the new tests for cells $(1,1)$ and $(2,2)$
and the new overall test exhibit very similar behavior under CSR independence.
Dixon's cell-specific test for cell $(2,2)$ is closest to the desired level.

\begin{table}[ht]
\centering
\begin{tabular}{|c||c|c||c|c||c|c|}
\hline
\multicolumn{7}{|c|}{Empirical significance levels under CSR independence} \\
\hline
sizes & \multicolumn{2}{|c||}{Dixon's} & \multicolumn{2}{|c||}{New}
& \multicolumn{2}{c|}{Overall} \\
\hline
$(n_1,n_2)$  & $\ah_{1,1}^D$ & $\ah_{2,2}^D$ & $\ah_{1,1}^N$ & $\ah_{2,2}^N$ & $\ah_D$ & $\ah_N$ \\
\hline
(10,10) & .0454$^c$ & .0465 & .0452$^c$ & .0459$^c$ & .0432$^c$ & .0484 \\
\hline
(10,30) & .0306$^c$ & .0485 & .0413$^c$ & .0420$^c$ & .0440$^c$ & .0434$^c$ \\
\hline
(10,50) & .0270$^c$ & .0464 & .0390$^c$ & .0396$^c$ & .0482 & .0408$^c$ \\
\hline
(30,30) & .0507 & .0505 & .0443$^c$ & .0442$^c$ & .0464 & .0453$^c$ \\
\hline
(30,50) & .0590$^\ell$ & .0522 & .0505 & .0510 & .0443$^c$ & .0512 \\
\hline
(50,50) & .0465 & .0469 & .0500 & .0502 & .0508 & .0506 \\
\hline
(50,100) & .0601$^\ell$  & .0533 & .0514 & .0515 & .0560$^{\ell}$ & .0525 \\
\hline
(100,100) & .0493  & .0463$^c$ & .0485 & .0486 & .0504 & .0489 \\
\hline
\end{tabular}
\caption{
\label{tab:cell-spec-null}
The empirical significance levels for Dixon's and new cell-specific and overall tests
in the two-class case under $H_o:CSR~~independence$ with
$N_{mc}=10000$, $n_1,n_2$ in $\{10,30,50,100\}$ at the nominal level of $\alpha=.05$.
$^c$: empirical size significantly less than .05; i.e., the test is conservative.
$^{\ell}$: empirical size significantly larger than .05; i.e., the test is liberal.
$\ah_{i,i}^D$ and $\ah_{i,i}^N$ are for the empirical significance
levels of Dixon's and the new cell-specific tests, respectively, for $i=1,2$;
$\ah_D$ is for Dixon's and $\ah_N$ is for the new overall tests of segregation.
}
\end{table}

\begin{figure}
\centering
Empirical Size Plots for the NNCT-Tests under CSR Independence of Two Classes\\
\rotatebox{-90}{ \resizebox{2. in}{!}{\includegraphics{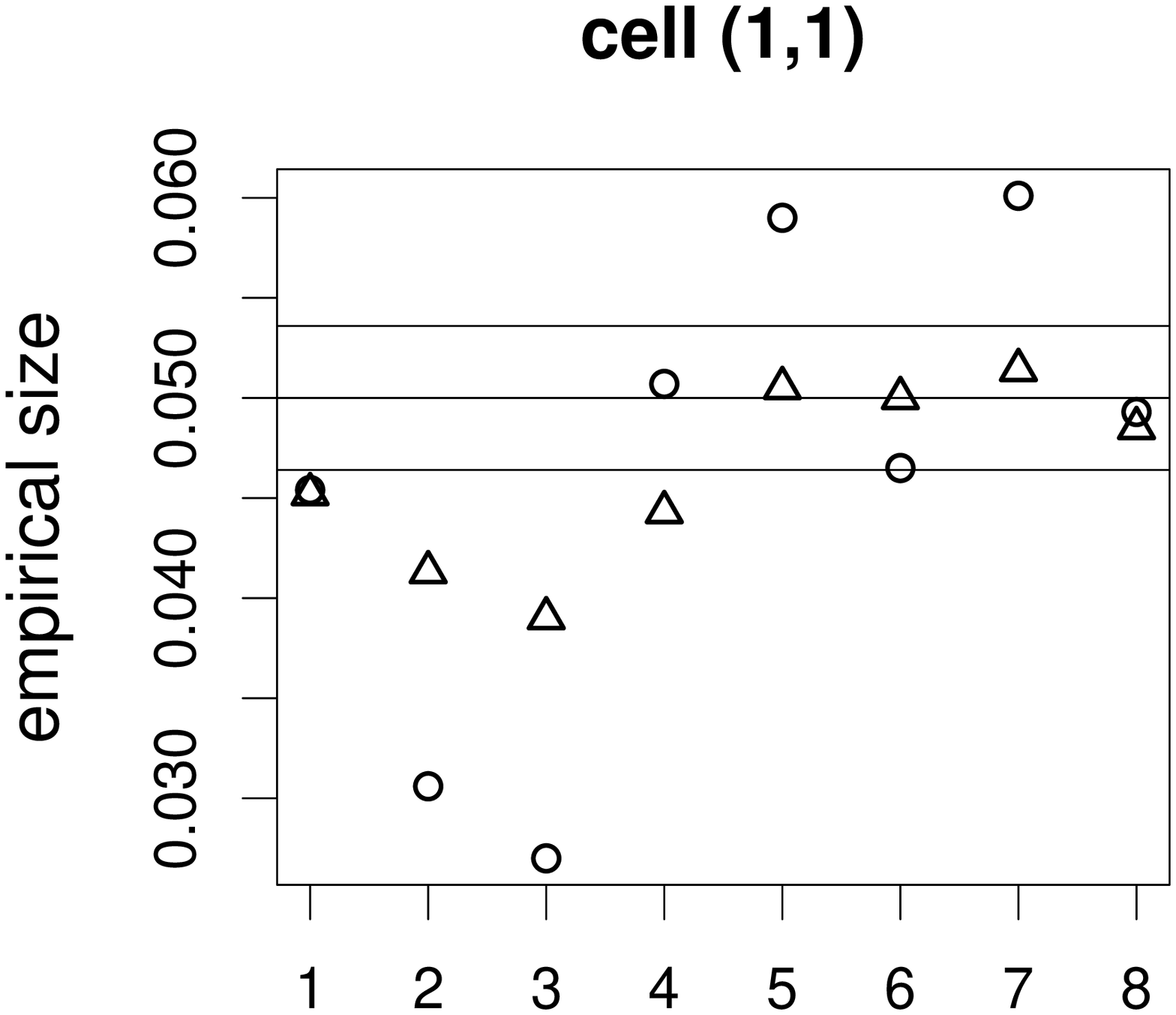} }}
\rotatebox{-90}{ \resizebox{2. in}{!}{\includegraphics{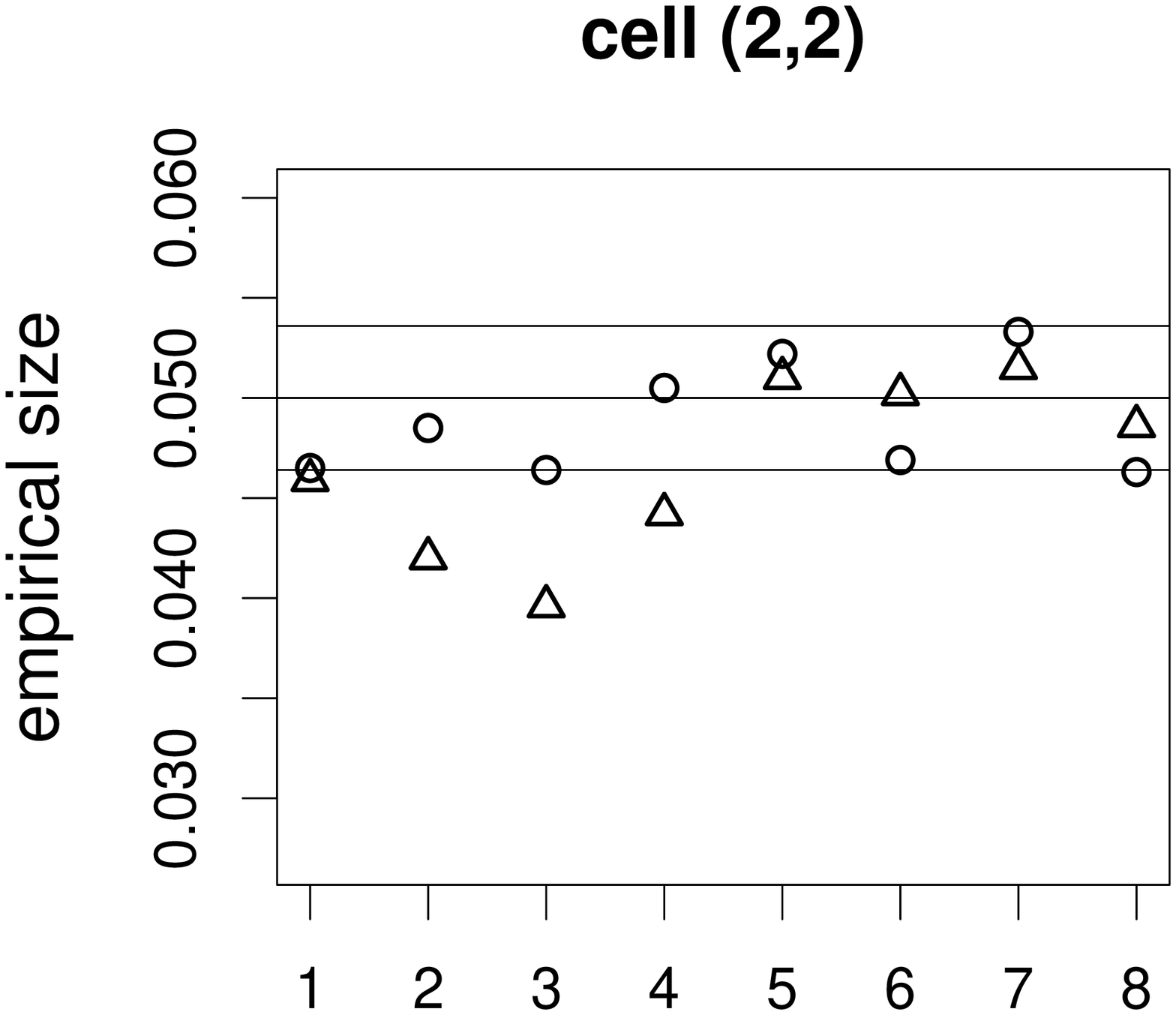} }}
\rotatebox{-90}{ \resizebox{2. in}{!}{\includegraphics{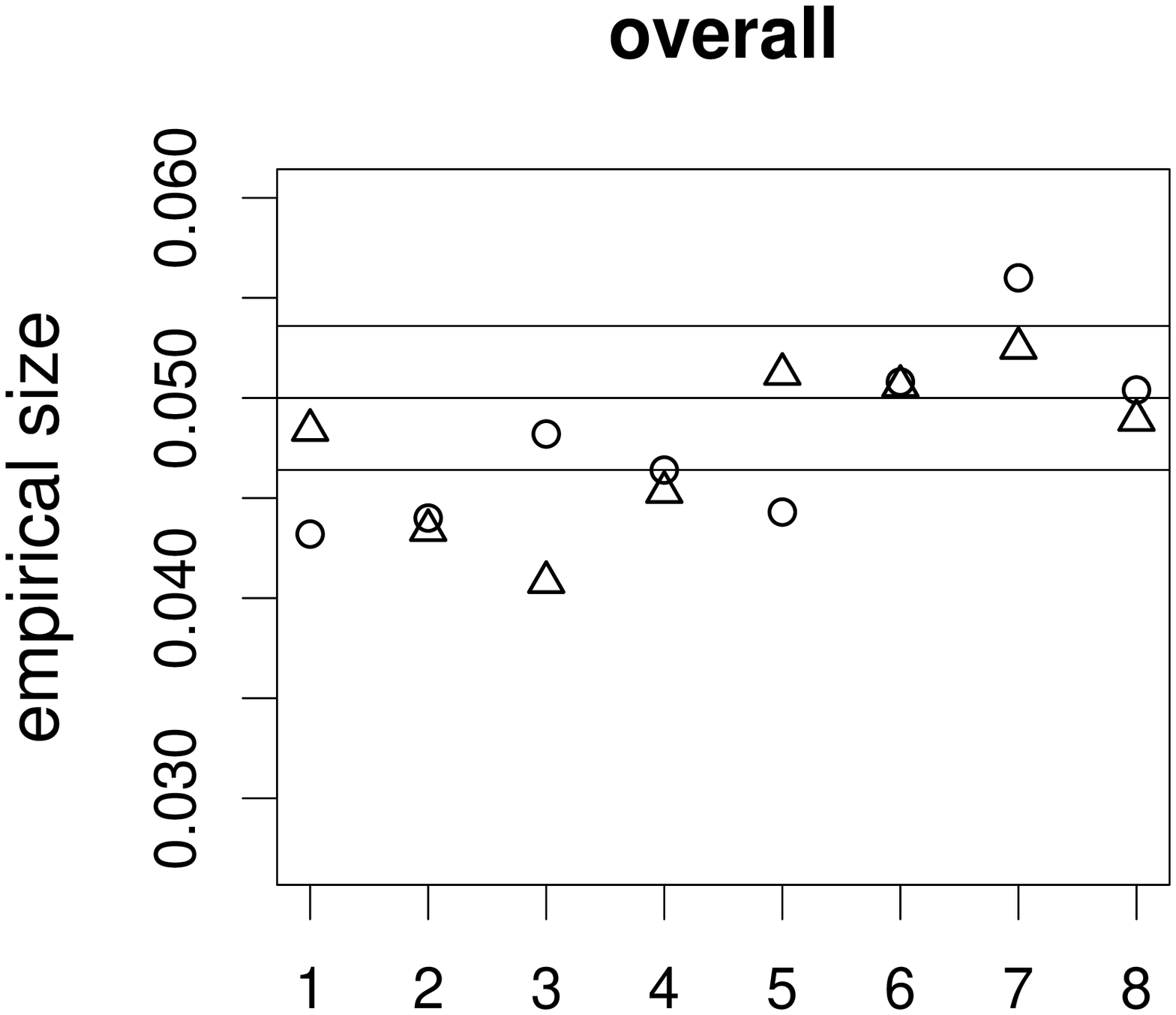} }}

\caption{
\label{fig:emp-size-CSR-2cl}
The empirical size estimates of the cell-specific tests for cells (1,1) (left),
cell (2,2) (middle), and overall test of segregation (right) under the CSR independence pattern
in the two-class case.
The empirical sizes for Dixon's and the new NNCT-tests are plotted in circles ($\circ$)
and triangles ($\triangle$), respectively.
The horizontal lines are located at .0464 (upper threshold for conservativeness),
.0500 (nominal level), and .0536 (lower threshold for liberalness).
The horizontal axis labels:
1=(10,10), 2=(10,30), 3=(10,50), 4=(30,30), 5=(30,50),
6=(50,50), 7=(50,100), 8=(100,100).
}
\end{figure}

\subsection{Empirical Significance Levels under RL of Two Classes}
\label{sec:RL-emp-sign-2Cl}
Recall that the segregation tests we consider are conditional under the CSR independence pattern.
To evaluate their empirical size performance better,
we also perform Monte Carlo simulations under the RL pattern,
for which the tests are not conditional.
For the RL pattern we consider three cases,
in each of which, we first determine the locations of points
and then assign labels to them randomly.

\noindent
\textbf{RL Case (1):}
First, we generate $n=(n_1+n_2)$ points iid $\U((0,1) \times (0,1))$
for some combinations of $n_1,n_2 \in \{10,30,50,100\}$.
The locations of these points are taken to be the fixed locations
for which we assign the labels randomly.
Thus, we simulate the RL pattern for the performance of the tests
under the null case.
For each sample size combination $(n_1,n_2)$,
we randomly choose $n_1$ points (without replacement) and label them
as $X$ and the remaining $n_2$ points as $Y$ points.
We repeat the RL procedure $N_{mc}=10000$ times for each sample size combination.
At each RL iteration, we construct the $2 \times 2$ NNCT,
and then compute the overall and cell-specific tests.
Out of these 10000 samples the number of significant results by each test is recorded.
The nominal significance level used in all these tests is $\alpha=.05$.
Based on these significant results, empirical sizes are calculated as
the ratio of number of significant test statistics to the number of Monte
Carlo replications, $N_{mc}$.

\noindent
\textbf{RL Case (2):}
We generate $n_1$ points iid $\U((0,2/3) \times (0,2/3))$ and
$n_2$ points iid $\U((1/3,1) \times (1/3,1))$
for some combinations of $n_1,n_2 \in \{10,30,50,100\}$.
The locations of these points are taken to be the fixed locations
for which we assign labels randomly.
The RL process is applied to these fixed points $N_{mc}=10000$
times for each sample size combination.
The empirical sizes for the tests are calculated similarly as in RL Case (1).

\noindent
\textbf{RL Case (3):}
We generate $n_1$ points iid $\U((0,1) \times (0,1))$ and
$n_2$ points iid $\U((2,3) \times (0,1))$
for some combinations of $n_1,n_2 \in \{10,30,50,100\}$.
RL procedure and the empirical sizes for the tests
are calculated similarly as in the previous RL Cases.

The locations for which the RL procedure is applied in RL Cases (1)-(3) are plotted
in Figure \ref{fig:RLcases} for $n_1=n_2=100$.
Observe that in RL Case (1), the set of points are iid $\U((0,1) \times (0,1))$,
i.e., it can be assumed to be from a Poison process in the unit square.
The set of locations are from two overlapping clusters in RL Case (2), and
from two disjoint clusters in RL Case (3).

\begin{figure}
\centering
\rotatebox{-90}{ \resizebox{2. in}{!}{\includegraphics{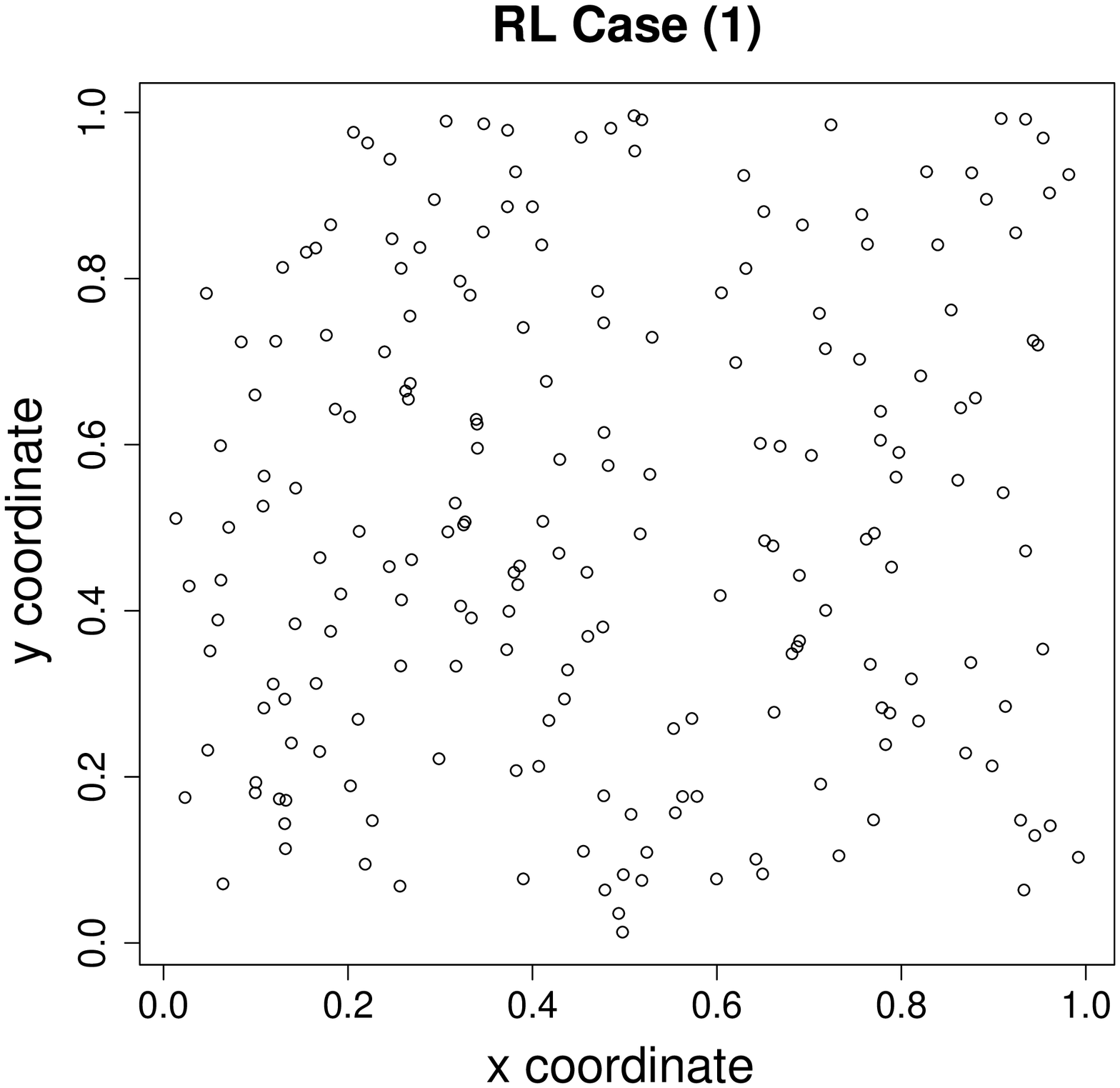} }}
\rotatebox{-90}{ \resizebox{2. in}{!}{\includegraphics{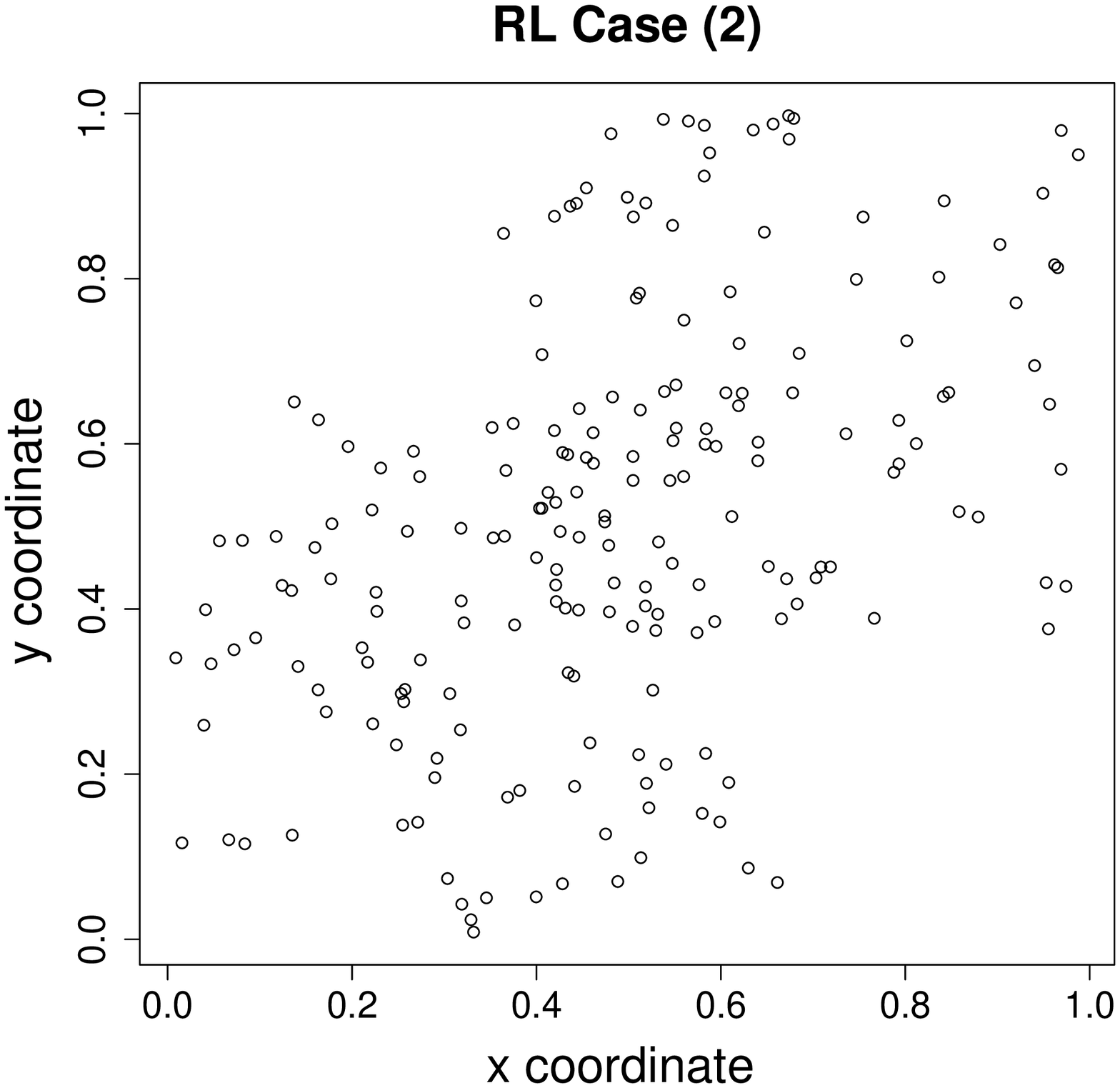} }}
\rotatebox{-90}{ \resizebox{2. in}{!}{\includegraphics{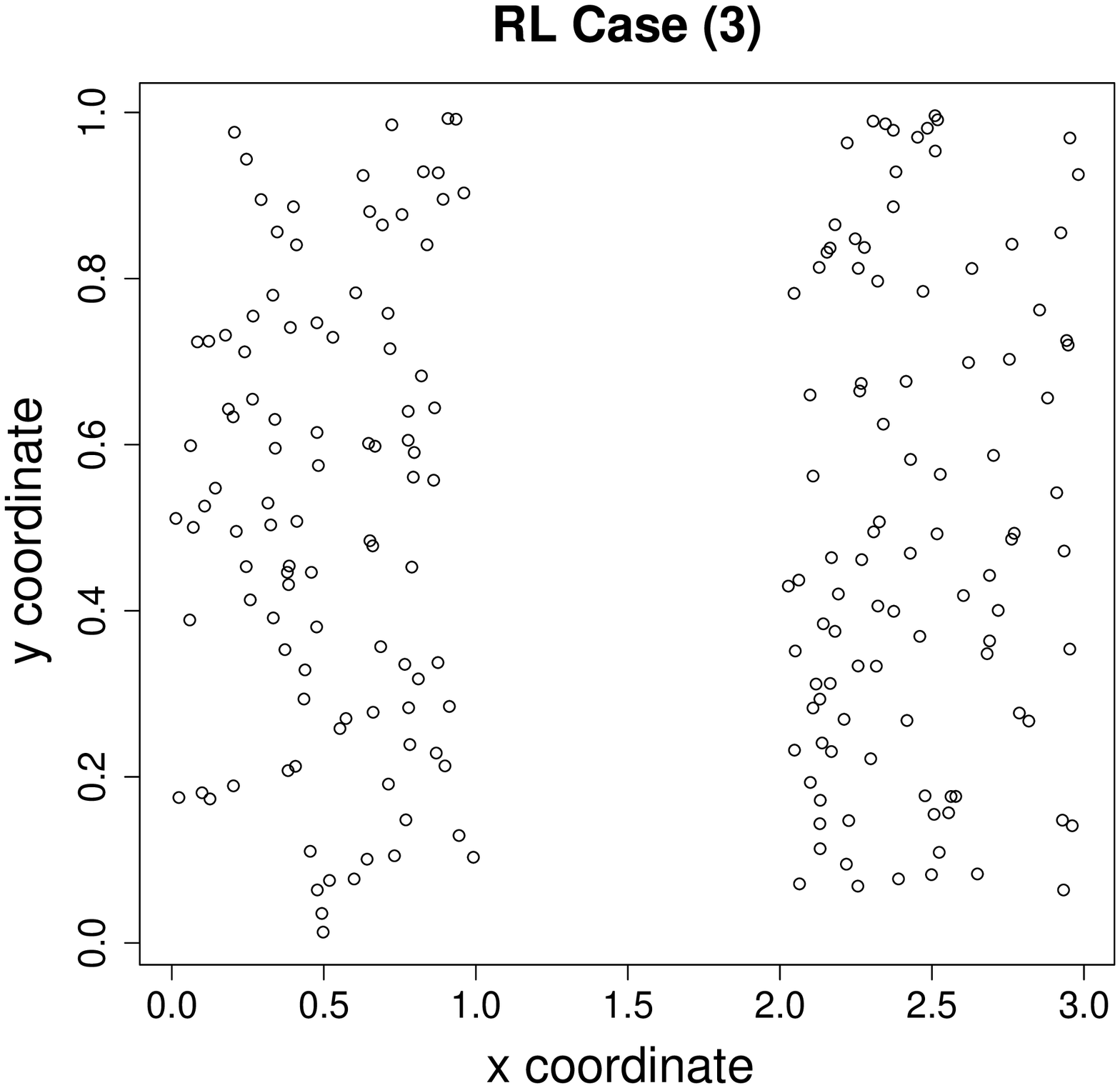} }}
 \caption{
\label{fig:RLcases}
The fixed locations for which RL procedure is applied for RL Cases (1)-(3) with $n_1=n_2=100$
in the two-class case. Notice that $x$-axis for RL Case (3) is differently scaled.
}
\end{figure}

We present the empirical significance
levels for the NNCT-tests in Table \ref{tab:cell-spec-null-RL},
where the empirical significance level labeling is as in Table \ref{tab:cell-spec-null}.
The empirical sizes are marked with $^c$ and $^{\ell}$
for conservativeness and liberalness as in Section \ref{sec:CSR-emp-sign-2Cl}.


\begin{table}[ht]
\centering
\begin{tabular}{|c||c|c|c||c|c|c||c|c|c|}
\hline
\multicolumn{10}{|c|}{Empirical significance levels under RL} \\
\hline
 & \multicolumn{3}{|c||}{RL Case (1)} & \multicolumn{3}{|c||}{RL Case (2)} &
 \multicolumn{3}{|c|}{RL Case (3)} \\
\hline
sizes & \multicolumn{2}{|c|}{cell} & \multicolumn{1}{|c||}{$C$} &
\multicolumn{2}{|c|}{cell} & \multicolumn{1}{|c||}{$C$} &
\multicolumn{2}{|c|}{cell} & \multicolumn{1}{|c|}{$C$} \\
\hline

$(n_1,n_2)$  & $\ah_{1,1}^D$ & $\ah_{2,2}^D$ & $\ah_D$
  & $\ah_{1,1}^D$ & $\ah_{2,2}^D$ & $\ah_D$
  & $\ah_{1,1}^D$ & $\ah_{2,2}^D$ & $\ah_D$\\
& $\ah_{1,1}^N$ & $\ah_{2,2}^N$ & $\ah_N$
  & $\ah_{1,1}^N$ & $\ah_{2,2}^N$ & $\ah_N$
  & $\ah_{1,1}^N$ & $\ah_{2,2}^N$ & $\ah_N$\\
\hline
(10,10) & .0604$^{\ell}$ & .0557$^{\ell}$ & .0349$^c$ & .0624$^{\ell}$ & .0657$^{\ell}$ & .0446$^c$
& .0444$^{c}$ & .0481 & .0404$^c$\\
 & .0359$^c$ & .0354$^c$ & .0409$^c$ & .0351$^c$ & .0357$^c$ & .0434$^c$  & .0345$^c$ & .0344$^c$ & .0421$^c$ \\
\hline
(10,30) & .0311$^c$ & .0699$^{\ell}$ & .0466 & .0297$^c$ & .0341$^c$ & .0327$^c$
& .0281$^c$ & .0447$^c$ & .0324$^c$\\
 & .0426$^c$ & .0391$^c$ & .0428$^c$ & .0364$^c$ & .0406$^c$ & .0366$^c$ & .0348$^c$ & .0321$^c$ & .0348$^c$ \\
\hline
(10,50) & .0264$^c$ & .0472 & .0507 & .0251$^c$ & .0384$^c$ & .0508 & .0260$^c$
& .0404$^c$ & .0500\\
 & .0424$^c$ & .0428$^c$ & .0437$^c$ & .0383$^c$ & .0390$^c$ & .0401$^c$ & .0390$^c$ & .0394$^c$ & .0408$^c$ \\
\hline
(30,30) & .0579$^{\ell}$ & .0547$^{\ell}$ & .0497 & .0513 & .0523 & .0469 & .0549$^{\ell}$
& .0553$^{\ell}$ & .0484\\
 & .0440$^c$ & .0429$^c$ & .0447$^c$ & .0468 & .0468 & .0471 & .0494 & .0494 & .0494 \\
\hline
(30,50) & .0621$^{\ell}$ & .0608$^{\ell}$ & .0444$^c$ & .0626$^{\ell}$ & .0594$^{\ell}$ & .0411$^c$
& .0677$^{\ell}$ & .0685$^{\ell}$ & .0445$^c$\\
 & .0454$^c$ & .0464 & .0469 & .0519 & .0506 & .0533 & .0513 & .0496 & .0525 \\
\hline
(50,50) & .0512 & .0524 & .0497 & .0509 & .0511 & .0501 & .0504 & .0506 & .0488\\
 & .0542$^{\ell}$ & .0531 & .0560$^{\ell}$ & .0439$^c$ & .0439$^c$ & .0446$^c$ & .0502 & .0493 & .0521 \\
\hline
(50,100) & .0625$^{\ell}$ & .0512 & .0482  & .0566$^{\ell}$ & .0421$^c$ & .0460$^c$
& .0590$^{\ell}$ & .0484 & .0479\\
 & .0496 & .0496 & .0518 & .0490 & .0490 & .0494 & .0499 & .0501 & .0501 \\
\hline
(100,100) & .0538$^{\ell}$ & .0534 & .0525 & .0439$^c$ & .0453$^c$ & .0505 & .0495
& .0476 & .0534\\
 & .0574$^{\ell}$ & .0571$^{\ell}$ & .0576$^{\ell}$ & .0484 & .0484 & .0483 & .0478 & .0478 & .0482 \\
\hline

\end{tabular}
\caption{
\label{tab:cell-spec-null-RL}
The empirical significance levels
in the two-class case under $H_o:RL$ for RL Cases (1)-(3) with
$N_{mc}=10000$, $n_1,n_2$ in $\{10,30,50,100\}$ at the nominal level of $\alpha=.05$.
($^c$: empirical size significantly less than .05; i.e., the test is conservative.
$^{\ell}$: empirical size significantly larger than .05; i.e., the test is liberal.
cell = cell-specific tests, $C$ = overall segregation test.)}
\end{table}

%
%
%

Observe that Dixon's cell-specific test for cell $(1,1)$
has the same trend under RL Cases (1)-(3):
extremely conservative when the observed cell
count is very likely to be $< 5$ (i.e., when $n_1 \leq 10$ and $n_1 \neq n_2$)
liberal for most other cases,
and close to being at the nominal size for large and equal sample sizes.
New cell-specific test for cell $(1,1)$ is conservative
for small samples and closer to the nominal level otherwise.
Moreover the new test fluctuates with smaller deviations from the nominal level
of .05 compared to Dixon's test.

Dixon's cell-specific test for cell $(2,2)$
is closer to the nominal level compare to that for cell $(1,1)$,
but is still conservative or liberal for very different sample sizes.
The new cell-specific test for cell $(2,2)$ is conservative
when  at least one sample is small (i.e., $n_i \leq 30$),
about the desired level otherwise.
Notice also that the new cell-specific tests for
both cells $(1,1)$ and $(2,2)$  have similar empirical size performance.

Dixon's overall test is conservative for small samples and $(n_1,n_2)=(30,50)$,
and about the desired level otherwise.
The new overall test is conservative when at least one sample
size is $\leq 10$, about the desired level otherwise for RL Cases (2) and (3).
For RL Case (1), it is conservative for small samples and liberal
for very different large samples,
about the desired level for similar size large samples.

Thus, under RL, for large samples the new cell-specific tests have better
empirical size performance compared to Dixon's cell-specific tests.
On the other hand the performance of the new overall test
depends on the RL Case, i.e., the allocation of the points
confounds the results of the overall tests.

Comparing Tables \ref{tab:cell-spec-null} and \ref{tab:cell-spec-null-RL},
we observe that the empirical sizes are not very similar
under the RL and CSR independence patterns.
Moreover, the performance of Dixon's cell-specific test for cell $(2,2)$
and the new overall test have different size performance
under each RL Case.
Although cell-specific test for cell $(1,1)$ is very similar for all RL and CSR independence Cases,
the other tests are not very similar,
and their sizes are closer to the nominal level under the CSR independence pattern
compared to those under RL Cases.
However, we can also conclude that the tests are usually conservative
when at least one sample is small,
regardless of whether the null case is RL or CSR independence.

\section{Empirical Significance Levels in the Three-Class Case}
\label{sec:monte-carlo-3Cl}
In this section, we provide the empirical significance levels
for Dixon's and the new overall and cell-specific tests of segregation
in the three-class case under RL and CSR independence patterns.

\subsection{Empirical Significance Levels under CSR Independence of Three Classes}
\label{sec:CSR-emp-sign-3Cl}
The symmetry in cell counts for
rows in Dixon's cell-specific tests
and columns in the new cell-specific tests occur only in the two-class case.
Therefore, in order to better evaluate the performance of cell-specific tests
in the absence of such symmetry,
we also consider the three-class case with classes $X$, $Y$, and $Z$ under CSR independence.
We generate $n_1,\,n_2,\,n_3$ points distributed independently uniformly
on the unit square $(0,1) \times (0,1)$
from classes $X,\,Y$, and $Z$, respectively.
That is, each data set of classes $X,\,Y$, and $Z$ enjoy within sample and
between sample independence.
We generate data points for some combinations of $n_1,n_2,n_3 \in \{10,30,50,100\}$;
and for each sample size combination, we generate data sets $X,\,Y$, and $Z$
for $N_{mc}=10000$ times.
The empirical sizes and the significance of their deviation from .05 are calculated
as in Section \ref{sec:CSR-emp-sign-2Cl}.

\begin{table}[t]
\centering
\small{
\begin{tabular}{|c||c|c|c||c|c|c||c|c|c||c|}
\hline
\multicolumn{11}{|c|}{Empirical significance levels for the NNCT-tests} \\
\hline
  & \multicolumn{9}{|c||}{cell-specific} & \multicolumn{1}{|c|}{overall} \\
\hline $(n_1,n_2,n_3)$  & $(1,1)$ & $(1,2)$ & $(1,3)$ & $(2,1)$
& $(2,2)$ & $(2,3)$ & $(3,1)$ & $(3,2)$ & $(3,3)$ &\\
\hline
(10,10,10) & .0277$^c$ & .0355$^c$ & .0337$^c$ & .0386$^c$ & .0283$^c$ & .0370$^c$ & .0371$^c$ & .0391$^c$ & .0250$^c$ & .0421$^c$\\
          & .0481 & .0447$^c$ & .0403$^c$ & .0463$^c$ & .0512 & .0456$^c$ & .0445$^c$ & .0457$^c$ & .0470 & .0459$^c$\\
\hline
(10,10,30) & .0464 & .0342$^c$ & .0260$^c$ & .0336$^c$ & .0428$^c$ & .0267$^c$ & .0455$^c$ & .0494 & .0477 & .0445$^c$\\
          & .0381$^c$ & .0428$^c$ & .0490 & .0425$^c$ & .0367$^c$ & .0495 & .0466 & .0468 & .0495 & .0445$^c$\\
\hline
(10,10,50) & .0661$^\ell$ & .0434$^c$ & .0416$^c$ & .0439$^c$ & .0667$^\ell$ & .0430$^c$ & .0505 & .0455$^c$ & .0505 & .0510\\
           & .0464 & .0394$^c$ & .0449$^c$ & .0408$^c$ & .0444$^c$ & .0449$^c$ & .0400$^c$ & .0441$^c$ & .0463$^c$ & .0543$^\ell$\\
\hline
(10,30,30) & .0657$^\ell$ & .0494 & .0520 & .0468 & .0425$^c$ & .0488 & .0511 & .0444$^c$ & .0402$^c$ & .0439$^c$\\
           & .0465 & .0432$^c$ & .0469 & .0462$^c$ & .0487 & .0506 & .0448$^c$ & .0487 & .0492 & .0453$^c$\\
\hline
(10,30,50) & .0367$^c$ & .0343$^c$ & .0566$^\ell$ & .0605$^\ell$ & .0539$^\ell$ & .0579$^\ell$ & .0452$^c$ & .0468 & .0544$^\ell$ & .0450$^c$ \\
           & .0407$^c$ & .0454$^c$ & .0486 & .0468 & .0488 & .0519 & .0497 & .0502 & .0502 & .0467\\
\hline
(30,30,30) & .0526 & .0508 & .0503 & .0487 & .0488 & .0499 & .0517 & .0458$^c$ & .0505 & .0497\\
           & .0479 & .0535 & .0520 & .0475 & .0455$^c$ & .0487 & .0542$^\ell$ & .0493 & .0485 & .0475\\
\hline
(10,50,50) & .0758$^\ell$ & .0548$^\ell$ & .0525 & .0322$^c$ & .0565$^\ell$ & .0457$^c$ & .0316$^c$ & .0442$^c$ & .0548$^\ell$ & .0517\\
           & .0515 & .0493 & .0491 & .0516 & .0527 & .0501 & .0519 & .0516 & .0517 & .0529\\
\hline
(30,30,50) & .0515 & .0535 & .0474 & .0566$^\ell$ & .0468 & .0442$^c$ & .0466 & .0532 & .0520 & .0463$^c$\\
           & .0516 & .0492 & .0485 & .0515 & .0469 & .0474 & .0495 & .0523 & .0513 & .0511\\
\hline
(30,50,50) & .0370$^c$ & .0606$^\ell$ & .0602$^\ell$ & .0440$^c$ & .0519 & .0424$^c$ & .0451$^c$ & .0424$^c$ & .0510 & .0486\\
           & .0463$^c$ & .0513 & .0493 & .0484 & .0525 & .0519 & .0494 & .0505 & .0482 & .0457$^c$\\
\hline
(50,50,50) & .0605$^\ell$ & .0514 & .0477 & .0503 & .0603$^\ell$ & .0483 & .0508 & .0480 & .0575$^\ell$ & .0497\\
           & .0520 & .0506 & .0521 & .0530 & .0504 & .0527 & .0539$^\ell$ & .0521 & .0450$^c$ & .0514\\
\hline
(50,50,100)& .0462$^c$ & .0444$^c$ & .0447$^c$ & .0421$^c$ & .0490 & .0405$^c$ & .0460$^c$ & .0458$^c$ & .0492 & .0488\\
           & .0466 & .0510 & .0535 & .0447$^c$ & .0481 & .0507 & .0527 & .0551$^\ell$ & .0499 & .0505\\
\hline
(50,100,100)& .0493 & .0614 & .0601 & .0505 & .0580$^\ell$ & .0540$^\ell$ & .0511 & .0554$^\ell$ & .0552$^\ell$ & .0495\\
            & .0463$^c$ & .0499 & .0475 & .0468 & .0523 & .0453$^c$ & .0522 & .0480 & .0507 & .0496\\
\hline
(100,100,100)& .0499 & .0522 & .0540$^\ell$ & .0571$^\ell$ & .0468 & .0525 & .0534 & .0508 & .0469 & .0456\\
             & .0533 & .0522 & .0514 & .0514 & .0451$^c$ & .0508 & .0513 & .0477 & .0473 & .0482\\
\hline
\end{tabular}
}
\caption{
\label{tab:cell-spec-null-3Cl}
The empirical significance
levels for the Dixon's cell-specific and overall tests (top) and for the new
version of the cell-specific and overall tests (bottom) in the three-class case
under $H_o:CSR~~independence$ with $N_{mc}=10000$, $n_1,n_2,n_3$ in $\{10,30,50,100\}$ at the
nominal level $\alpha=.05$.
$^c$: The empirical level is significantly smaller than .05;
$^\ell$: The empirical level is significantly larger than .05.}
\end{table}

%
%
%

We present the empirical significance
levels for the cell-specific tests in Table \ref{tab:cell-spec-null-3Cl},
where the estimated levels for Dixon's test
are provided in the top, while for the new version in the bottom
for cell $(i,j) \in \{(1,1),(1,2),(1,3),\ldots,(3,3)\}$.
Notice that when at least one class is small (i.e., $n_i \leq 10$)
tests are usually conservative, with the Dixon's cell-specific tests
being the most conservative.
The empirical sizes for the new cell-specific tests are closer to the nominal level
for all sample size combinations, while Dixon's cell-specific
tests fluctuate around .05 with larger deviations.
In the three-class case, both of the overall tests
exhibit similar performance in terms of empirical size,
with Dixon's test being slightly more conservative for small samples.
Thus, Type I error rates of the new cell-specific tests are
more robust to the differences in sample sizes (i.e., relative abundance)
and are closer to .05 compared to Dixon's cell-specific tests.

\subsection{Empirical Significance Levels under RL of Three Classes}
\label{sec:RL-emp-sign-3Cl}
To remove the confounding effect of conditional nature of the tests under CSR independence,
we also perform Monte Carlo simulations under the RL pattern.
For RL with 3 classes, we consider two cases.
In each case,
we first determine the locations of points and then assign labels to
them randomly.

For the RL pattern we consider three cases, in each of which, we
first determine the locations of points and then assign labels to
them randomly.

\noindent
\textbf{RL Case (1):}
First, we generate $n_1+n_2+n_3$ points iid $\U((0,1) \times (0,1))$
for some combinations of $n_1,n_2,n_3 \in \{10,30,50,100\}$.
The locations of these points are taken to be the fixed locations
for which we assign the labels randomly.
Thus, we simulate the RL pattern for the performance of the tests
under the null case.
For each sample size combination $(n_1,n_2,n_3)$
we pick $n_1$ points (without replacement) and label them as $X$,
pick $n_2$ points from the remaining points (without replacement) and label them as $Y$ points,
and label the remaining $n_3$ points as $Z$ points.
We repeat the RL procedure $N_{mc}=10000$ times for each sample size combination.
At each RL iteration, we construct the $3 \times 3$ NNCT for classes $X$, $Y$, and $Z$
and then compute the test statistics.
Out of these 10000 samples the number of significant tests by the tests is recorded.
The nominal significance level of .05 is used in all these tests.
Based on the number of significant results, empirical sizes are calculated as before.

\noindent
\textbf{RL Case (2):}
We generate $n_1$ points iid $\U((0,1) \times (0,1))$,
$n_2$ points iid $\U((2,3) \times (0,1))$,
and $n_3$ points iid $\U((1,2) \times (2,3))$
for some combinations of $n_1,n_2,n_3 \in \{10,30,50,100\}$.
RL procedure is performed
and the empirical sizes for the tests
are calculated similarly as in RL Case (1).

The locations for which the RL procedure is applied in RL Cases (1) and (2) are plotted
in Figure \ref{fig:RL3Clcases} for $n_1=n_2=n_3=100$.
In RL Case (1), the locations of the points can be assumed to
be from a Poisson process in the unit square.
In RL Case (2), the locations of the points are from three disjoint clusters.

\begin{figure}
\centering
\rotatebox{-90}{ \resizebox{2.5 in}{!}{\includegraphics{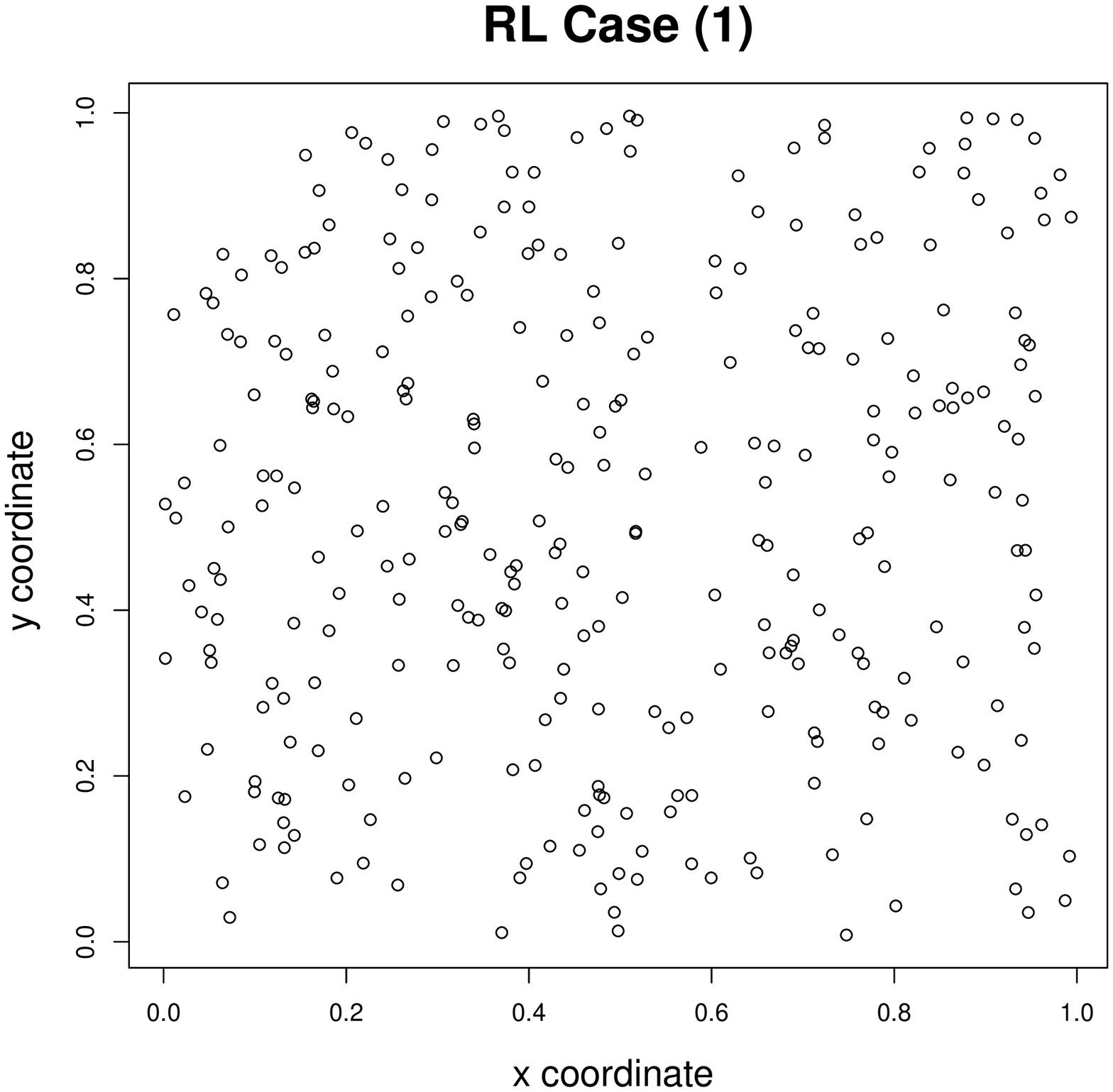} }}
\rotatebox{-90}{ \resizebox{2.5 in}{!}{\includegraphics{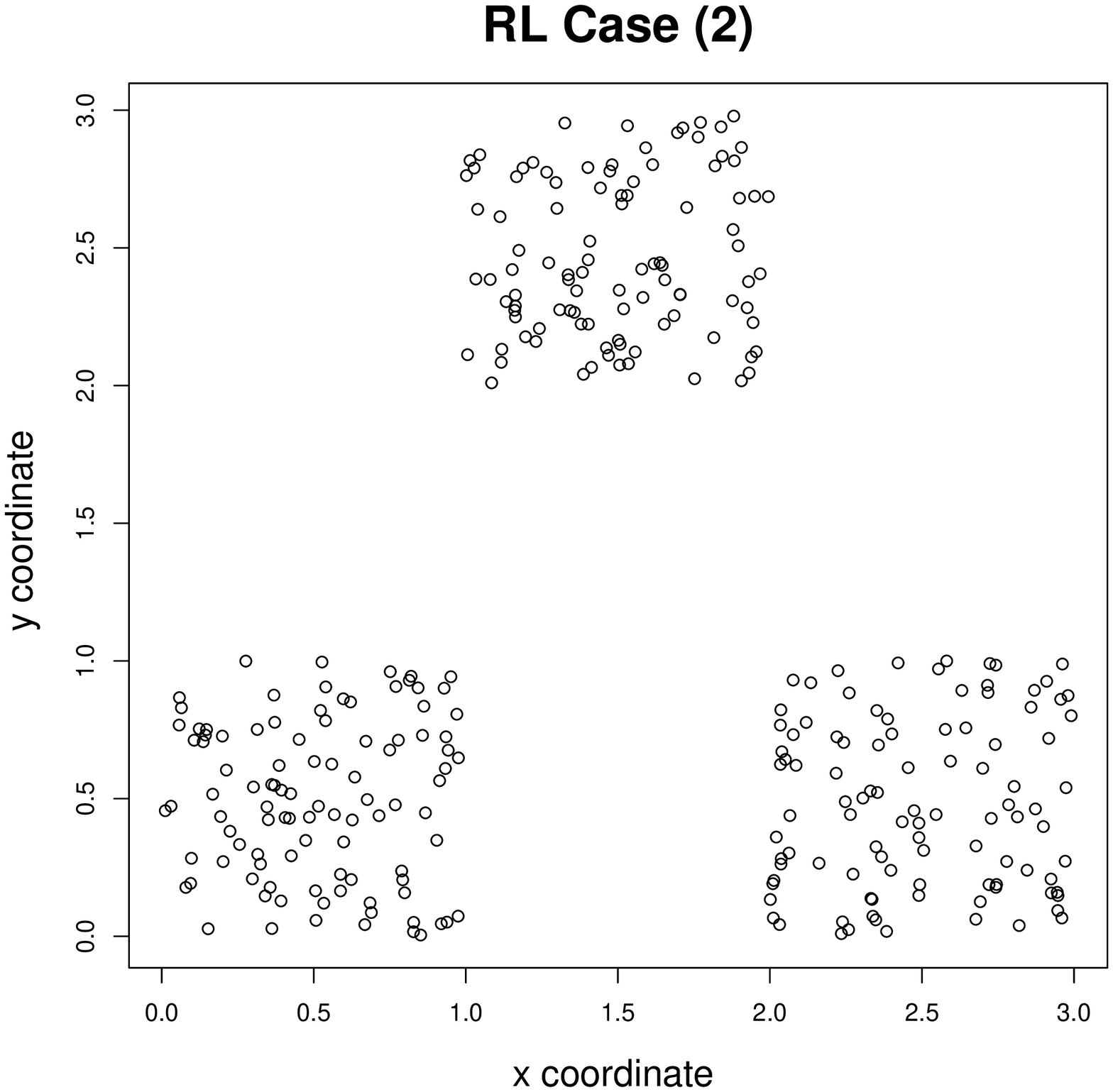} }}
 \caption{
\label{fig:RL3Clcases}
The fixed locations for which RL procedure is applied for RL Cases (1) and (2) with $n_1=n_2=n_3=100$
in the two-class case. Notice that $x$-axis for RL Case (2) is differently scaled.
}
\end{figure}

\begin{table}[ht]
\centering
\small{
\begin{tabular}{|c||c|c|c||c|c|c||c|c|c||c|}
\hline
\multicolumn{11}{|c|}{Empirical significance levels for the RL pattern} \\
\hline
  & \multicolumn{9}{|c||}{cell} & \multicolumn{1}{|c|}{$C$} \\
\hline $(n_1,n_2,n_3)$  & $(1,1)$ & $(1,2)$ & $(1,3)$ & $(2,1)$
& $(2,2)$ & $(2,3)$ & $(3,1)$ & $(3,2)$ & $(3,3)$ &\\
\hline
\multicolumn{11}{|c|}{RL Case (1)} \\
\hline
(10,10,10)  & .0239$^c$ & .0343$^c$ & .0365$^c$ & .0320$^c$ & .0234$^c$ & .0342$^c$ & .0292$^c$ & .0330$^c$ & .0244$^c$ & .0377$^c$ \\
 & .0469 & .0422$^c$ & .0418$^c$ & .0411$^c$ & .0471 & .0440$^c$ & .0385$^c$ & .0452$^c$ & .0452$^c$ & .0422$^c$\\
\hline
(10,10,30)  & .0498 & .0361$^c$ & .0273$^c$ & .0399$^c$ & .0488 & .0287$^c$ & .0430$^c$ & .0517 & .0432$^c$ & .0455$^c$ \\
 & .0421$^c$ & .0429$^c$ & .0447$^c$ & .0459$^c$ & .0425$^c$ & .0451$^c$ & .0486 & .0485 & .0515 & .0480\\
\hline
(10,10,50)  & .0676$^\ell$ & .0415$^c$ & .0423$^c$ & .0418$^c$ & .0630$^\ell$ & .0394$^c$ & .0390$^c$ & .0386$^c$ & .0416$^c$ & .0477 \\
 & .0475 & .0404$^c$ & .0487 & .0382$^c$ & .0408$^c$ & .0415$^c$ & .0453$^c$ & .0434$^c$ & .0462$^c$ & .0511 \\
\hline
(10,30,30)  & .0737$^\ell$ & .0510 & .0505 & .0533 & .0458$^c$ & .0433$^c$ & .0533 & .0449$^c$ & .0435$^c$ & .0477 \\
 & .0483 & .0450$^c$ & .0415$^c$ & .0478 & .0505 & .0479 & .0462$^c$ & .0506 & .0495 & .0495 \\
\hline
(10,30,50)  & .0389$^c$ & .0409$^c$ & .0612$^\ell$ & .0660$^\ell$ & .0418$^c$ & .0625$^\ell$ & .0531 & .0545$^\ell$ & .0590$^\ell$ & .0518 \\
 & .0444$^c$ & .0535 & .0539$^\ell$ & .0521 & .0471 & .0555$^\ell$ & .0547$^\ell$ & .0533 & .0537$^\ell$ & .0547$^\ell$ \\
\hline
(30,30,30)  & .0562$^\ell$ & .0318$^c$ & .0317$^c$ & .0371$^c$ & .0580$^\ell$ & .0328$^c$ & .0323$^c$ & .0309$^c$ & .0602$^\ell$ & .0471 \\
 & .0517 & .0520 & .0544$^\ell$ & .0553$^\ell$ & .0481 & .0537$^\ell$ & .0539$^\ell$ & .0516 & .0486 & .0496 \\
\hline
(10,50,50)  & .0758$^\ell$ & .0561$^\ell$ & .0498 & .0303$^c$ & .0620$^\ell$ & .0355$^c$ & .0301$^c$ & .0346$^c$ & .0617$^\ell$ & .0533 \\
 & .0505 & .0533 & .0463$^c$ & .0533 & .0467 & .0487 & .0513 & .0479 & .0491 & .0526 \\
\hline
(30,30,50)  & .0493 & .0400$^c$ & .0505 & .0407$^c$ & .0479 & .0550$^\ell$ & .0586$^\ell$ & .0622$^\ell$ & .0398$^c$ & .0489 \\
 & .0489 & .0532 & .0554$^\ell$ & .0494 & .0463$^c$ & .0559$^\ell$ & .0529 & .0541$^\ell$ & .0526 & .0516 \\
\hline
(30,50,50)  & .0396$^c$ & .0609$^\ell$ & .0649$^\ell$ & .0379$^c$ & .0481 & .0443$^c$ & .0406$^c$ & .0454$^c$ & .0518 & .0473\\
 & .0475 & .0515 & .0503 & .0491 & .0483 & .0487 & .0545$^\ell$ & .0511 & .0519 & .0483 \\
\hline
(50,50,50)  & .0483 & .0453$^c$ & .0454$^c$ & .0468 & .0521 & .0468 & .0436$^c$ & .0510 & .0500 & .0469\\
 & .0474 & .0489 & .0471 & .0476 & .0487 & .0511 & .0440$^c$ & .0493 & .0445$^c$ & .0466 \\
\hline
(50,50,100)  & .0478 & .0479 & .0388$^c$ & .0459$^c$ & .0495 & .0405$^c$ & .0464 & .0460$^c$ & .0509 & .0450$^c$\\
 & .0490 & .0514 & .0471 & .0489 & .0511 & .0502 & .0517 & .0518 & .0419$^c$ & .0461$^c$ \\
\hline
(50,100,100)  & .0520 & .0603$^\ell$ & .0601$^\ell$ & .0490 & .0557$^\ell$ & .0504 & .0467 & .0501 & .0491 & .0476 \\
 & .0481 & .0483 & .0501 & .0514 & .0505 & .0466 & .0532 & .0465 & .0467 & .0488 \\
\hline
(100,100,100)  & .0499 & .0559$^\ell$ & .0539$^\ell$ & .0563$^\ell$ & .0470 & .0551$^\ell$ & .0513 & .0539$^\ell$ & .0501 & .0510\\
 & .0484 & .0515 & .0495 & .0537$^\ell$ & .0498 & .0534 & .0494 & .0521 & .0496 & .0519 \\
\hline

\multicolumn{11}{|c|}{RL Case (2)} \\
\hline
(10,10,10) & .0227$^c$ & .0302$^c$ & .0312$^c$ & .0349$^c$ & .0220$^c$ & .0327$^c$ & .0339$^c$ & .0329$^c$ & .0217$^c$ & .0372$^c$ \\
 & .0466 & .0391$^c$ & .0408$^c$ & .0415$^c$ & .0469 & .0437$^c$ & .0426$^c$ & .0417$^c$ & .0471 & .0431$^c$ \\
\hline
(10,10,30) & .0488 & .0338$^c$ & .0217$^c$ & .0304$^c$ & .0475 & .0249$^c$ & .0461$^c$ & .0474 & .0426$^c$ & .0470 \\
 & .0397$^c$ & .0459$^c$ & .0472 & .0441$^c$ & .0408$^c$ & .0500 & .0482 & .0499 & .0466 & .0447$^c$ \\
\hline
(10,10,50) & .0657$^\ell$ & .0429$^c$ & .0469 & .0414$^c$ & .0631$^\ell$ & .0412$^c$ & .0443$^c$ & .0459$^c$ & .0501 & .0501 \\
 & .0448$^c$ & .0404$^c$ & .0468 & .0389$^c$ & .0445$^c$ & .0446$^c$ & .0457$^c$ & .0408$^c$ & .0464 & .0529 \\
\hline
(10,30,30) & .0677$^\ell$ & .0487 & .0492 & .0472 & .0429$^c$ & .0420$^c$ & .0505 & .0399$^c$ & .0377$^c$ & .0420$^c$ \\
 & .0483 & .0399$^c$ & .0442$^c$ & .0429$^c$ & .0519 & .0492 & .0468 & .0525 & .0484 & .0449$^c$ \\
\hline
(10,30,50) & .0414$^c$ & .0387$^c$ & .0551$^\ell$ & .0619$^\ell$ & .0599$^\ell$ & .0541$^\ell$ & .0378$^c$ & .0392$^c$ & .0479 & .0529 \\
 & .0422$^c$ & .0479 & .0537$^\ell$ & .0508 & .0499 & .0454$^c$ & .0566$^\ell$ & .0461$^c$ & .0505 & .0528 \\
\hline
(10,50,50) & .0752$^\ell$ & .0518 & .0512 & .0334$^c$ & .0674$^\ell$ & .0395$^c$ & .0341$^c$ & .0393$^c$ & .0642$^\ell$ & .0507 \\
 & .0530 & .0475 & .0472 & .0500 & .0526 & .0522 & .0501 & .0513 & .0526 & .0530 \\
\hline
(30,30,30) & .0607$^\ell$ & .0440$^c$ & .0500 & .0473 & .0571$^\ell$ & .0468 & .0496 & .0462$^c$ & .0622$^\ell$ & .0472 \\
 & .0512 & .0468 & .0463$^c$ & .0456$^c$ & .0430$^c$ & .0472 & .0494 & .0509 & .0509 & .0495 \\
\hline
(30,30,50) & .0445$^c$ & .0538$^\ell$ & .0459$^c$ & .0598$^\ell$ & .0485 & .0457$^c$ & .0466 & .0455$^c$ & .0678$^\ell$ & .0436$^c$ \\
 & .0462$^c$ & .0458$^c$ & .0500 & .0488 & .0463$^c$ & .0490 & .0501 & .0469 & .0545$^\ell$ & .0451$^c$ \\
\hline
(30,50,50) & .0360$^c$ & .0623$^\ell$ & .0620$^\ell$ & .0371$^c$ & .0472$^c$ & .0437$^c$ & .0379$^c$ & .0441$^c$ & .0504 & .0475 \\
 & .0480 & .0520 & .0493 & .0495 & .0477 & .0524 & .0505 & .0523 & .0514 & .0491 \\
\hline
(50,50,50) & .0534 & .0515 & .0508 & .0462$^c$ & .0531 & .0479 & .0496 & .0501 & .0510 & .0513 \\
 & .0512 & .0561$^\ell$ & .0521 & .0499 & .0534 & .0499 & .0542$^\ell$ & .0540 & .0512 & .0462$^c$ \\
\hline
(50,50,100) & .0462$^c$ & .0480 & .0429$^c$ & .0497 & .0464 & .0414$^c$ & .0458$^c$ & .0473 & .0512 & .0497 \\
 & .0468 & .0480 & .0528 & .0477 & .0485 & .0512 & .0533 & .0524 & .0536 & .0497 \\
\hline
(50,100,100) & .0510 & .0638$^\ell$ & .0631 & .0483 & .0544$^\ell$ & .0539$^\ell$ & .0471 & .0531 & .0480 & .0516 \\
 & .0489 & .0525 & .0519 & .0532 & .0478 & .0481 & .0518 & .0483 & .0474 & .0511 \\
\hline
(100,100,100) & .0477 & .0549$^\ell$ & .0525 & .0499 & .0473 & .0571$^\ell$ & .0525 & .0520 & .0449$^c$ & .0457$^c$ \\
 & .0491 & .0518 & .0494 & .0487 & .0479 & .0541$^\ell$ & .0489 & .0534 & .0478 & .0469 \\
\hline
\end{tabular}
}
\caption{
\label{tab:cell-spec-null-RL-3Cl}
The empirical significance levels in the three-class case
under $H_o:RL$ for RL Cases (1) and (2) with
$N_{mc}=10000$, $n_1,n_2,n_3$ in $\{10,30,50,100\}$ at the nominal level of $\alpha=.05$.
($^c$: the empirical size is significantly smaller than .05; i.e., the test is conservative.
$^{\ell}$: the empirical size is significantly larger than .05; i.e., the test is liberal.
cell = cell-specific test, $C$ = overall test.)
}
\end{table}

We present the empirical significance
levels for the NNCT-tests in Table \ref{tab:cell-spec-null-RL-3Cl},
where the empirical significance level labeling is as in Table \ref{tab:cell-spec-null-3Cl}.
The empirical sizes are marked with $^c$ and $^{\ell}$
for conservativeness and liberalness as in Section \ref{sec:CSR-emp-sign-2Cl}.
Observe that under both RL Cases,
the new cell-specific tests are closer to the nominal level,
and are more robust to differences in sample sizes.
The overall tests exhibit similar performance under each RL Case,
with sizes for Dixon's overall test being slightly smaller
than those for the new overall test for most sample sizes.

Comparing Tables \ref{tab:cell-spec-null-3Cl} and \ref{tab:cell-spec-null-RL-3Cl},
we observe that, although the empirical sizes are not similar
for the RL and CSR independence patterns,
the trend is similar.
That is, the new cell-specific tests are at about the desired level
for most sample sizes, and more robust to the differences in sample sizes
compared to Dixon's cell-specific tests.
Overall tests have similar size performance under both RL Cases.

\begin{remark}
\label{rem:MC-emp-size}
\textbf{Main Result of Monte Carlo Simulations for Empirical Sizes:}
\cite{dixon:1994} recommends Monte Carlo randomization
test when some cell count(s) are smaller than 10 in a NNCT
for his cell-specific tests and when some cell counts are less than 5 for
his overall tests and we concur with his suggestion.
We extend his suggestion to the new cell-specific test for cell $(i,j)$
when sum of column $j$ is $< 10$ which happens less frequently
than cell $(i,j)$ being $< 10$.

Dixon's and new overall tests exhibit similar performance in terms
of empirical sizes: for small samples they are usually conservative and
are about the nominal level otherwise.

Thus, when sample sizes are small
(hence the corresponding cell counts are $< 5$ for overall test
and $<10$ for cell-specific tests),
the asymptotic approximation of the tests may not appropriate,
especially for Dixon's cell-specific tests.
In this case,
the power comparisons should be carried out using the Monte Carlo
critical values.
On the other hand, for large samples,
the power comparisons can be made using the asymptotic
or Monte Carlo critical values.

Furthermore, Dixon's cell-specific and overall tests are confounded
by the differences in the relative abundance of the classes.
On the other hand, the new cell-specific tests
are more robust to differences in sample sizes (i.e., relative abundance)
and less sensitive to the cell counts they pertain to.
$\square$
\end{remark}

\begin{remark}
\label{rem:MC-crit-value}
\textbf{Monte Carlo Critical Values:}
When sample sizes are small so that some cell counts
or column sums are expected to be $< 5$ with a high probability,
then it will not be appropriate to use
the asymptotic approximation hence the asymptotic critical values
for the overall and cell-specific tests of segregation (see Remark \ref{rem:MC-emp-size}).
In order to better evaluate the empirical power performance of the tests,
for each sample size combination, we record the test statistics at each Monte Carlo
simulation under the CSR independence cases of Sections \ref{sec:CSR-emp-sign-2Cl} and \ref{sec:CSR-emp-sign-3Cl}.
We find the 95$^{th}$ percentiles of the recorded test statistics at each sample size combination
(not presented) and use them as ``Monte Carlo critical values" for
the power estimation in the following sections.
For example, for Dixon's cell-specific test for cell $(1,1)$
in the two-class case for $(n_1,n_2)=(30,50)$,
the $Z^D_{1,1}$ values are recorded for $(n_1,n_2)=(30,50)$
under the CSR independence pattern as in Section \ref{sec:CSR-emp-sign-2Cl},
then the 95$^{th}$ percentile of these statistics is used
as the Monte Carlo critical value for $(n_1,n_2)=(30,50)$.
That is, under a segregation or association alternative with $(n_1,n_2)=(30,50)$,
a calculated test statistic is deemed significant if it is larger than
this Monte Carlo critical value.
$\square$
\end{remark}

\section{Finite Sample Performance of NNCT-Tests under Various Poisson and
Inhomogeneous Point Processes}
\label{sec:other-process}
In this section, we provide the finite sample performance
of the NNCT-tests under point patterns that are different from RL or CSR independence.
In particular, we will consider various versions of Poisson cluster processes
and some other inhomogeneous processes (\cite{diggle:2003}).

\textbf{First Version of Poisson Cluster Process ($PCP1(n_p,n_1,n_2,\sigma)$):}
In this process, first we generate $n_p$ parents iid on the unit square, $(0,1)\times(0,1)$
then for each parent $n_1/n_p$ offsprings are generated for sample $X$ and
$n_2/n_p$ for sample $Y$ from radially symmetric Gaussian distribution
with parameter $\sigma$.
Hence we generate $n_1$ $X$ and $n_2$ $Y$ points, respectively.
In the first case,
we use the same parent set for both $X$ and $Y$ points.
In the second case,
we use different parent sets for each of $X$ and $Y$ points.

\textbf{Second Version of Poisson Cluster Process ($PCP2(n_p,n_1,n_2,\sigma)$):}
In this process, we generate $n_p$ parents and $n_1$ $X$
and $n_2$ $Y$ offsprings as in the first version PCP1,
except the offsprings are randomly allocated amongst the parents.

For both versions of the above Poisson cluster processes,
we take $\sigma \in \{0.05,.10,.20\}$ and $(n_1,n_2) \in \{(30,30),(30,50),\\
(50,50)\}$.

\textbf{Matern Cluster Process ($MCP(\kappa,r,\mu)$):}
In this process, first we generate a Poisson point process
of ``parent'' points with intensity $\kappa$.
Then each parent point is replaced by a random cluster of points.
The number of points in each cluster are random with a Poisson($\mu$) distribution,
and the points are placed independently and uniformly inside a disc of radius $r$
centered on the parent point.
The parent points are not restricted to lie in the unit square;
the parent process is effectively the uniform Poisson process on the infinite plane.
We consider $\kappa=5$, $r \in \{.05,.10,.20\}$ for both $X$ and $Y$ points
and $\mu=n_1/5$ for $X$ points and $\mu=n_2/5$ for $Y$ points.
In case 1, we use the same parents for both $X$ and $Y$ offsprings,
while in case 2, we generate different sets of parents with $\kappa=5$.
For each of the above cases,
we take $(n_1,n_2) \in \{(50,50),(50,100),(100,100)\}$.
For more on Matern cluster processes, see (\cite{matern:1986} and \cite{waagepetersen:2007}).

\textbf{Inhomogeneous Poisson Cluster Process ($IPCP(\lambda(x,y))$):}
In this process,
the intensity of the Poisson process is set to be
$\lambda(x,y)$ which is a function of $(x,y)$.
We generate a realization of the inhomogeneous Poisson process
with intensity function $\lambda(x,y)$ at spatial location $(x,y)$
inside the unit square by random ``thinning''.
That is, we first generate a uniform Poisson process of intensity $\lambda(x,y)$,
then randomly delete or retain each point,
independently of other points,
with retention probability $p(x,y) = \lambda(x,y)/\ell_{max}$
where $\ell_{max}=\sup_{(x,y) \in (0,1)\times(0,1)}\lambda(x,y)$.

We take $\lambda(x,y)=n_1 \sqrt{x+y}$ for sample $X$.
Then for sample $Y$,
we  take $\lambda(x,y)=n_2 \sqrt{x+y}$ in case 1,
$\lambda(x,y)=n_2 \sqrt{x\,y}$ in case 2,
and $\lambda(x,y)=n_2 |x-y|$ in case 3.
That is, in case 1 $X$ and $Y$ points are from the same inhomogeneous Poisson process;
in cases 2 and 3, they are from different processes.
For each of the above cases,
we take $(n_1,n_2) \in \{(50,50),(50,100),(100,100)\}$.
For more on inhomogeneous Poisson cluster processes,
see (\cite{diggle:2003} and \cite{baddeley:2005}).
The rejection rates of the NNCT-tests are provided
in Table \ref{tab:rejection-rates}.
Observe that under PCP1 with same parents,
the rejection rates are slightly (but significantly) larger than 0.05.
Hence under PCP1, the two classes are slightly segregated,
so they do not satisfy randomness in the NN structure.
Under PCP1 with different parents,
the two classes are strongly segregated.
Under PCP2 with the same parents,
the two classes satisfy randomness in the NN structure,
while under PCP2 with different parents,
the two classes are strongly segregated.
Notice that under these implementations of PCP,
the rejection rates decrease as $\sigma$ increases;
i.e., the level of segregation is inversely related to $\sigma$.
Under MCP with the same parents,
the two classes satisfy randomness in NN structure;
but with different parents, the classes are strongly segregated.
Furthermore, as $r$ increases, the level of segregation decreases
under MCP with different parents.
Under IPCP patterns,
the two classes satisfy randomness in NN structure
as long as the density functions are same or similar
(see cases 1 and 2);
but if the density functions are very different,
we observe moderate segregation between the two classes.
Notice also that this segregation is detected
better by the new NNCT-tests.

\begin{sidewaystable}[ht]
\centering
\begin{tabular}{|c||c|c||c|c||c|c||c|c||c|c||c|c||c|c||c|c||c|c|}
\hline
\multicolumn{19}{|c|}{Rejection Rates of the NNCT-Tests Under Various PCP and Inhomogeneous Patterns} \\
\hline
& \multicolumn{18}{|c|}{$PCP1(n_p,n_1,n_2,\sigma,(0,1)\times(0,1))$ with $n_p=5$,
$n_c=n_1/n_p$ for sample $X$ and $n_c=n_2/n_p$ for sample $Y$(same parent set for $X$ and $Y$)}\\
\hline
& \multicolumn{6}{|c||}{case 1: $\sigma=.025$} & \multicolumn{6}{|c||}{case 2: $\sigma=.05$} &
\multicolumn{6}{|c|}{case 3: $\sigma=.10$}\\
\hline
sizes & \multicolumn{2}{|c|}{Dixon's} & \multicolumn{2}{|c|}{New} & \multicolumn{2}{|c||}{Overall}
& \multicolumn{2}{|c|}{Dixon's} & \multicolumn{2}{|c|}{New} & \multicolumn{2}{|c||}{Overall}
& \multicolumn{2}{|c|}{Dixon's} & \multicolumn{2}{|c|}{New} & \multicolumn{2}{|c|}{Overall} \\
\hline
$(n_1,n_2)$ & $\ah_{1,1}^D$ & $\ah_{2,2}^D$ & $\ah_{1,1}^N$ & $\ah_{2,2}^N$ & $\ah_D$ & $\ah_N$
& $\ah_{1,1}^D$ & $\ah_{2,2}^D$ & $\ah_{1,1}^N$ & $\ah_{2,2}^N$ & $\ah_D$ & $\ah_N$
& $\ah_{1,1}^D$ & $\ah_{2,2}^D$ & $\ah_{1,1}^N$ & $\ah_{2,2}^N$ & $\ah_D$ & $\ah_N$ \\
\hline
$(30,30)$ & .0814 & .0716 & .0726 & .0719 & .0709 & .0735 & .0779 & .0701 & .0707 & .0717
& .0675 & .0727 & .0719 & .0656 & .0630 & .0619 & .0611 & .0631\\
\hline
$(30,50)$ & .0702 & .0728 & .0620 & .0615 & .0571 & .0637 & .0676 & .0684 & .0598 & .0600
& .0582 & .0613 & .0627 & .0634 & .0533 & .0537 & .0538 & .0543\\
\hline
$(50,50)$ & .0624 & .0614 & .0693 & .0694 & .0615 & .0691 & .0605 & .0602 & .0677 & .0677
& .0608 & .0674 & .0556 & .0541 & .0598 & .0599 & .0543 & .0600\\
\hline
& \multicolumn{18}{|c|}{(different parent sets for $X$ and $Y$)}\\
\hline
$(30,30)$ & .9993 & 1.000 & 1.000 & 1.000 & 1.000 & 1.000 & .9880 & .9882 & .9959 & .9958
& .9922 & .9957 & .7641 & .7622 & .8527 & .8522 & .7806 & .8522\\
\hline
$(30,50)$ & 1.000 & .9997 & 1.000 & 1.000 & 1.000 & 1.000 & .9967 & .9950 & .9985 & .9985
& .9964 & .9985 & .8702 & .8271 & .9058 & .9064 & .8555 & .9075\\
\hline
$(50,50)$ & .9999 & .9999 & 1.000 & 1.000  & 1.000 & 1.000 & .9987 & .9988 & .9996 & .9996
& .9993 & .9996 & .8985 & .8984 & .9525 & .9531 & .9232 & .9532\\
\hline
\hline
& \multicolumn{18}{|c|}{$PCP2(n_p,n_1,n_2,\sigma,(0,1)\times(0,1))$ with $n_p=5$,
$n_c=n_1/n_p$ for sample $X$ and $n_c=n_2/n_p$ for sample $Y$ (same parent set for $X$ and $Y$)}\\
\hline
$(30,30)$ & .0509 & .0498 & .0465 & .0475 & .0472 & .0470 & .0523 & .0509 & .0459 & .0470
& .0475 & .0469 & .0512 & .0485 & .0428 & .0438 & .0448 & .0440\\
\hline
$(30,50)$ & .0587 & .0545 & .0483 & .0492 & .0443 & .0499 & .0595 & .0522 & .0492 & .0494
& .0453 & .0509 & .0609 & .0573 & .0502 & .0496 & .0460 & .0510\\
\hline
$(50,50)$ & .0479 & .0499 & .0534 & .0536 & .0483 & .0537 & .0482 & .0487 & .0526 & .0527
& .0474 & .0530 & .0476 & .0466 & .0516 & .0525 & .0487 & .0529\\
\hline
& \multicolumn{18}{|c|}{(different parent sets for $X$ and $Y$)}\\
\hline
$(30,30)$ & .9993 & .9998 & .9999 & 1.000 & .9998 & 1.000 & .9884 & .9878 & .9953 & .9953
& .9910 & .9953 & .7779 & .7812 & .8624 & .8633 & .8007 & .8627\\
\hline
$(30,50)$ & 1.000 & .9999 & 1.000 & 1.000 & .9999 & 1.000 & .9976 & .9945 & .9991 & .9991
& .9977 & .9991 & .8854 & .8445 & .9169 & .9177 & .8677 & .9187\\
\hline
$(50,50)$ & 1.000 & 1.000 & 1.000 & 1.000 & 1.000 & 1.000 & .9987 & .9991 & .9997 & .9997
& .9995 & .9997 & .9182 & .9112 & .9607 & .9600 & .9354 & .9602\\
\hline
\hline
& \multicolumn{18}{|c|}{$MCP(\kappa,r,\mu,(0,1)\times(0,1))$ with $\kappa=5$
for both $X$ and $Y$ samples (same parent set for $X$ and $Y$)}\\
\hline
& \multicolumn{6}{|c||}{case 1: $r=.05$} & \multicolumn{6}{|c||}{case 2: $r=.10$} &
\multicolumn{6}{|c|}{case 3: $r=.20$}\\
\hline
$(50,50)$ & .0494 & .0487 & .0478 & .0483 & .0495 & .0487 & .0530 & .0540 & .0582 & .0580
& .0504 & .0585 & .0514 & .0487 & .0527 & .0519 & .0476 & .0524\\
\hline
$(50,100)$ & .0495 & .0463 & .0466 & .0471 & .0418 & .0473 & .0452 & .0496 & .0481 & .0485
& .0457 & .0487 & .0502 & .0516 & .0489 & .0489 & .0459 & .0497\\
\hline
$(100,100)$ & .0458 & .0508 & .0498 & .0499 & .0471 & .0501 & .0527 & .0495 & .0536 & .0534
& .0498 & .0536 & .0507 & .0496 & .0500 & .0501 & .0493 & .0504\\

\hline
& \multicolumn{18}{|c|}{(different parent sets for $X$ and $Y$)}\\
\hline
$(50,50)$ & .9983 & .9985 & .9997 & .9997 & .9996 & .9997 & .9887 & .9899 & .9954 & .9954
& .9938 & .9954 & .8019 & .8026 & .8873 & .8874 & .8341 & .8881\\
\hline
$(50,100)$ & .9992 & .9981 & .9994 & .9994 & .9993 & .9994 & .9963 & .9927 & .9979 & .9979
& .9973 & .9979 & .9087 & .8674 & .8674 & .9438 & .9149 & .9443\\
\hline
$(100,100)$ & .9992 & .9994 & .9998 & .9998 & .9998 & .9998 & .9988 & .9987 & .9998 & .9998
& .9997 & .9998 & .9581 & .9579 & .9804 & .9805 & .9705 & .9805\\
\hline
\hline
& \multicolumn{18}{|c|}{$IPCP(\lambda(x,y)=n_i\,f_i(x,y),(0,1)\times(0,1))$ with $i=1,2$ for $X$
and $Y$ points, respectively}\\
\hline
& \multicolumn{6}{|c||}{case 1: $f_1(x,y)=f_2(x,y)=\sqrt{x+y}$ }
& \multicolumn{6}{|c||}{case 2: $f_1(x,y)=\sqrt{x+y}$, $f_2(x,y)=\sqrt{x\,y}$} &
\multicolumn{6}{|c|}{case 3: $f_1(x,y)=\sqrt{(x+y)}$, $f_2(x,y)=|x-y|$}\\
\hline
$(50,50)$ & .0541 & .0495 & .0507 & .0507 & .0511 & .0512 & .0487 & .0518 & .0525 & .0527 & .0499
& .0534 & .0617 & .1018 & .1042 & .1048 & .0797 & .1059\\
\hline
$(50,100)$ & .0471 & .0465 & .0459 & .0458 & .0440 & .0463 & .0515 & .0540 & .0563 & .0561 & .0524
& .0564 & .0936 & .1225 & .1408 & .1420 & .0962 & .1423\\
\hline
$(100,100)$ & .0469 & .0516 & .0491 & .0490 & .0496 & .0491 & .0513 & .0572 & .0591 & .0592 & .0532
& .0596 & .0819 & .1326 & .1421 & .1415 & .1071 & .1421\\
\hline
\end{tabular}
\caption{ \label{tab:rejection-rates}
The rejection rates for the NNCT-tests under various patterns
different from CSR and RL.
PCP: Poisson Cluster Process, MCP: Matern Cluster Process,
IPCP: Inhomogeneous Poisson Cluster Process.
See Section \ref{sec:other-process} for details on these point processes.
}
\end{sidewaystable}

\section{Empirical Power Analysis in the Two-Class Case}
\label{sec:emp-power-2Cl}
We consider three cases for each of segregation and association alternatives
in the two-class case.

\subsection{Empirical Power Analysis under Segregation of Two Classes}
\label{sec:power-comp-seg-2Cl}
For the segregation alternatives, we generate
$X_i \stackrel{iid}{\sim} \U((0,1-s)\times(0,1-s))$ and
$Y_j \stackrel{iid}{\sim} \U((s,1)\times(s,1))$
for $i=1,\ldots,n_1$ and $j=1,\ldots,n_2$.
Notice the level of segregation is determined by the magnitude of $s \in (0,1)$.
We consider the following three segregation alternatives:
\begin{equation}
\label{eqn:seg-alt-2Cl}
H_S^I: s=1/6, \;\;\; H_S^{II}: s=1/4, \text{ and } H_S^{III}: s=1/3.
\end{equation}

Observe that, from $H_S^I$ to $H_S^{III}$ (i.e., as $s$ increases), the segregation gets stronger
in the sense that $X$ and $Y$ points tend to form one-class clumps or clusters
more and more frequently.
We calculate the power estimates using the asymptotic critical values
based on the standard normal distribution
for the cell-specific tests
and the corresponding $\chi^2$-distributions for the overall tests and using the Monte Carlo critical values.

\begin{table}[ht]
\centering
\begin{tabular}{|c|c||c|c||c|c||c|c|}
\hline
\multicolumn{8}{|c|}{Empirical power estimates under
the segregation alternatives} \\
\hline
& sizes & \multicolumn{2}{|c||}{Dixon's} & \multicolumn{2}{|c||}{New}
& \multicolumn{2}{c|}{Overall} \\
\hline
& $(n_1,n_2)$  & $\bh_{1,1}^D$ & $\bh_{2,2}^D$ & $\bh_{1,1}^N$ & $\bh_{2,2}^N$
& $\bh_D$ & $\bh_N$ \\

\hline \hline
 & 10,10) & .0734 & .0698 & .1068 & .1060 & .0775 & .1086\\
\cline{2-8}
 & 10,30) & .1436 & .1540 & .1977 & .2019 & .1414 & .1997\\
\cline{2-8}
 & 10,50) & .1639 & .1615 & .2465 & .2491 & .2193 & .2497\\
\cline{2-8}
 & 30,30) & .2883 & .2783 & .3898 & .3894 & .2904 & .3891\\
\cline{2-8}
 & 30,50) & .4491 & .4045 & .5228 & .5243 & .3911 & .5270\\
\cline{2-8}
 & 50,50) & .5091 & .5016 & .6786 & .6793 & .5546 & .6811\\
\cline{2-8}
 & 50,100) & .7686 & .6689 & .8417 & .8420 & .7425 & .8423\\
\cline{2-8}\raisebox{10.ex}[0pt]{$H_S^{I}$}
 & 100,100) & .8761 & .8730 & .9564 & .9567 & .9121 & .9568\\
\hline \hline
 & 10,10) & .2057 & .2044 & .3280 & .3270 & .2305 & .3279\\
\cline{2-8}
 & 10,30) & .4601 & .4133 & .5725 & .5793 & .4555 & .5750\\
\cline{2-8}
 & 10,50) & .5420 & .4477 & .6747 & .6794 & .6174 & .6803\\
\cline{2-8}
 & 30,30) & .7783 & .7769 & .8939 & .8938 & .8141 & .8933\\
\cline{2-8}
 & 30,50) & .9262 & .8775 & .9619 & .9626 & .9126 & .9627\\
\cline{2-8}
 & 50,50) & .9543 & .9551 & .9938 & .9936 & .9777 & .9935\\
\cline{2-8}
 & 50,100) & .9977 & .9866 & .9994 & .9994 & .9975 & .9994\\
\cline{2-8} \raisebox{10.ex}[0pt]{$H_S^{II}$}
 & 100,100) & .9998 & .9999 & 1.000 & 1.000 & 1.000 & 1.000\\
\hline \hline
 & 10,10) & .5144 & .5121 & .7324 & .7320 & .5817 & .7296\\
\cline{2-8}
 & 10,30) & .8873 & .7833 & .9402 & .9425 & .8787 & .9409\\
\cline{2-8}
 & 10,50) & .9353 & .8002 & .9699 & .9711 & .9528 & .9713\\
\cline{2-8}
 & 30,30) & .9929 & .9915 & .9990 & .9990 & .9969 & .9990\\
\cline{2-8}
 & 30,50) & .9999 & .9979 & 1.000 & 1.000 & .9997 & 1.000\\
\cline{2-8}
 & 50,50) & .9999 & 1.000 & 1.000 & 1.000 & 1.000 & 1.000\\
\cline{2-8}
 & 50,100) & 1.000 & 1.000 & 1.000 & 1.000 & 1.000 & 1.000\\
\cline{2-8} \raisebox{10.ex}[0pt]{$H_S^{III}$}
 & 100,100) & 1.000 & 1.000 & 1.000 & 1.000 & 1.000 & 1.000\\
\hline
\end{tabular}
\caption{ \label{tab:emp-power-seg} The empirical power estimates
for the tests under the segregation alternatives, $H_S^I$,
$H_S^{II}$, and $H_S^{III}$ in the two-class case with
$N_{mc}=10000$, for some combinations of $n_1,n_2 \in \{10,30,50\}$ at $\alpha=.05$.
The power estimates that are not significantly different between Dixon's
and the new cell specific tests are marked with an asterisk (*).
For all others the larger power estimate is significantly larger than the other at $\alpha=.05$.
$\bh_D$ and $\bh_N$ stand for empirical power estimates for Dixon's and new overall tests,
respectively.
$\bh^D_{ii}$ and $\bh^N_{ii}$ stand for empirical power estimates for
Dixon's and cell-specific tests, respectively, for cell $(i,i)$ with $i=1,2$.
}
\end{table}

%
%

The power estimates based on the asymptotic critical values
are presented in Table \ref{tab:emp-power-seg}.
We omit the power estimates of the cell-specific tests for cells $(1,2)$ and $(2,1)$,
since $\bh^D_{1,1}=\bh^D_{1,2}$ and $\bh^D_{2,1}=\bh^D_{2,2}$;
likewise $\bh^N_{1,1}=\bh^N_{2,1}$ and $\bh^N_{1,2}=\bh^N_{2,2}$.
Observe that, for both cell-specific tests,
as $n=(n_1+n_2)$ gets larger, the power estimates get larger;
for the same $n=(n_1+n_2)$ values, the power estimate is larger
for classes with similar sample sizes;
and as the segregation gets stronger, the power estimates get larger
at each sample size combination.
For both cells $(1,1)$ and $(2,2)$,
the new cell-specific tests have higher power estimates
compared to those of Dixon's.
Furthermore, the new overall test has higher power estimates
compared to Dixon's overall test.

\begin{figure}
\centering
Empirical Power Estimates of the NNCT-Tests under $H_S$\\
\rotatebox{-90}{ \resizebox{2. in}{!}{\includegraphics{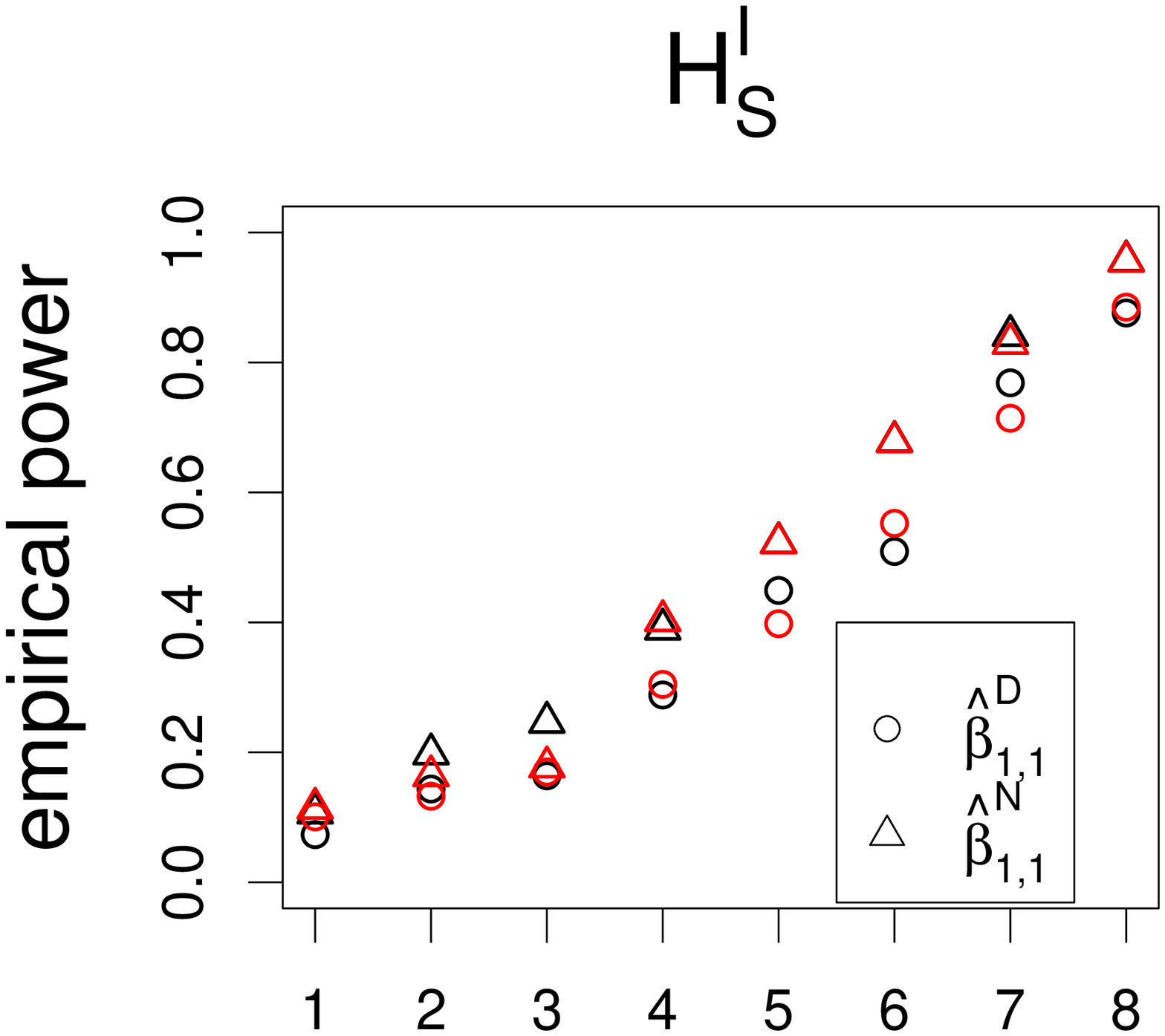} }}
\rotatebox{-90}{ \resizebox{2. in}{!}{\includegraphics{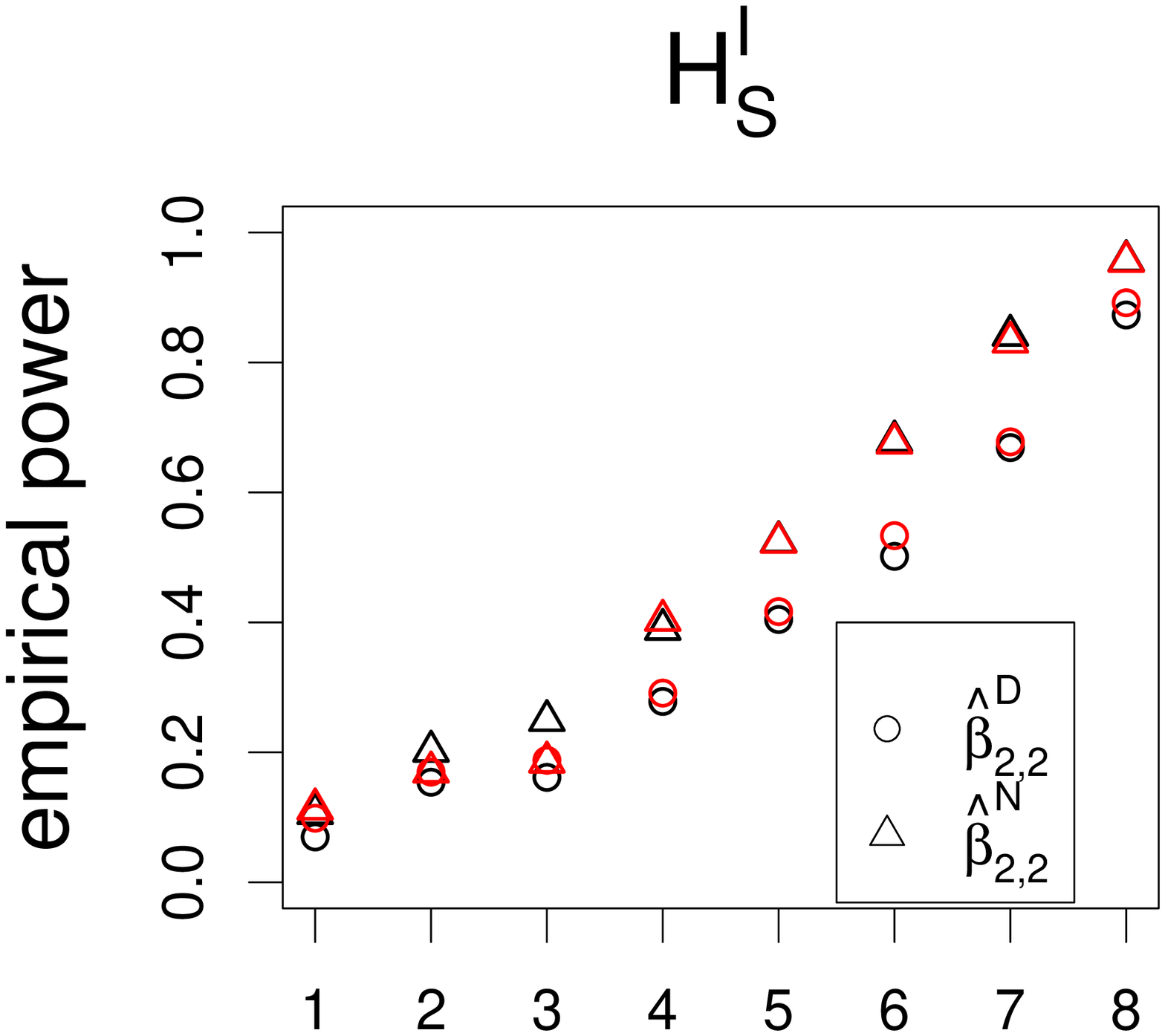} }}
\rotatebox{-90}{ \resizebox{2. in}{!}{\includegraphics{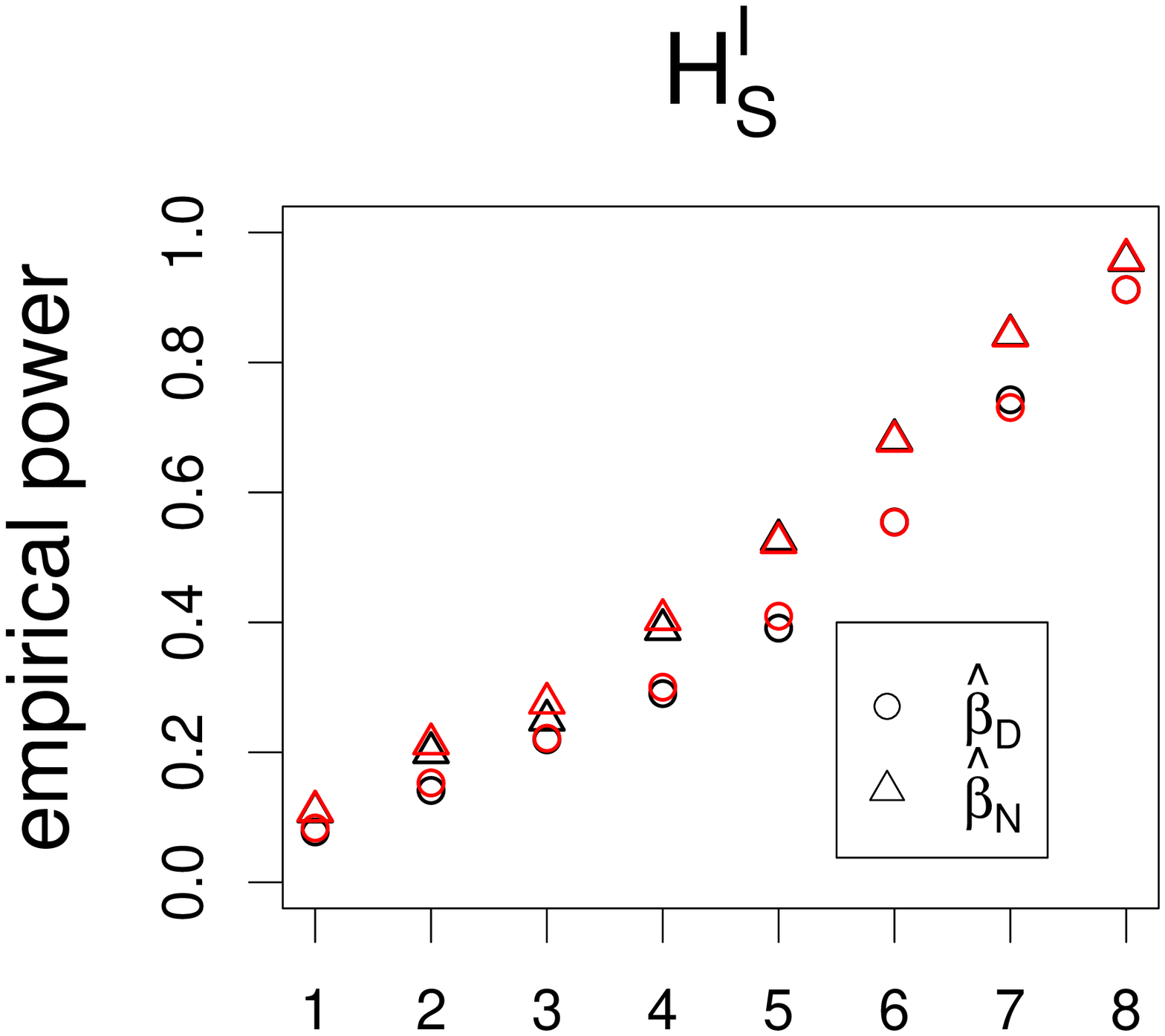} }}

\caption{
\label{fig:power-seg-MC-2cl}
The empirical power estimates for Dixon's tests (circles ($\circ$))
based on asymptotic critical values (black) and Monte Carlo critical values (red)
and new tests (triangles ($\triangle$))
based on asymptotic critical values (black) and Monte Carlo critical values (red)
under the segregation alternative $H^I_S$
in the two-class case.
The horizontal axis labels:
1=(10,10), 2=(10,30), 3=(10,50), 4=(30,30), 5=(30,50),
6=(50,50), 7=(50,100), 8=(100,100).
}
\end{figure}

The power estimates based on the asymptotic and Monte Carlo critical values
under $H^I_S$ are plotted in Figure \ref{fig:power-seg-MC-2cl}.
Observe that the power estimates based on Monte Carlo critical values
are very similar to
but tend to be slightly larger compared to the ones using the asymptotic critical values.
However, this difference do not influence the trend in the power estimates,
that is new tests tend to have higher power
based on either asymptotic critical values or Monte Carlo critical values.
Hence we omit the power estimates based on Monte Carlo critical values
under other segregation alternatives.

Considering the empirical significance levels and power estimates,
for small samples we recommend Monte Carlo randomization tests;
for large samples,
the new version of the cell-specific tests in the two-class case
when testing against the segregation alternatives,
as they are at the desired level for more sample size combinations
and have higher power for each cell.
Likewise, we recommend the new overall test over the use of Dixon's overall test
for the segregation alternatives.

\subsection{Empirical Power Analysis under Association of Two Classes}
\label{sec:power-comp-assoc-2Cl}
For the association alternatives, we consider three cases also.
In each case, first we generate $X_i \stackrel{iid}{\sim} \U((0,1)\times(0,1))$ for $i=1,2,\ldots,n_1$.
Then we generate $Y_j$ for $j=1,2,\ldots,n_2$ as follows.
For each $j$, we pick an $i$ randomly, then generate $Y_j$ as
$X_i+R_j\,\left( \cos T_j, \sin T_j \right)'$ where
$R_j \stackrel{iid}{\sim} \U(0,r)$ with $r \in (0,1)$ and
$T_j \stackrel{iid}{\sim} \U(0,2\,\pi)$.
In the pattern generated, appropriate choices of
$r$ will imply association between classes $X$ and $Y$.
That is, it will be  more likely to have $(X,Y)$ or $(Y,X)$
NN pairs than same-class NN pairs (i.e., $(X,X)$ or $(Y,Y)$).
The three values of $r$ we consider constitute
the following association alternatives;
\begin{equation}
\label{eqn:assoc-alt-2Cl}
H_A^{I}: r=1/4,\;\;\; H_A^{II}: r=1/7, \text{ and } H_A^{III}: r=1/10.
\end{equation}

Observe that, from $H_A^I$ to $H_A^{III}$ (i.e., as $r$ decreases),
the association gets stronger
in the sense that $X$ and $Y$ points tend to occur
together more and more frequently.
By construction, for similar sample sizes the association between $X$ and $Y$ are
at about the same degree as association between $Y$ and $X$.
For very different samples, larger sample is associated with the smaller
but the abundance of the larger sample confounds its association with the smaller.

\begin{table}[ht]
\centering
\begin{tabular}{|c|c||c|c||c|c||c|c|}
\hline
\multicolumn{8}{|c|}{Empirical power estimates under
the association alternatives} \\
\hline
& sizes & \multicolumn{2}{|c||}{Dixon's} & \multicolumn{2}{|c||}{New}
& \multicolumn{2}{c|}{Overall} \\
\hline
& $(n_1,n_2)$  & $\bh_{1,1}^D$ & $\bh_{2,2}^D$ & $\bh_{1,1}^N$ & $\bh_{2,2}^N$
& $\bh_D$ & $\bh_N$ \\

\hline \hline
& (10,10) & .1349 & .1776 & .1638 & .1689 & .1105 & .1792\\
\cline{2-8}
& (10,30) & .0002 & .4366 & .2575 & .2728 & .3007 & .2838\\
\cline{2-8}
& (10,50) & .0002 & .4947 & .0686 & .1071 & .3318 & .1536\\
\cline{2-8}
& (30,30) & .1413 & .2434 & .2110 & .2134 & .1697 & .2138\\
\cline{2-8}
& (30,50) & .1833 & .3984 & .3268 & .3314 & .2903 & .3335\\
\cline{2-8}
& (50,50) & .1149 & .2421 & .2151 & .2181 & .1738 & .2181\\
\cline{2-8}
& (50,100) & .1813 & .4411 & .3448 & .3497 & .3410 & .3534\\
\cline{2-8} \raisebox{10.ex}[0pt]{$H_A^{I}$}
& (100,100) & .0853 & .2309 & .1720 & .1740 & .1677 & .1750\\
\hline \hline
& (10,10) & .2499 & .2569 & .2898 & .2900 & .1834 & .3006\\
\cline{2-8}
& (10,30) & .0000 & .6463 & .4919 & .5123 & .4956 & .5255\\
\cline{2-8}
& (10,50) & .0000 & .7062 & .1959 & .2699 & .5500 & .3418\\
\cline{2-8}
& (30,30) & .4053 & .4457 & .5267 & .5293 & .4141 & .5294\\
\cline{2-8}
& (30,50) & .4896 & .6957 & .7196 & .7239 & .6332 & .7258\\
\cline{2-8}
& (50,50) & .4034 & .4961 & .5824 & .5848 & .4616 & .5854\\
\cline{2-8}
& (50,100) & .5527 & .7991 & .8003 & .8043 & .7559 & .8070\\
\cline{2-8} \raisebox{10.ex}[0pt]{$H_A^{II}$}
& (100,100) & .3868 & .5575 & .5944 & .5957 & .5013 & .5981\\
\hline \hline
& (10,10) & .3038 & .2918 & .3475 & .3471 & .2222 & .3554\\
\cline{2-8}
& (10,30) & .0000 & .7364 & .6115 & .6290 & .6003 & .6407\\
\cline{2-8}
& (10,50) & .0000 & .7907 & .2885 & .3718 & .6512 & .4522\\
\cline{2-8}
& (30,30) & .6092 & .6011 & .7308 & .7301 & .6157 & .7319\\
\cline{2-8}
& (30,50) & .7211 & .8491 & .9052 & .9072 & .8386 & .9082\\
\cline{2-8}
& (50,50) & .6842 & .6891 & .8289 & .8302 & .7285 & .8299\\
\cline{2-8}
& (50,100) & .8024 & .9442 & .9631 & .9640 & .9433 & .9648\\
\cline{2-8} \raisebox{10.ex}[0pt]{$H_A^{III}$}
& (100,100) & .7207 & .7973 & .8831 & .8828 & .8030 & .8834\\
\hline
\end{tabular}
\caption{ \label{tab:emp-power-assoc} The empirical power estimates
for the tests under the association alternatives $H_A^I$,
$H_A^{II}$, and $H_A^{III}$ in the two-class case with
$N_{mc}=10000$, for some combinations of $n_1,n_2 \in \{10,30,50\}$ at $\alpha=.05$.
The power estimates that are not significantly different between Dixon's
and the new cell specific tests are marked with an asterisk (*).
For all others the larger power estimate is significantly larger than the other at $\alpha=.05$.}
\end{table}

%
%

The empirical power estimates are presented in Table \ref{tab:emp-power-assoc}.
Observe that the power estimates
increase as the association gets stronger at each sample size combination
and the power estimates increase as the equal sample sizes increase
and as the very different sample sizes increase under each association alternative.

Dixon's cell-specific
test for cell $(1,1)$ has extremely poor performance
for very different small samples (i.e., $n_1 \leq 10$ and $n_1 \neq n_2$).
On the other hand, for larger samples,
the empirical power estimates get larger as association gets stronger
at each sample size combination.
When samples are large, class $Y$ is more associated with class $X$
if $n_2 > n_1$ and this is reflected in the empirical power estimates.
The power estimates for the new cell-specific test for cell $(1,1)$
increase as the association gets stronger and equal sample sizes increase.
Both tests have the lowest power estimates for $(n_1,n_2)=(10,50)$,
since cell counts and column sums could be very small for this sample size combination.

Dixon's cell-specific test for cell $(2,2)$
has higher power estimates under weak association
compared to those of the new cell-specific test.
When association gets stronger, power estimates for Dixon's cell-specific
test for cell $(2,2)$ has higher power for smaller samples
and lower power for larger samples compared to the new cell-specific tests.
The new cell-specific test has the worst performance for $(n_1,n_2)=(10,50)$,
in which case, column sums could be small.

Dixon's overall test has similar power as the new overall test
for smaller samples;
and new overall test has higher power estimates for larger samples.

Furthermore, empirical power estimates based on Monte Carlo critical values
exhibit similar behavior hence not presented.

Considering the empirical significance levels and power estimates,
for small samples we recommend Monte Carlo randomization or simulation approach;
for larger samples
we recommend both Dixon's and new overall and cell-specific tests
for testing against the association alternatives,
as it will not be very likely to know the degree of association a priori.

\section{Empirical Power Analysis in the Three-Class Case}
\label{sec:emp-power-3Cl}
We consider three cases for each of segregation and association alternatives in the
three-class case.

\subsection{Empirical Power Analysis under Segregation of Three Classes}
\label{sec:power-comp-seg-3Cl}
For the segregation alternatives, we generate
$X_i \stackrel{iid}{\sim} \U((0,1-2s)\times(0,1-2s))$,
$Y_j \stackrel{iid}{\sim} \U((2s,1)\times(2s,1))$, and
$Z_{\ell} \stackrel{iid}{\sim} \U((s,1-s)\times(s,1-s))$
for $i=1,\ldots,n_1$, $j=1,\ldots,n_2$, and $\ell=1,\ldots,n_3$.
Notice that the level of segregation is determined by the magnitude of $s \in (0,1/2)$.
We consider the following three segregation alternatives:
\begin{equation}
\label{eqn:seg-alt-3Cl}
H_{S_1}: s=1/12, \;\;\; H_{S_2}: s=1/8, \text{ and } H_{S_3}: s=1/6.
\end{equation}


Observe that, from $H_{S_1}$ to $H_{S_3}$ (i.e., as $s$ increases), the segregation gets stronger
in the sense that $X$, $Y$, and $Z$ points tend to form one-class clumps or clusters more frequently.
Furthermore, for each segregation alternative, $X$ and $Y$ are more segregated
compared to $Z$ and $X$ or $Z$ and $Y$.

\begin{figure}
\centering
Empirical Power Estimates of Cell-Specific Tests under $H_S$\\
\rotatebox{-90}{ \resizebox{2. in}{!}{\includegraphics{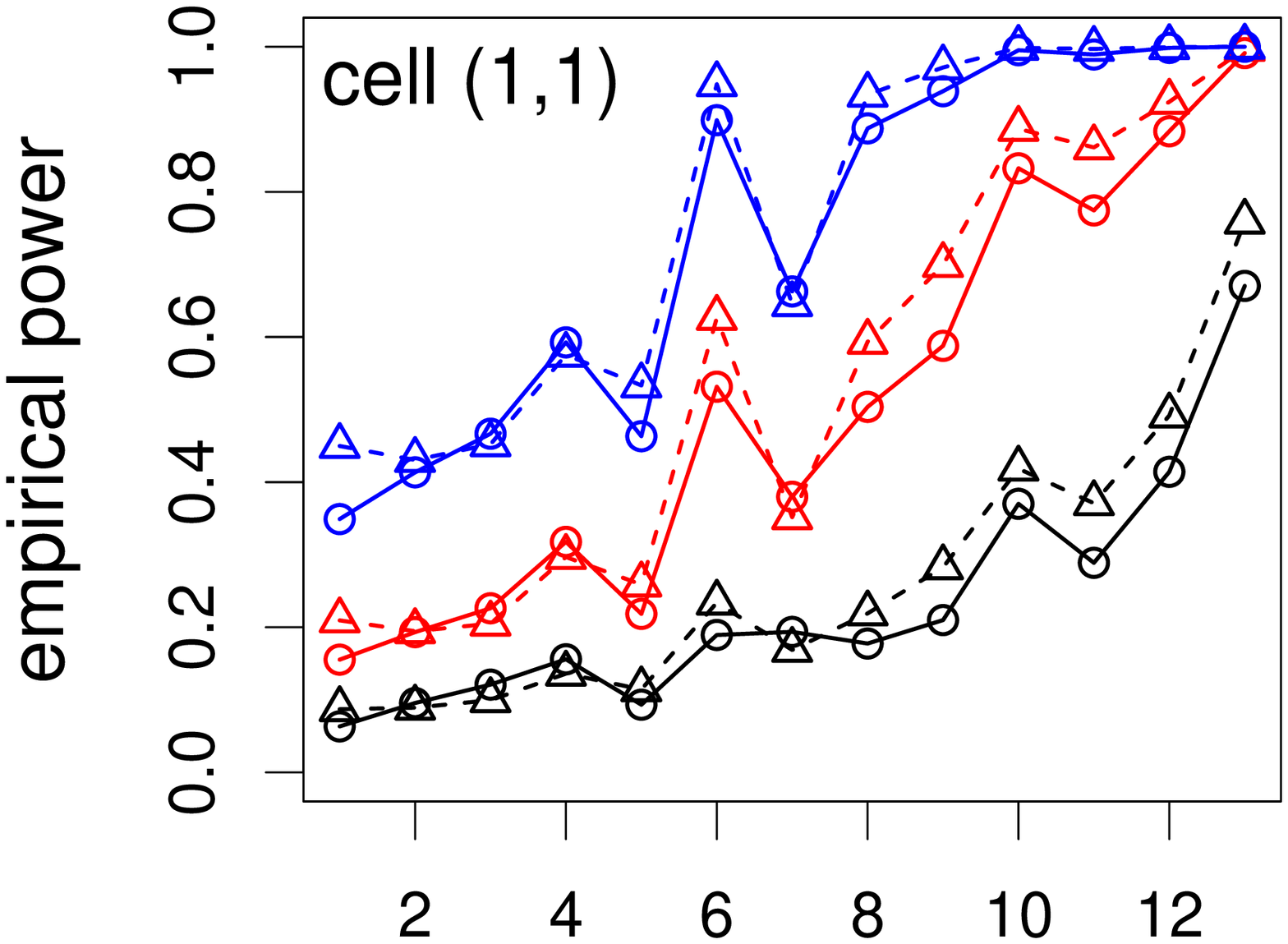} }}
\rotatebox{-90}{ \resizebox{2. in}{!}{\includegraphics{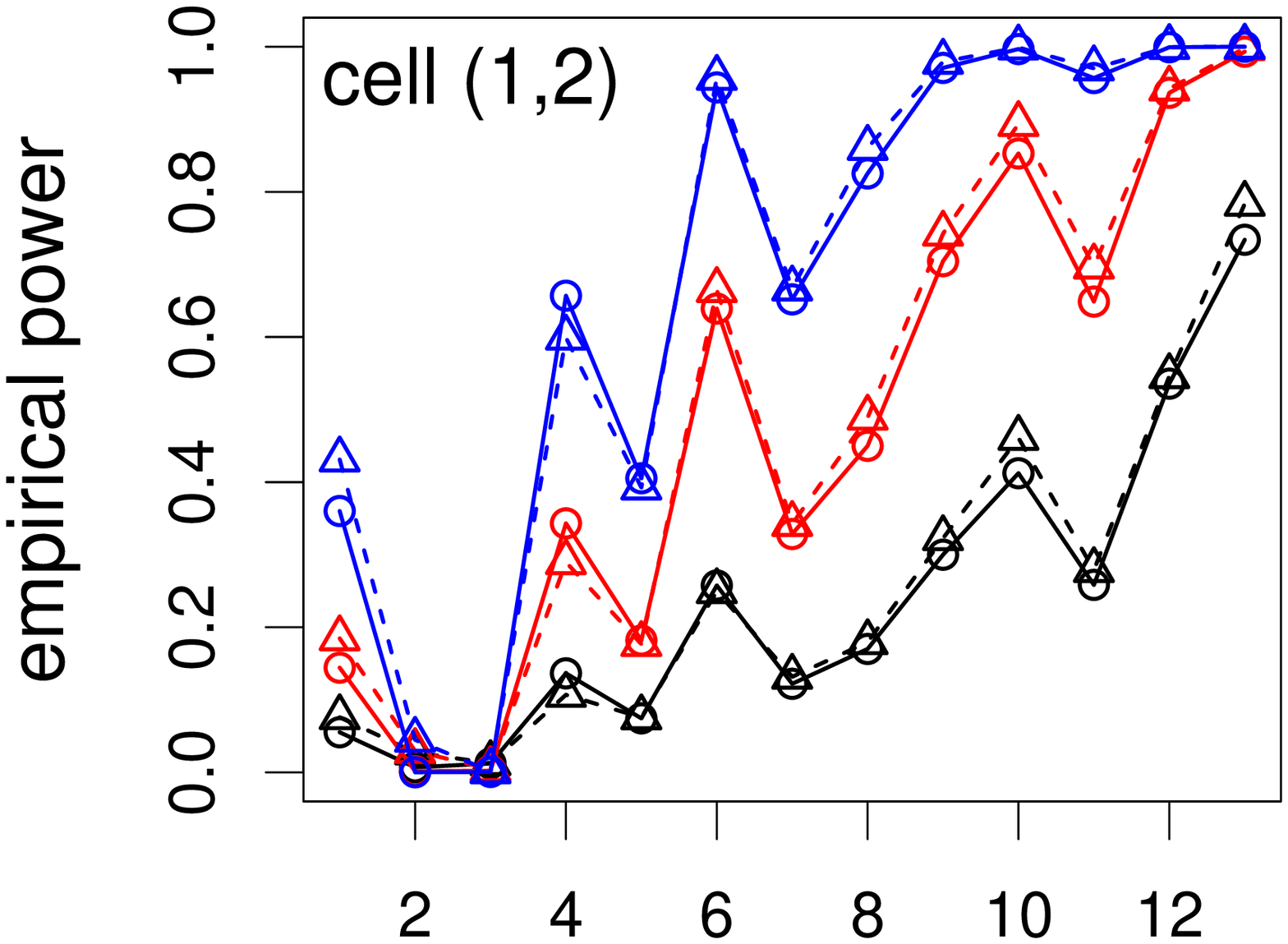} }}
\rotatebox{-90}{ \resizebox{2. in}{!}{\includegraphics{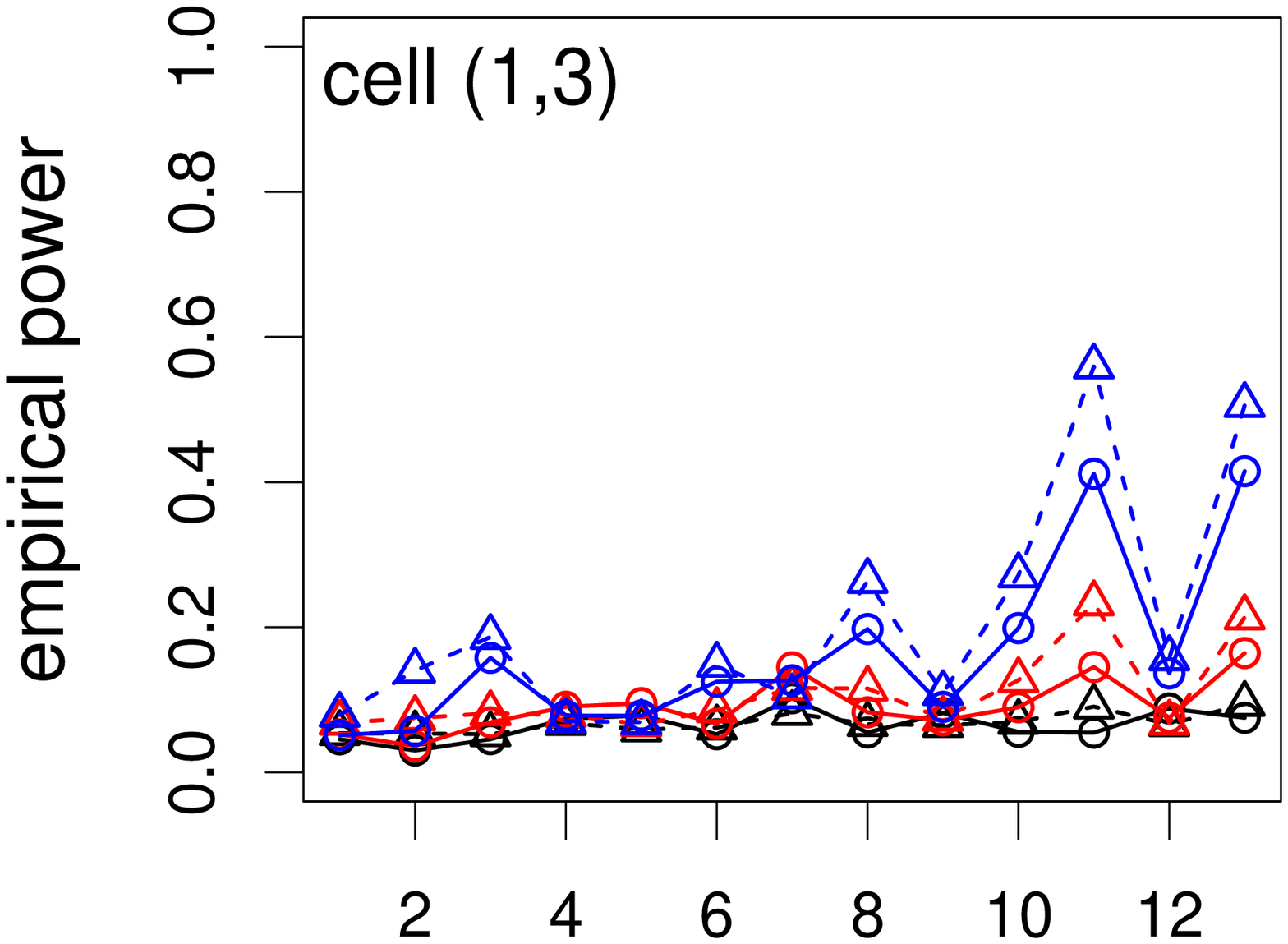} }}

\rotatebox{-90}{ \resizebox{2. in}{!}{\includegraphics{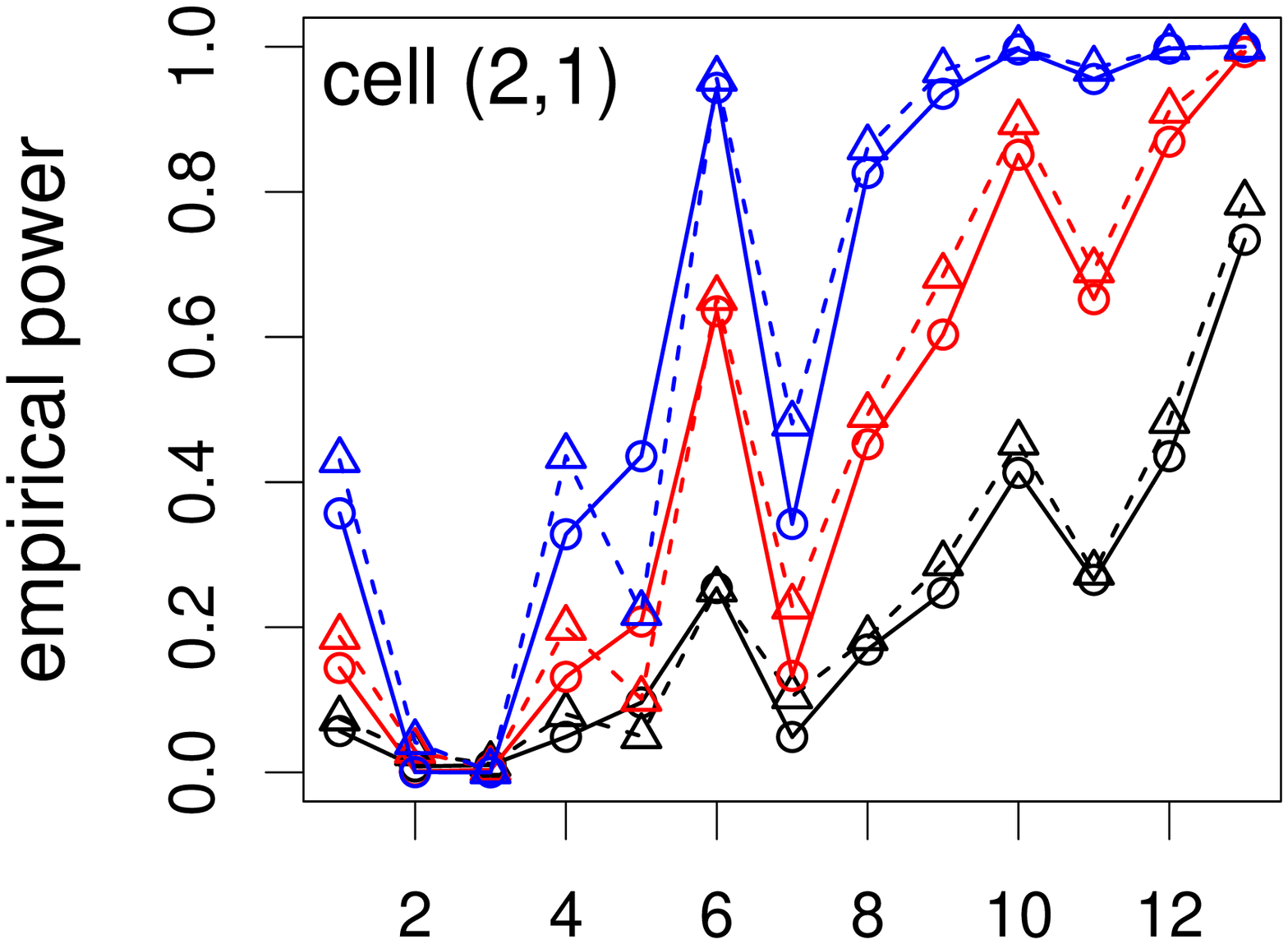} }}
\rotatebox{-90}{ \resizebox{2. in}{!}{\includegraphics{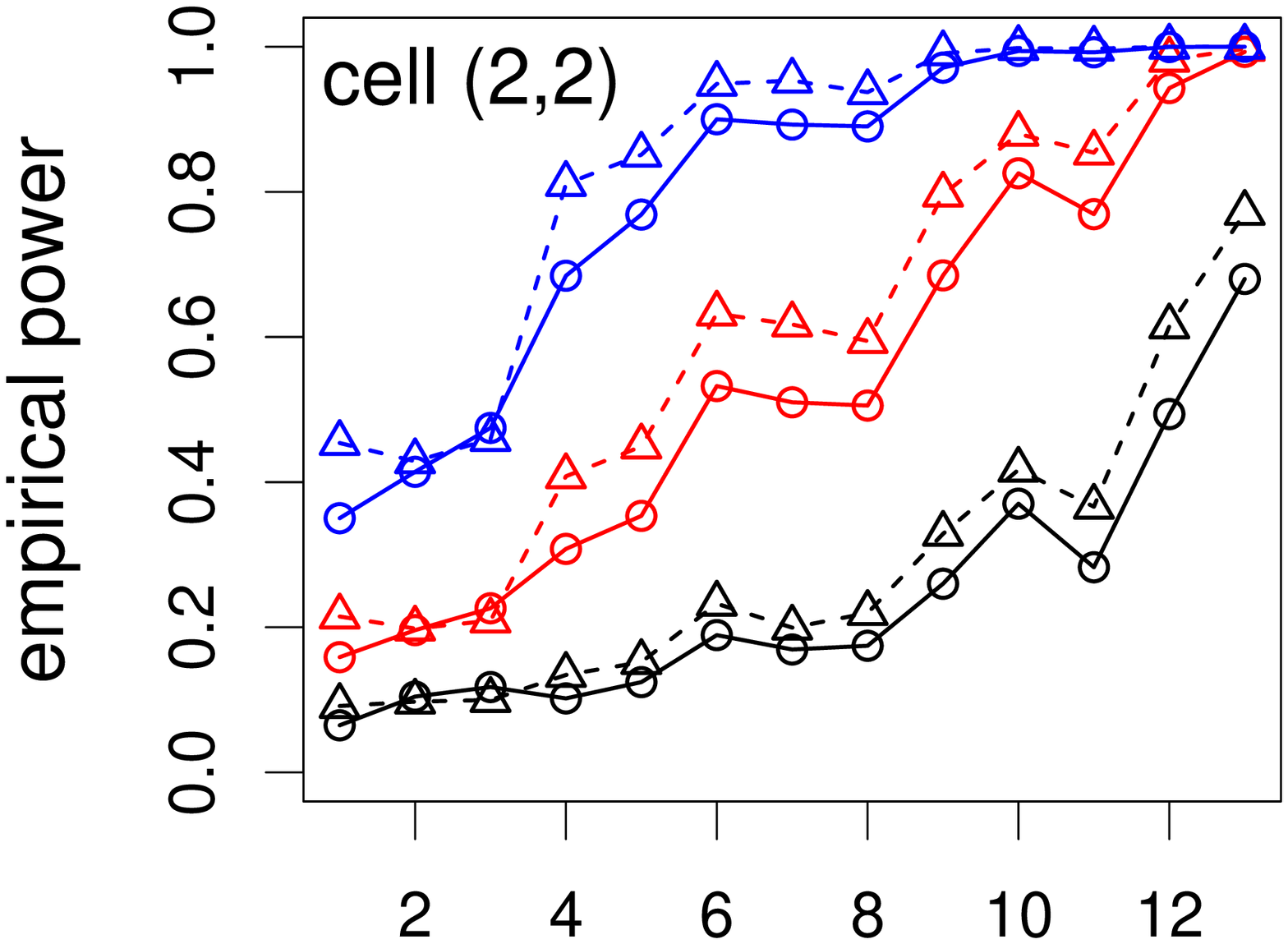} }}
\rotatebox{-90}{ \resizebox{2. in}{!}{\includegraphics{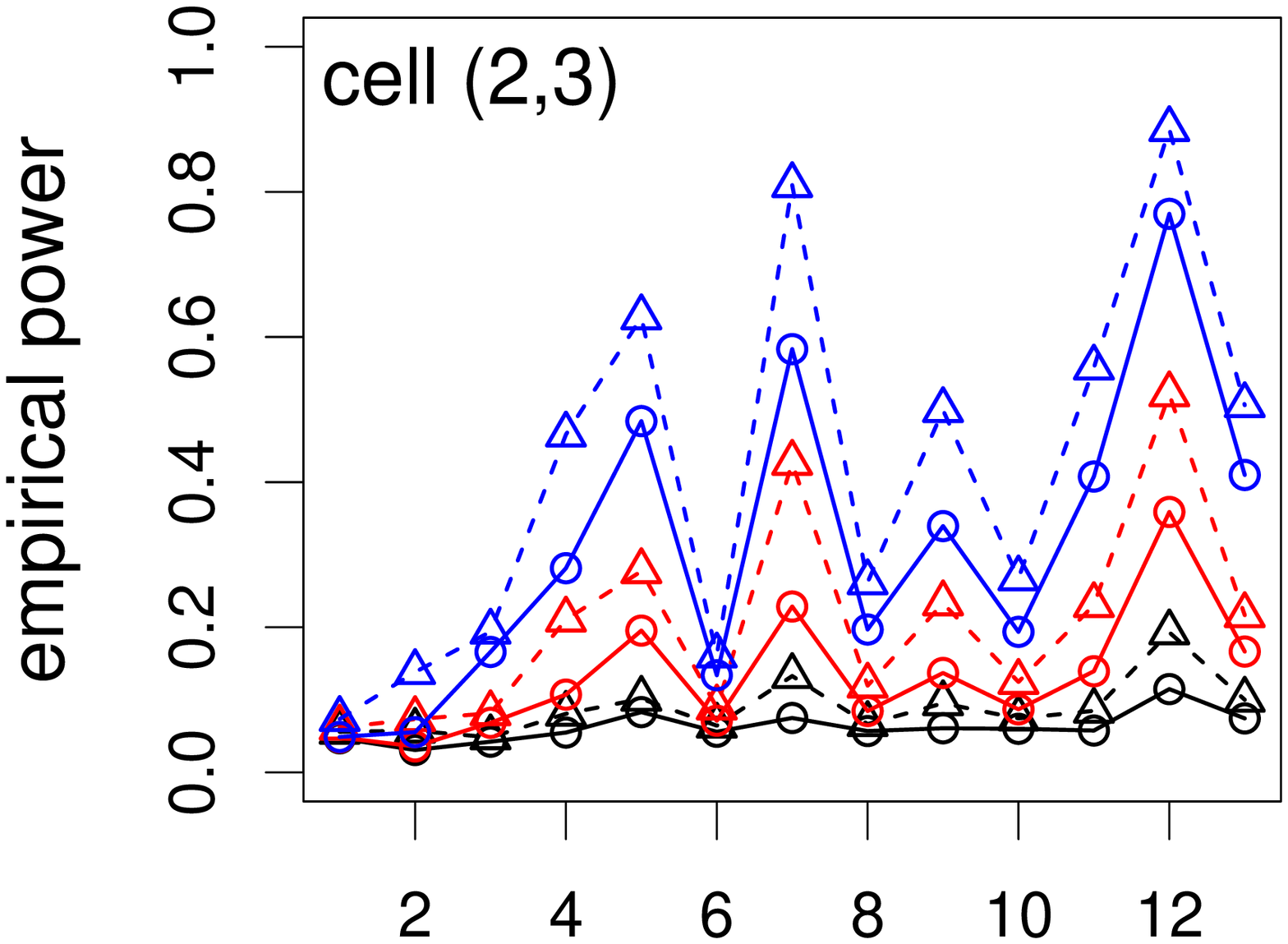} }}

\rotatebox{-90}{ \resizebox{2. in}{!}{\includegraphics{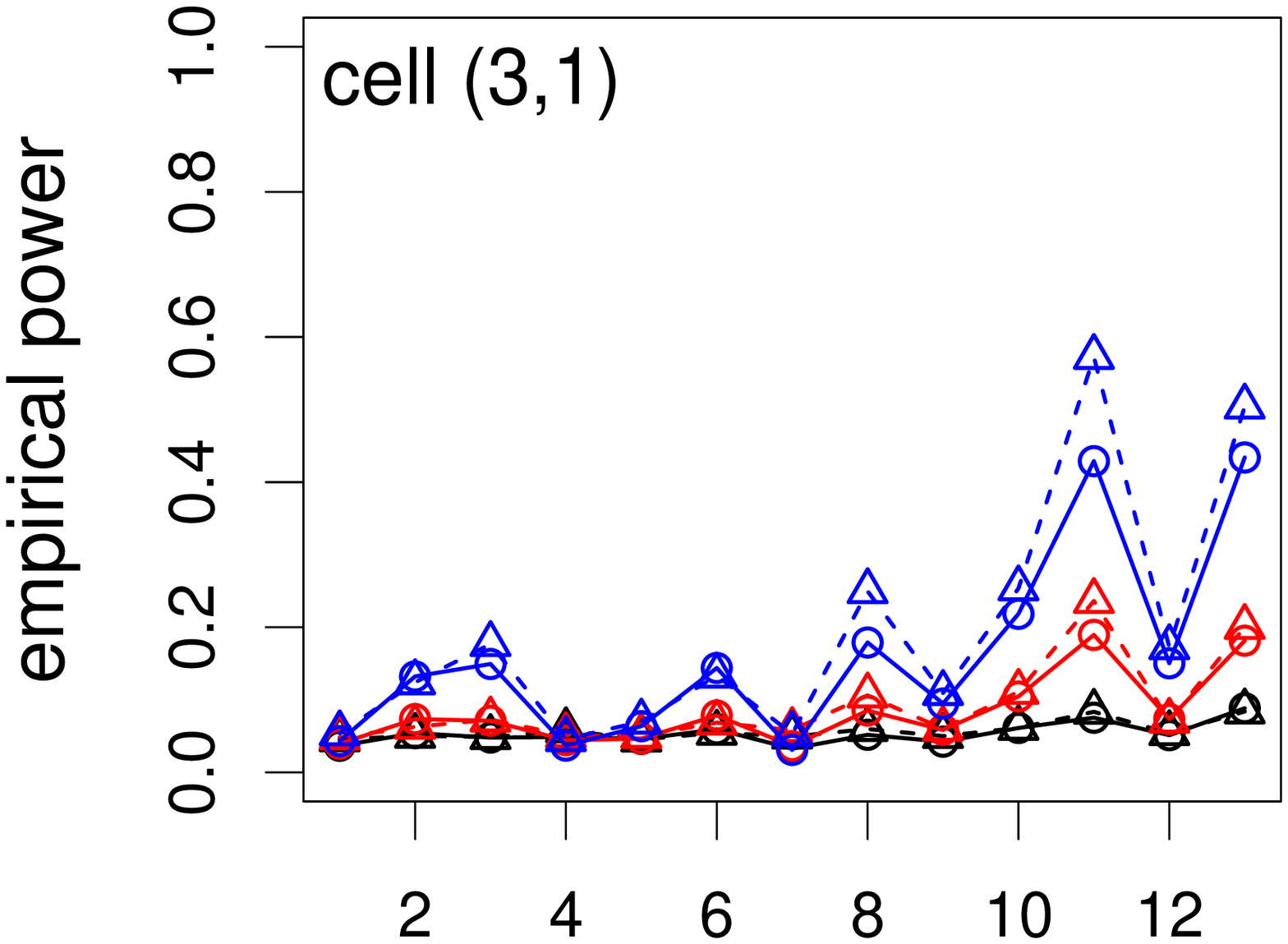} }}
\rotatebox{-90}{ \resizebox{2. in}{!}{\includegraphics{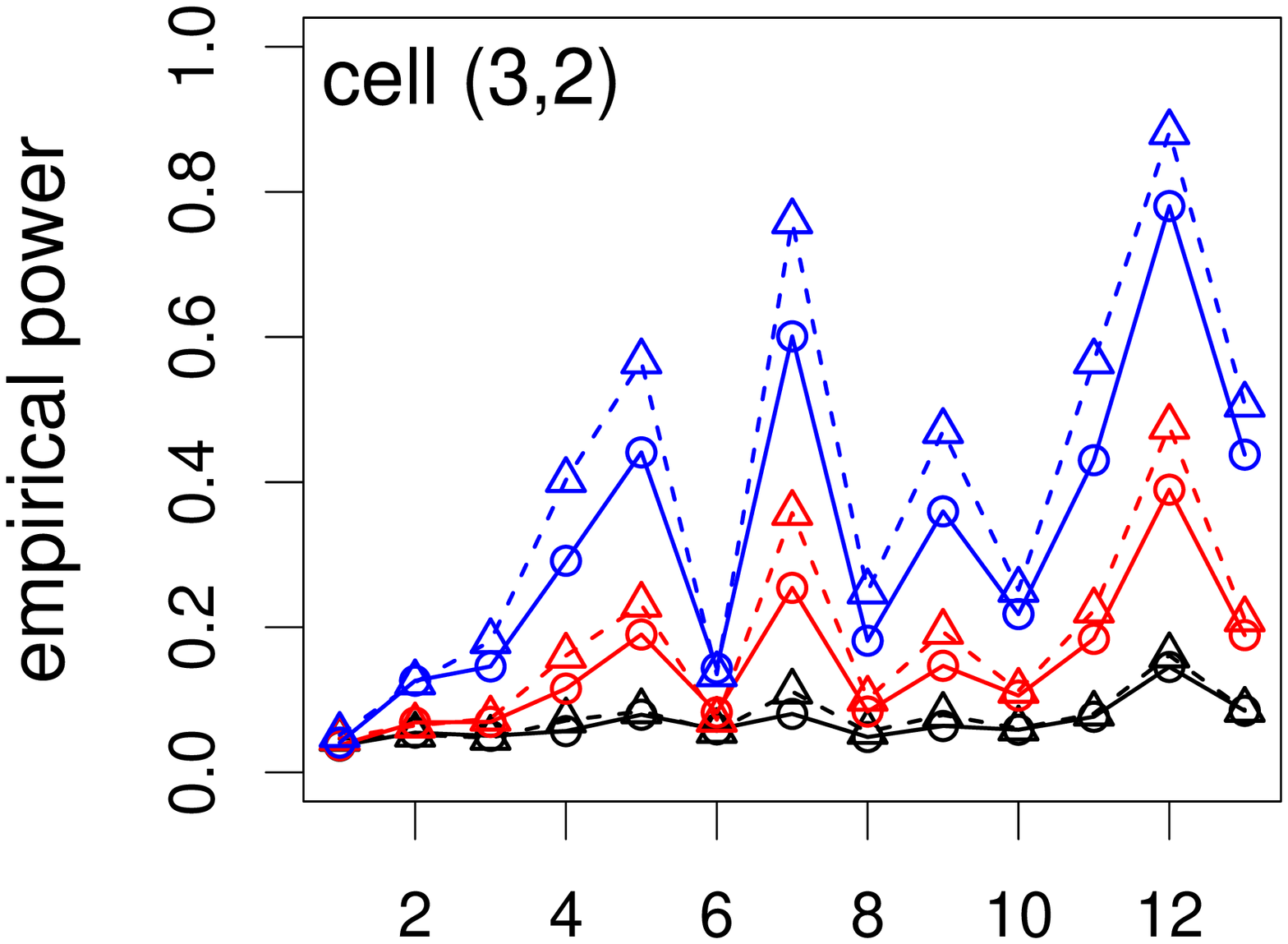} }}
\rotatebox{-90}{ \resizebox{2. in}{!}{\includegraphics{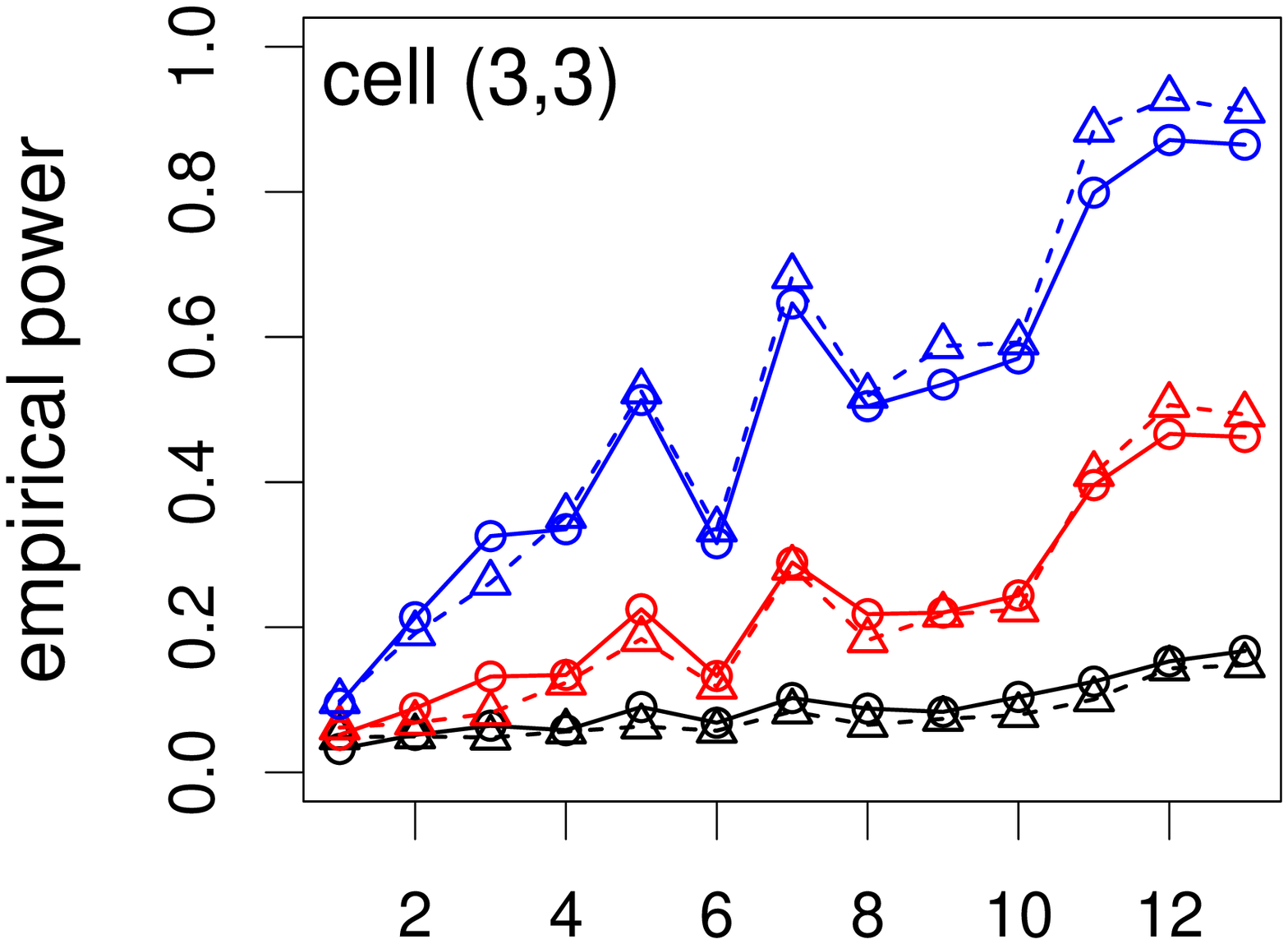} }}
 \caption{
\label{fig:power-seg-cell-3cl}
The empirical power estimates of Dixon's cell-specific tests (circles ($\circ$))
and the new cell-specific tests (triangles ($\triangle$)) under the segregation alternatives
$H_{S_1}$ (black), $H_{S_2}$ (red), and $H_{S_3}$ (blue) in the three-class case.
The horizontal axis labels are:
1=(10,10,10), 2=(10,10,30), 3=(10,10,50), 4=(10,30,30), 5=(10,30,50),
6=(30,30,30), 7=(10,50,50), 8=(30,30,50), 9=(30,50,50), 10=(50,50,50),
11=(50,50,100), 12=(50,100,100), 13=(100,100,100).
Notice that they are arranged in the increasing order for the first and then the second entries.
The size values for discrete sample size combinations are joined
by piecewise straight lines for better visualization.
}
\end{figure}

\begin{figure}[t]
\centering
Empirical Power Estimates of Overall Tests under $H_S$\\
\rotatebox{-90}{ \resizebox{2.5 in}{!}{\includegraphics{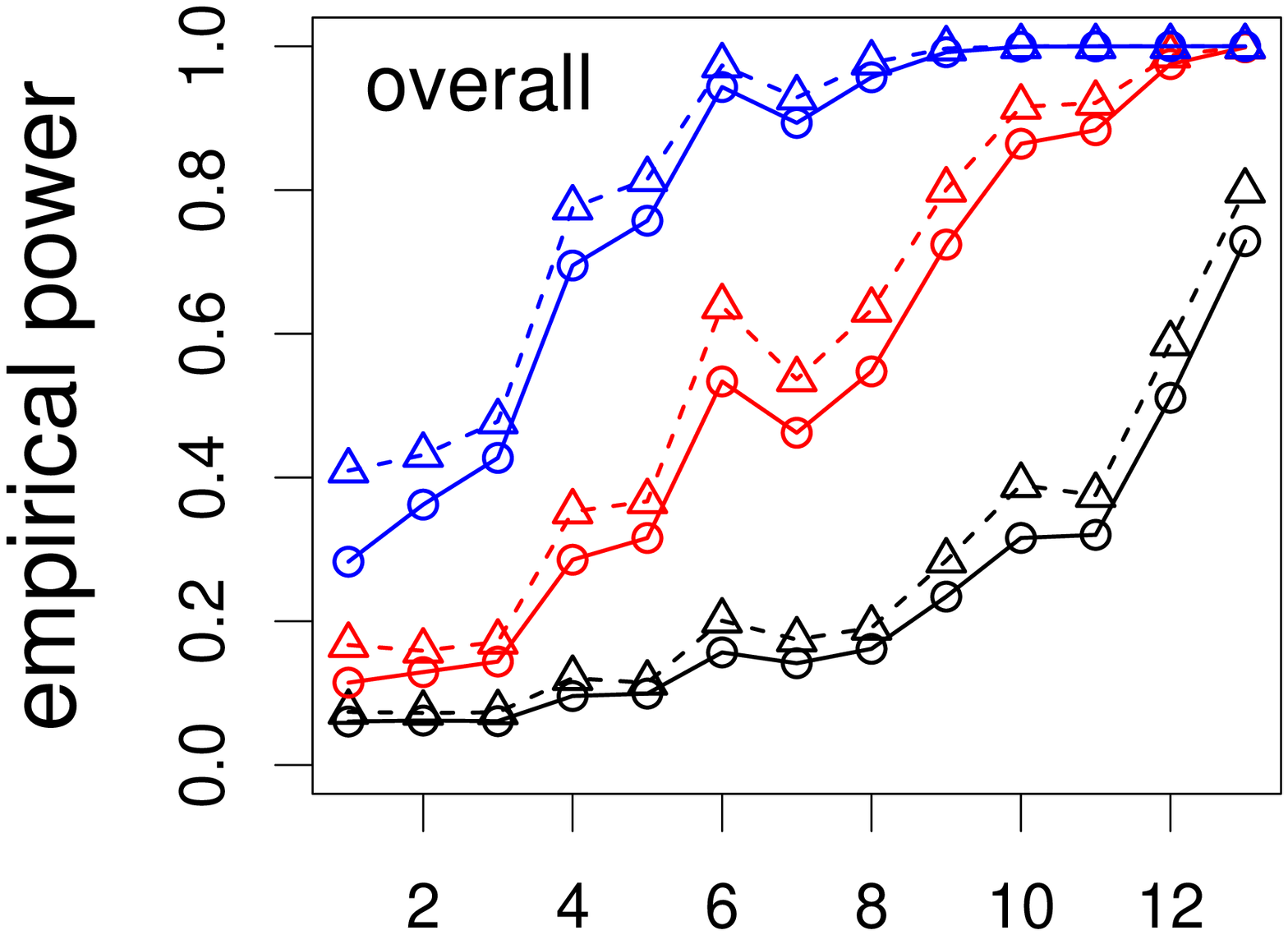} }}

\caption{
\label{fig:power-seg-overall-3cl}
The empirical power estimates of Dixon's overall test (circles ($\circ$))
and the new overall test (triangles ($\triangle$)) under the segregation alternatives
$H_{S_1}$ (black), $H_{S_2}$ (red), and $H_{S_3}$ (blue) in the three-class case.
The horizontal axis labels are as in Figure \ref{fig:power-seg-cell-3cl}.
}
\end{figure}

We plot the empirical power estimates for the NNCT-tests
in Figures \ref{fig:power-seg-cell-3cl} and \ref{fig:power-seg-overall-3cl}.
The test statistics are mostly positive for diagonal cells which implies segregation of classes and
are mostly negative for off-diagonal cells which implies lack of association between classes.
For both cell-specific tests for the diagonal cells $(i,i)$ for $i=1,2,3$,
as equal sample sizes get larger, the power estimates get larger
under each segregation alternative;
and as the segregation gets stronger, the power estimates get larger
for each cell at each sample size combination.
The higher degree of segregation between $X$ and $Y$ is reflected in cells $(1,1)$ and $(2,2)$.
Furthermore, since the sample sizes satisfy $n_1 \leq n_2$ in our simulation study, cell $(2,2)$
power estimates tend to be larger.
Since class $Z$ is less segregated from the other two classes,
cell $(3,3)$ power estimates tend to be lower than the other diagonal cell statistics.
Notice also that off-diagonal cells are more severely affected
by the differences in the sample sizes.

The higher degree of segregation between classes $X$ and $Y$ can also
be observed in cells $(1,2)$ and $(2,1)$ power estimates,
since more segregation of these classes imply higher negative values
in these cells' test statistics.
The lesser degree of segregation between classes $X$ and $Z$
can  be observed in cells $(1,3)$ and $(3,1)$, as they
yield much lower power estimates compared to the other cells.
Although $Y$ and $Z$ are segregated in the same degree as $X$ and $Z$,
the power estimates for cells $(2,3)$ and $(3,2)$ are larger than
those for cells $(1,3)$ and $(3,1)$, since $(n_1+n_3) \leq (n_2+n_3)$ in our simulation study
and larger sample sizes imply higher power under the same degree of segregation.

Furthermore, the power estimates for the new cell-specific tests
tend to be higher for each cell under each segregation alternative
for each sample size combination.
In summary, in the three-class case,
new cell-specific tests have better performance in terms of power.

The performance of the overall tests are similar to the
performance of cell-specific tests for the diagonal cells:
power estimates increase as the segregation gets stronger;
power estimates increase as the sample sizes increase;
and new overall test has higher power than Dixon's overall test.

The empirical power estimates based on the Monte Carlo critical values
yield similar results, hence not presented.

Considering the empirical significance levels and power estimates,
for small samples we recommend Monte Carlo randomization for these tests;
for larger samples
we recommend the new versions of the overall and cell-specific tests for
testing against the segregation alternatives,
as they either have about the same power as
or have larger power than Dixon's tests.
Furthermore, if one wants to see the level
of segregation between pairs of classes,
we recommend using the diagonal cells, $(i,i)$ for $i=1,2,3$
as they are more robust to the differences in class sizes
(i.e., relative abundance) and more sensitive to the level of segregation.

\subsection{Empirical Power Analysis under Association of Three Classes}
\label{sec:power-comp-assoc-3Cl}
For the association alternatives, we also consider three cases.
In each case, first we generate $X_i \stackrel{iid}{\sim} \U((0,1)\times(0,1))$ for $i=1,2,\ldots,n_1$.
Then we generate $Y_j$ and $Z_{\ell}$ for $j=1,2,\ldots,n_2$ and $\ell=1,2,\ldots,n_3$
as follows.
For each $j$, we pick an $i$ randomly, then generate
$R^Y_j \stackrel{iid}{\sim} \U(0,r_y)$ with $r_y \in (0,1)$
and
$T_j \stackrel{iid}{\sim} \U(0,2\,\pi)$
set
$Y_j:=X_i+R^Y_j\,(\cos T_j, \sin T_j)'$.
Similarly,
for each $\ell$, we pick an $i$ randomly, then generate
$R^Z_{\ell}\stackrel{iid}{\sim} \U(0,r_z)$ with $r_z \in (0,1)$
and
$U_{\ell}\stackrel{iid}{\sim} \U(0,2\,\pi)$ and set
$Z_{\ell}:=X_i+R^Z_{\ell}\,(\cos U_{\ell}, \sin U_{\ell})'$.

In the pattern generated, appropriate choices of
$r_y$ (and $r_z$) values will imply association between classes $X$ and $Y$ (and $X$ and $Z$).
The three association alternatives are
\begin{equation}
\label{eqn:assoc-alt-3Cl}
H_{A_1}: r_y=1/7,\,r_z=1/10,\;\;\; H_{A_2}: r_y=1/10,\,r_z=1/20,\;\;\; H_{A_3}: r_y=1/13,\,r_z=1/30.
\end{equation}
Observe that, from $H_{A_1}$ to $H_{A_3}$ (i.e., as $r_y$ and $r_z$ decrease),
the association between $X$ and $Y$ gets stronger in the sense that
$Y$ points tend to be found more and more frequently around the $X$ points.
Likewise for $X$ and $Z$ points.
Furthermore, by construction,
classes $X$ and $Z$ are more associated compared to classes $X$ and $Y$.

\begin{figure}[ht]
\centering
Empirical Power Estimates of Cell-Specific Tests under $H_A$\\
\rotatebox{-90}{ \resizebox{2. in}{!}{\includegraphics{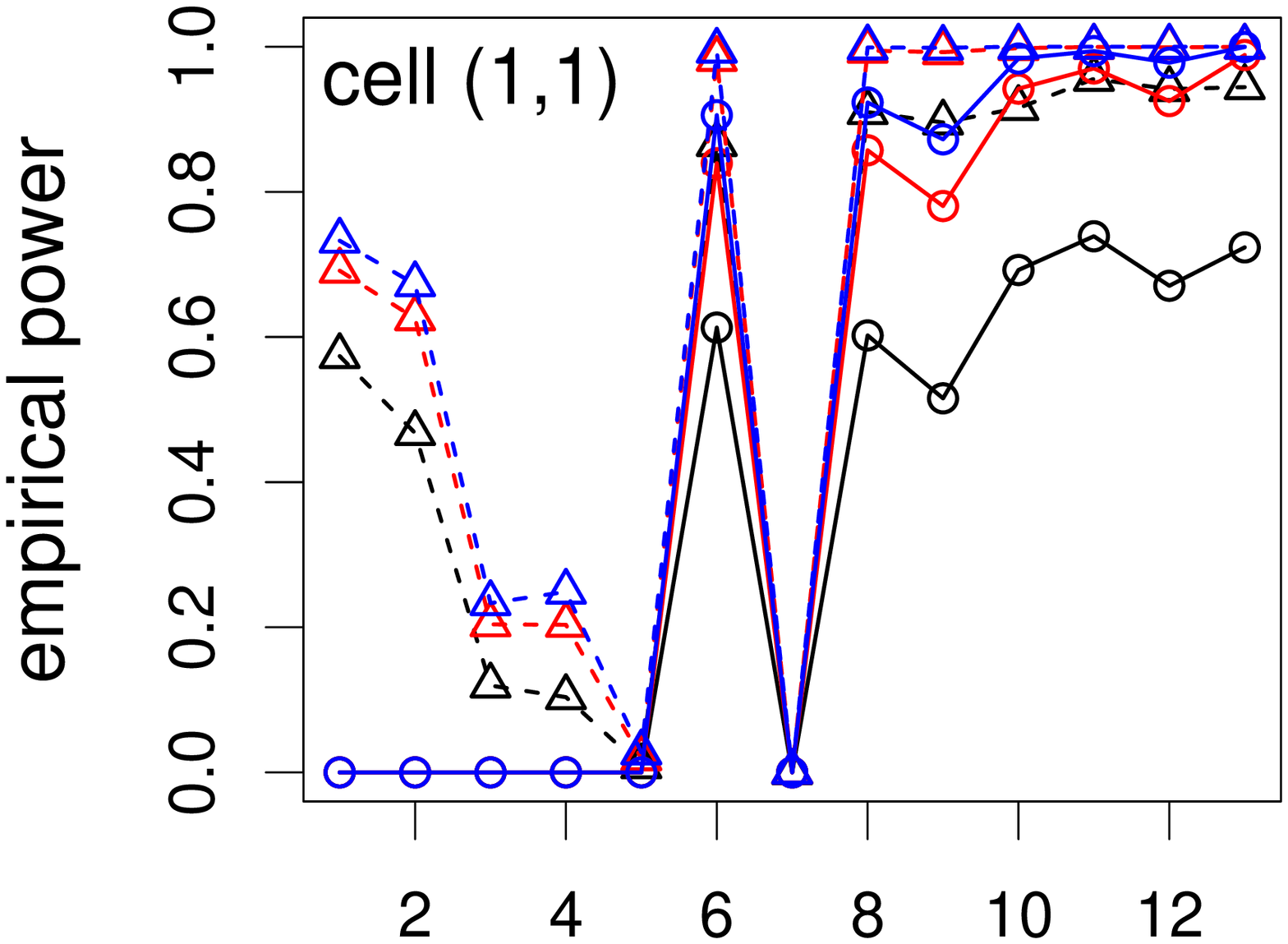} }}
\rotatebox{-90}{ \resizebox{2. in}{!}{\includegraphics{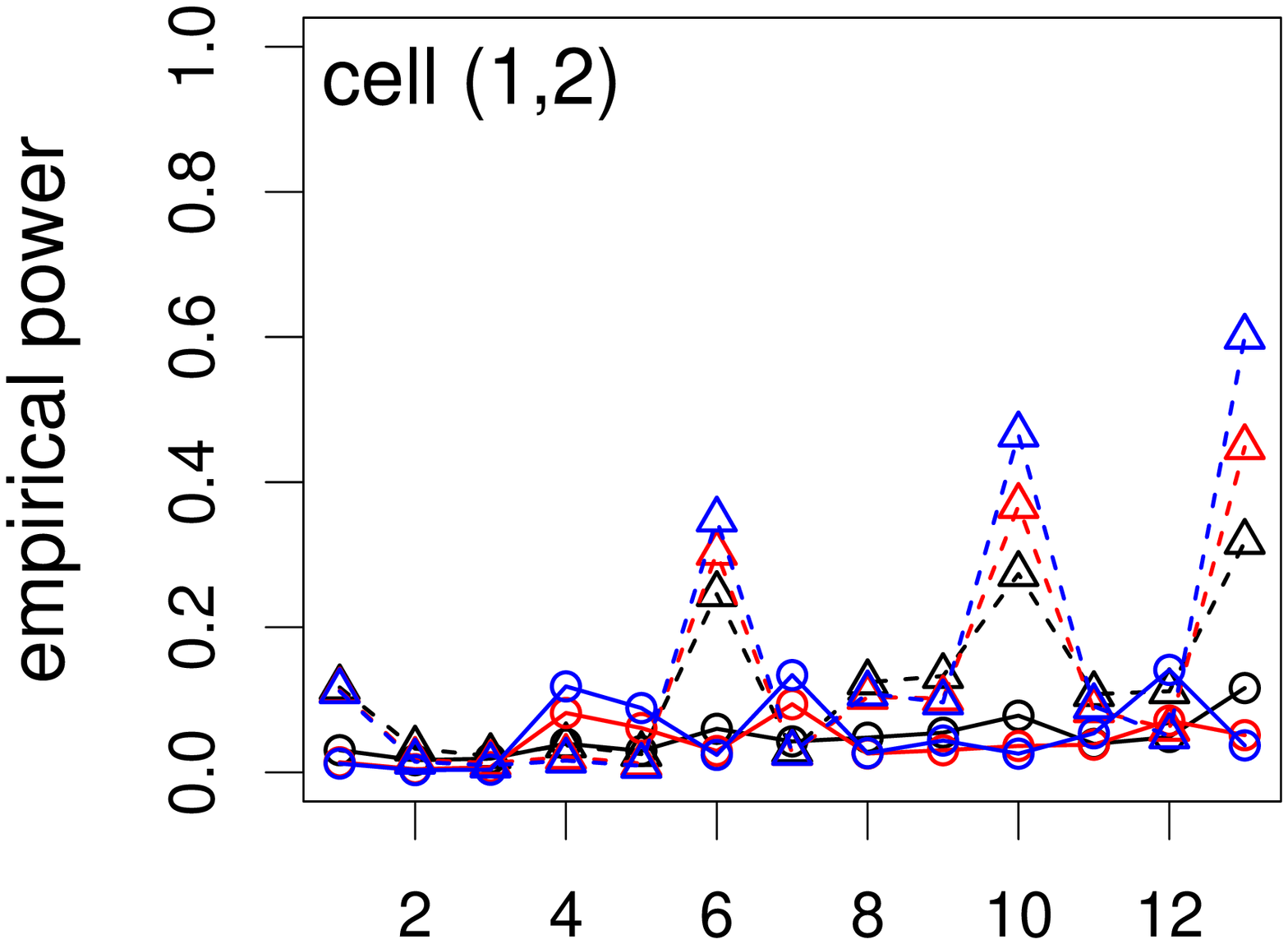} }}
\rotatebox{-90}{ \resizebox{2. in}{!}{\includegraphics{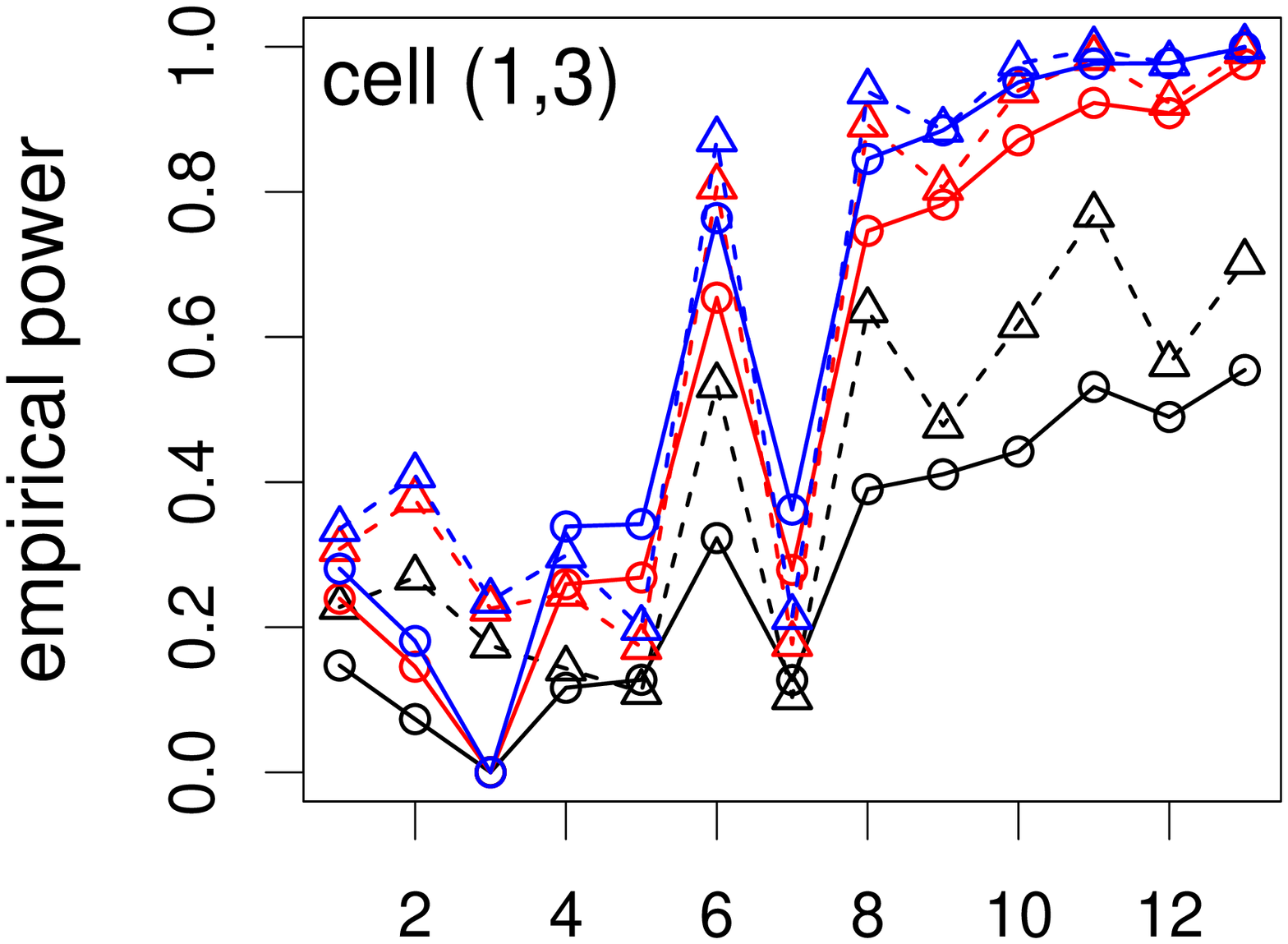} }}

\rotatebox{-90}{ \resizebox{2. in}{!}{\includegraphics{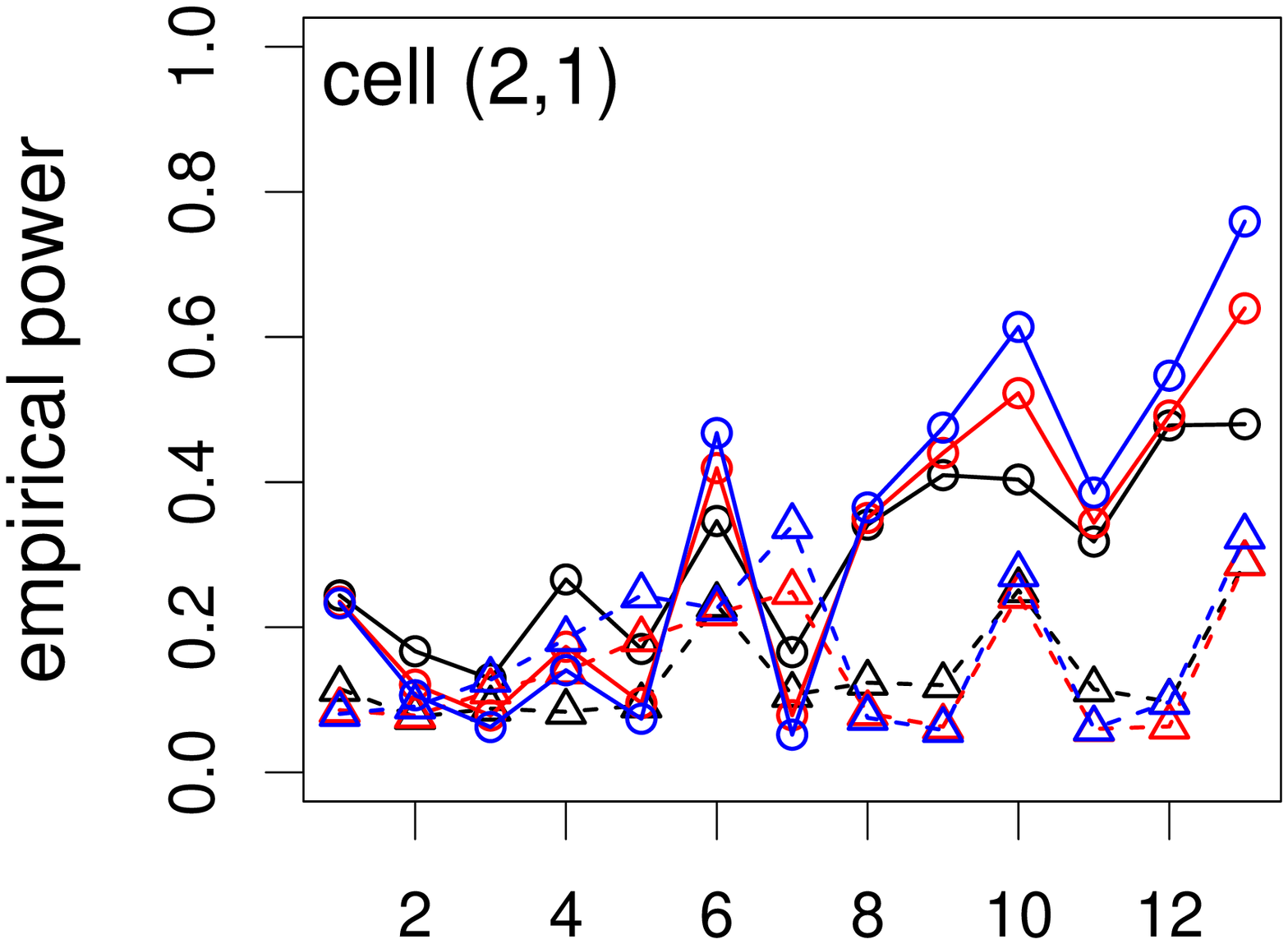} }}
\rotatebox{-90}{ \resizebox{2. in}{!}{\includegraphics{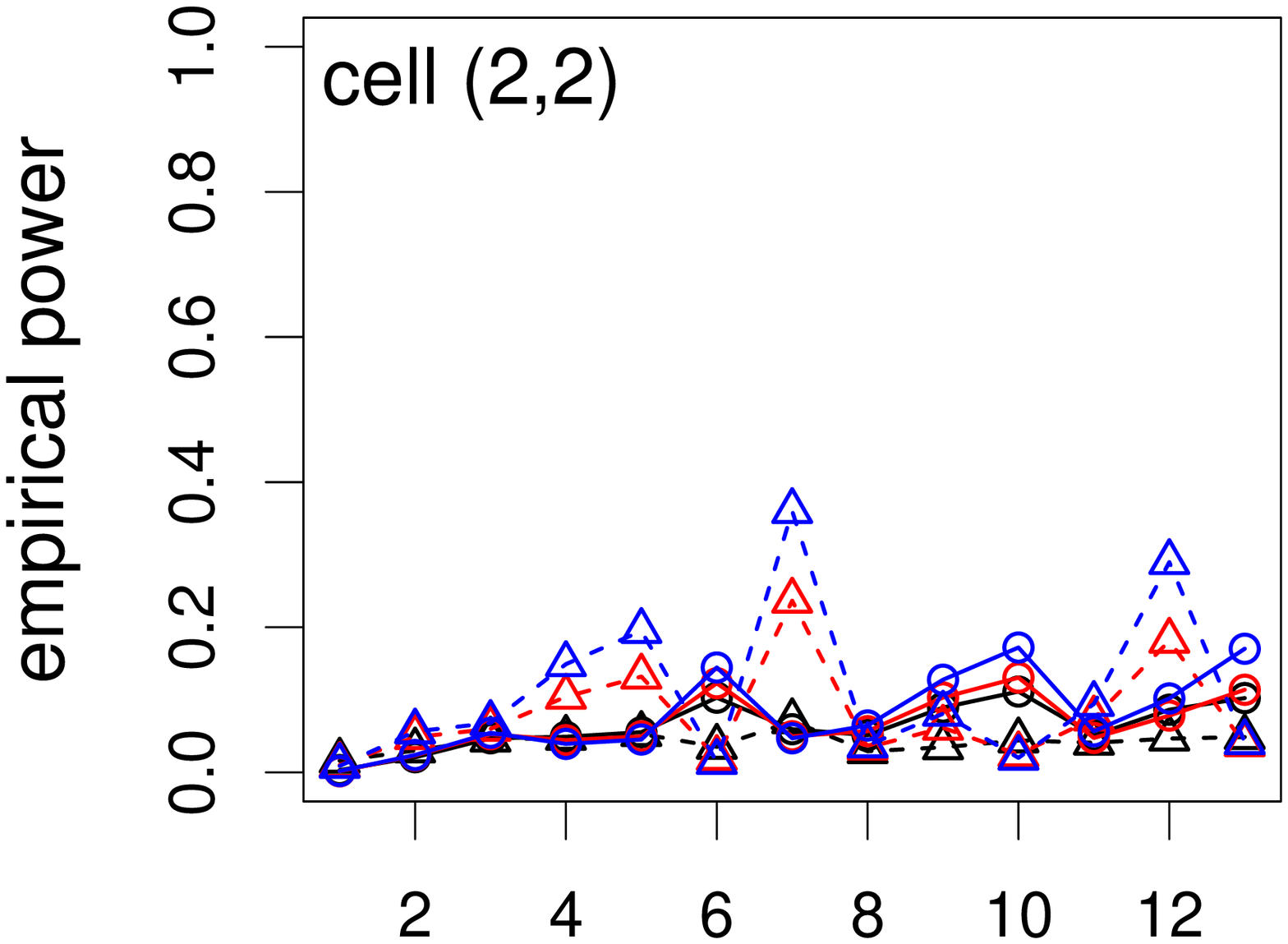} }}
\rotatebox{-90}{ \resizebox{2. in}{!}{\includegraphics{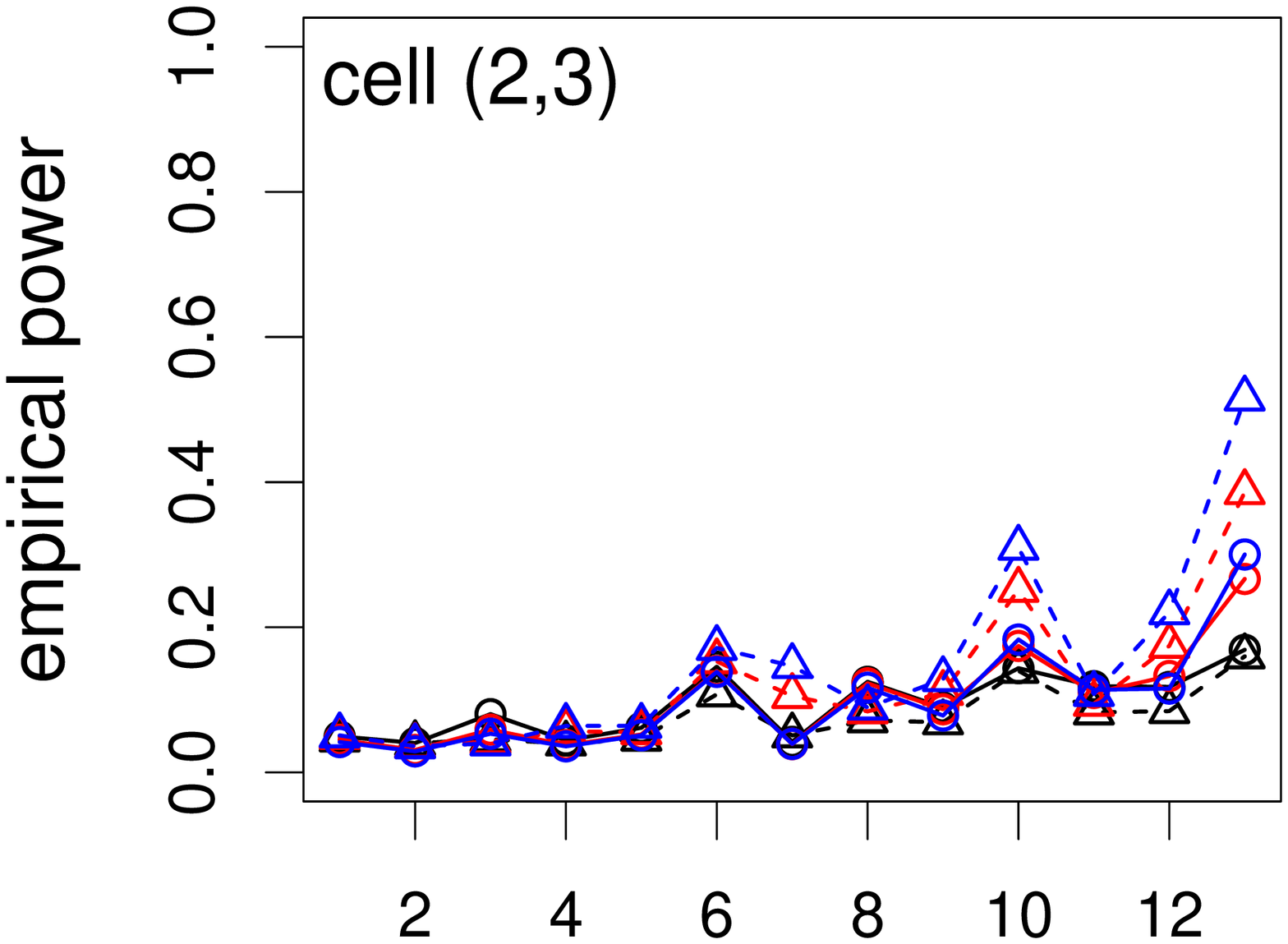} }}

\rotatebox{-90}{ \resizebox{2. in}{!}{\includegraphics{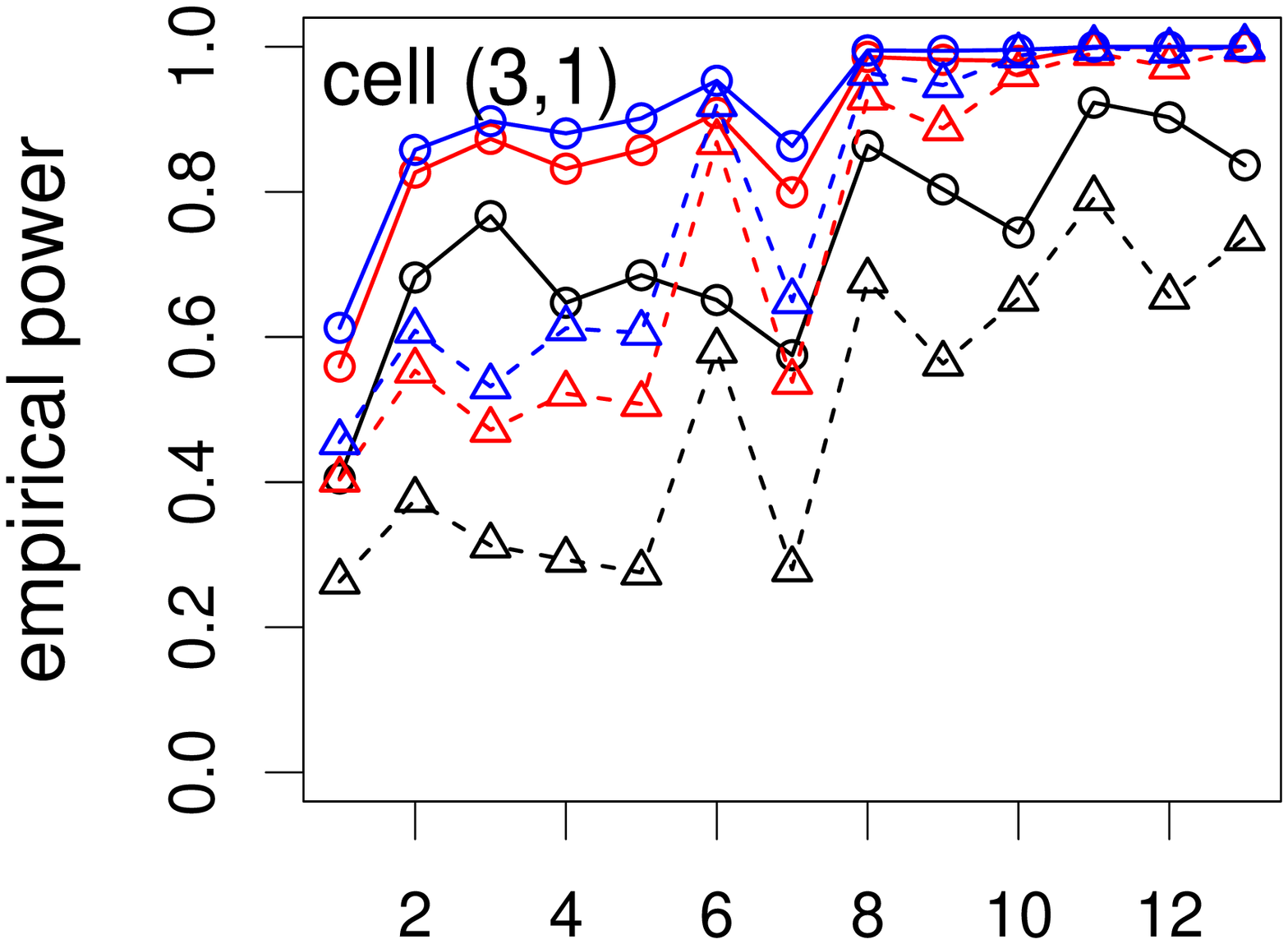} }}
\rotatebox{-90}{ \resizebox{2. in}{!}{\includegraphics{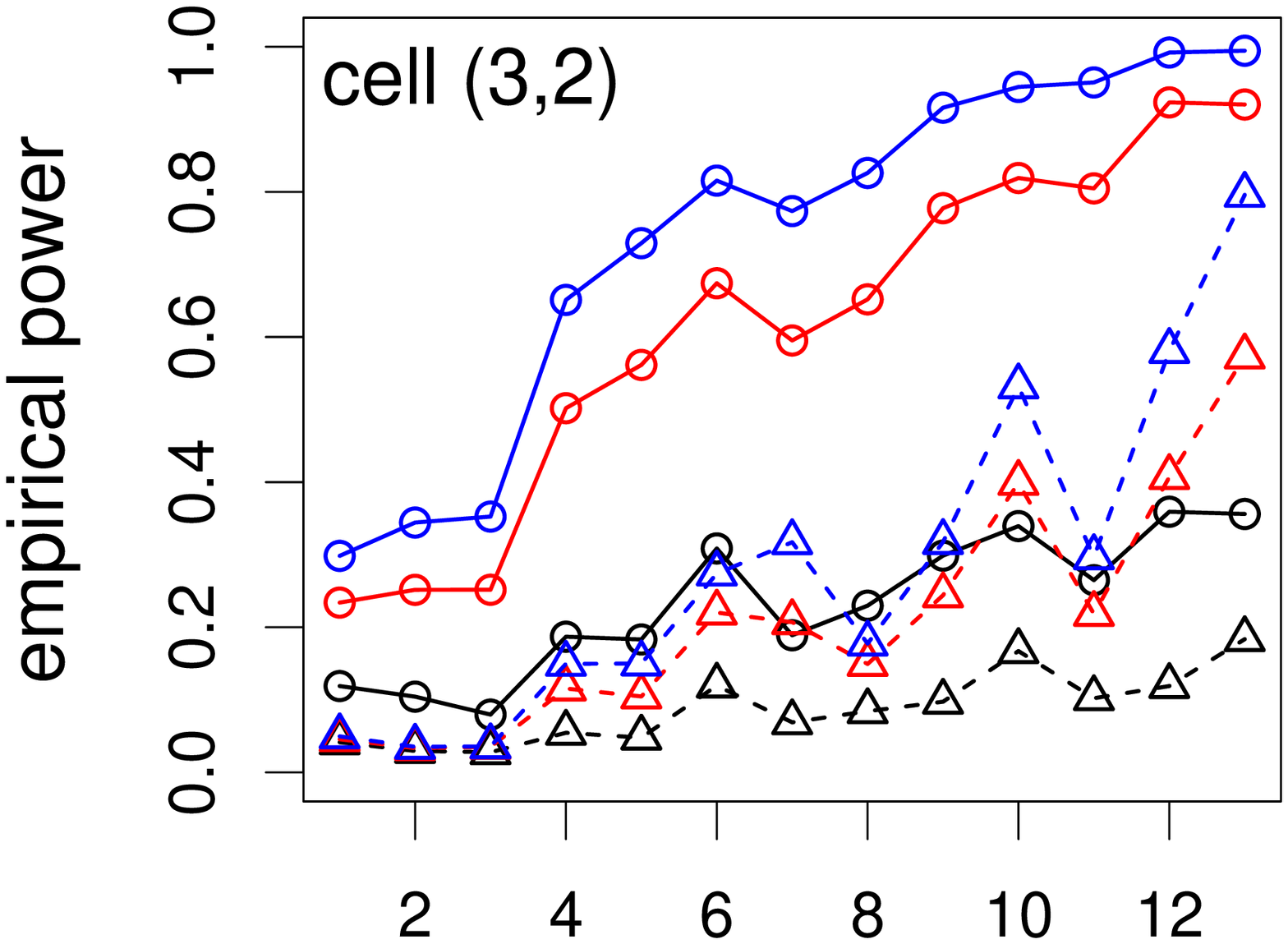} }}
\rotatebox{-90}{ \resizebox{2. in}{!}{\includegraphics{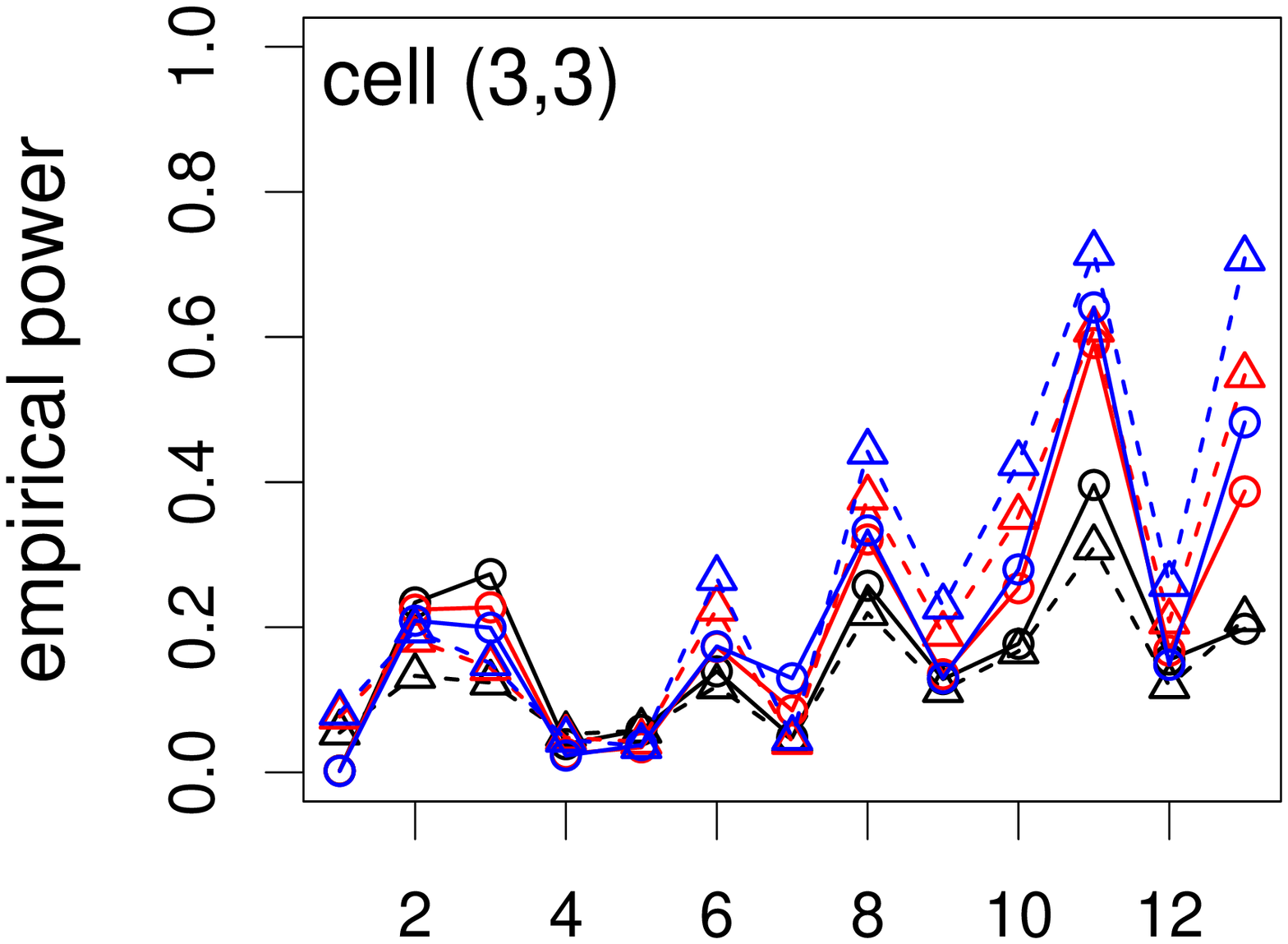} }}
 \caption{
\label{fig:power-assoc-cell-3cl}
The empirical power estimates of Dixon's cell-specific tests (circles ($\circ$))
and the new cell-specific tests (triangles ($\triangle$)) under the association alternatives
$H_{A_1}$ (black), $H_{A_2}$ (red), and $H_{A_3}$ (blue) in the three-class case.
The horizontal axis labels are as in Figure \ref{fig:power-seg-cell-3cl}.
}
\end{figure}

\begin{figure}[t]
\centering
Empirical Power Estimates of Overall Tests under $H_A$\\
\rotatebox{-90}{ \resizebox{2.5 in}{!}{\includegraphics{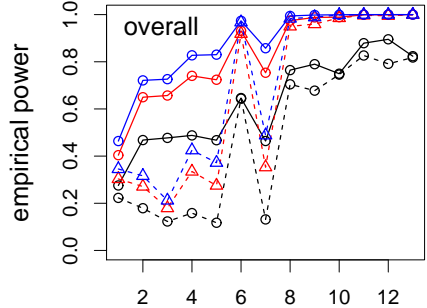} }}

\caption{
\label{fig:power-assoc-overall-3cl}
The empirical power estimates of Dixon's overall test (circles ($\circ$))
and the new overall test (triangles ($\triangle$)) under the association alternatives
$H_{A_1}$ (black), $H_{A_2}$ (red), and $H_{A_3}$ (blue) in the three-class case.
The horizontal axis labels are as in Figure \ref{fig:power-seg-cell-3cl}.
}
\end{figure}

The power estimates for the NNCT-tests
are plotted in Figures \ref{fig:power-assoc-cell-3cl} and \ref{fig:power-assoc-overall-3cl}.
The test statistics tend to be negative for the diagonal cells,
which implies lack of segregation for the classes;
positive for cells $(1,2), (2,1), (1,3)$, and $(3,1)$,
which implies association between classes $X$ and $Y$
and association between classes $X$ and $Y$;
negative for cells $(2,3)$ and $(3,2)$,
which implies lack of association (and perhaps mild segregation) between classes $Y$ and $Z$.

At each sample size combination,
as the association gets stronger,
the power estimates increase.
Further, the higher degree of association between $X$ and
$Z$ compared to that of $X$ and $Y$ are reflected in higher power
estimates for cell $(1,3)$ compared to cell $(1,2)$.
For the same reason, power estimates for cell $(3,1)$
are higher than those for cell $(2,1)$.

For cells $(1,2)$, $(2,1)$, $(1,3)$, and $(3,1)$
the power estimates get larger as the equal sample sizes increase.
The new cell-specific tests have higher power for cells $(1,2)$ and $(1,3)$
and Dixon's cell-specific tests have higher power for cells $(2,1)$ and $(3,1)$.
By construction the classes $Y$ and $Z$ are not associated,
instead they can be viewed as mildly segregated from each other.
The test statistics for cells $(2,3)$ and $(3,2)$ are negative to indicate
such segregation or lack of association between classes $Y$ and $Z$.
However, cell $(3,2)$ power estimates are much larger than cell $(3,2)$,
which implies that class $Z$ can be viewed as more segregated from class $Y$.
As for the diagonal cells,
the higher power estimates for cell $(1,1)$ for larger samples
are indicative of high degree of association of classes $Z$ and $Y$  with class $X$.
Such association, by construction, is barely reflected
in cells $(2,2)$ and $(3,3)$.

In summary, for cells $(i,j)$ with $i < j$,
the new tests have higher power,
while for cells $(i,j)$ with $i > j$, Dixon's tests have higher power.
For the diagonal cells both versions of the cell-specific tests
have about the same power performance.
So we recommend both versions of cell-specific tests to be applied in a given situation
and the results compared and interpreted carefully.

The power estimates of the overall tests
tend to increase as the association gets stronger at
each sample size combination;
as the sample sizes tend to increase,
except for the sudden decrease at $(n_1,n_2,n_3)=(10,50,50)$,
in which case cell $(1,1)$ counts and column 1 sums tend to be very small.
Furthermore, Dixon's overall test has higher power compared to the new overall test.
So Dixon's overall test is recommended for the association alternative
over the new overall test.

The empirical power estimates based on the Monte Carlo
critical values yield similar results, hence not presented.

\begin{remark}
\label{rem:MC-power}
\textbf{Main Result of Monte Carlo Power Analysis:}
Based on the recommendations made in Remark \ref{rem:MC-emp-size},
when at least one sample size is small
(in the sense that some cell count is $< 5$),
we recommend Monte Carlo randomization for the NNCT-tests.
For large samples, one can use asymptotic or Monte Carlo versions of the NNCT-tests.
In Sections \ref{sec:power-comp-seg-2Cl}, \ref{sec:power-comp-assoc-2Cl},
\ref{sec:power-comp-seg-3Cl}, and \ref{sec:power-comp-assoc-3Cl},
we observe that under the segregation alternatives,
the new cell-specific and new overall
tests have higher power
compared to Dixon's cell-specific and overall tests.
Under the association alternatives,
we observe that for cells with the associated class is the NN class,
the new cell-specific tests have higher power,
while for cells with the associated class as the base class,
Dixon's cell-specific tests have higher power,
and for diagonal cells Dixon's and new cell-specific tests
have similar power.
Additionally, Dixon's overall test has higher power than the new overall
test for association.
Thus we recommend both of the new and Dixon's NNCT-tests
under the association alternatives.
$\square$
\end{remark}

\section{Examples}
\label{sec:examples}
We illustrate the tests on two ecological data sets:
Pielou's Douglas-fir/ponderosa pine data (\cite{pielou:1961})
and a swamp tree data (\cite{good:1982}}).

\subsection{Pielou's Data}
\label{sec:pielou-data}
\cite{pielou:1961} used a completely mapped data set that is
comprised of two tree species:
Douglas-fir trees (\emph{Pseudotsuga menziesii} formerly \emph{P. taxifolia})
and ponderosa pine (\emph{Pinus ponderosa}) from a region in British Columbia.
Her data set was also used by
Dixon as an illustrative example (\cite{dixon:1994}).
The question of interest is the type of spatial interaction
between the two tree species.
The corresponding $2 \times 2$ NNCT and
the percentages for each cell are provided in Table \ref{tab:NNCT-pielou}.
The cell percentages are with respect to the sample sizes of each species,
for example, 86 \% of Douglas-firs have NNs from
Douglas firs and remaining 15 \% of Douglas-firs have NNs from ponderosa pines.
The row and column percentages are marginal percentages with respect to the
total sample size.
The percentage values are suggestive of segregation for both species.

\begin{table}[ht]
\centering

\begin{tabular}{cc|cc|c}
\multicolumn{2}{c}{}& \multicolumn{2}{c}{NN}& \\
\multicolumn{2}{c}{}&    D.F. &  P.P.   &   sum  \\
\hline
&D.F.&    137  &   23    &   160  \\
\raisebox{1.5ex}[0pt]{base}
&P.P. &    38 &  30    &   68  \\
\hline
&sum     &    175   & 53             &  228  \\
\end{tabular}
\hspace{0.33 in}
\begin{tabular}{cc|cc|c}
\multicolumn{2}{c}{}& \multicolumn{2}{c}{NN}& \\
\multicolumn{2}{c}{}&    D.F. &  P.P.   & \\
\hline
&D.F.&    86 \%  &   15 \%    &  70 \%  \\
\raisebox{1.5ex}[0pt]{base}
&P.P. &    56 \% &   44 \%    &  30 \%  \\
\hline
& &    77 \% &   23 \%    &  100 \%  \\
\end{tabular}

\caption{ \label{tab:NNCT-pielou}
The NNCT for Pielou's data (left) and the corresponding percentages (right),
where the cell percentages are with respect to the row sums (i.e., species sizes)
and the marginal percentages are with respect to the total sample size.
D.F. = Douglas-firs, P.P. = ponderosa pines.}
\end{table}

\begin{table}[ht]
\centering

\begin{tabular}{|c|c|c|}
\multicolumn{3}{c}{Dixon's NNCT-Tests}\\
\hline
\multicolumn{3}{c}{Overall test}\\
\hline
\multicolumn{3}{c}{ $C_D=19.67$  ($\pasy =.0001$)} \\
\hline
\hline
\multicolumn{3}{c}{Cell-specific tests}\\
\hline
\multicolumn{1}{c|}{}&    D.F. &  P.P.     \\
\hline
D.F.&    4.36  &   -4.36   \\
 &    ($<.0001$)  &   ($<.0001$)    \\
\hline
P.P. &    -2.29 & 2.29      \\
 &    (.0221)  &   (.0221)     \\
\hline
\end{tabular}
\hspace{0.33 in}
\begin{tabular}{|c|c|c|}
\multicolumn{3}{c}{The New NNCT-Tests}\\
\hline
\multicolumn{3}{c}{Overall test}\\
\hline
\multicolumn{3}{c}{ $C_N=13.11$  ($\pasy = .0003$)} \\
\hline
\hline
\multicolumn{3}{c}{Cell-specific tests}\\
\hline
\multicolumn{1}{c|}{}&    D.F. &  P.P.     \\
\hline
D.F.&    3.63  &   -3.61   \\
 &    (.0003)  &   (.0003)    \\
\hline
P.P. &    -3.63 & 3.61      \\
 &    (.0003)  &   (.0003)    \\
\hline
\end{tabular}

\caption{
\label{tab:test-stat-pielou}
The overall and cell-specific test statistics for Dixon's NNCT-tests (left)
and the new NNCT-tests (right) and the corresponding $p$-values (in parenthesis)
based on asymptotic approximation for Pielou's data.
D.F. = Douglas-firs, P.P. = ponderosa pines;
$C_D$ and $C_N$ stand for the value of Dixon's and new overall test statistic,
respectively. $\pasy$ stands for the $p$-value based on the asymptotic
approximation of the tests.}
\end{table}

The raw data are not available, hence we can not perform Monte Carlo simulation
nor randomization versions of the tests.
Fortunately, \cite{pielou:1961} provided $Q=162$ and $R=134$,
hence we can calculate the test statistics and use the asymptotic approximation
for these tests.
The overall and cell-specific test statistics and the corresponding $p$-values (in parentheses)
based on the asymptotic approximation, denoted by $\pasy$,
are provided in Table \ref{tab:test-stat-pielou}.
Although the locations of the tree species are not known,
they can be viewed a priori resulting
from different processes rather than some process affecting a posteriori
the individuals of a single population.
So the more appropriate null hypothesis is CSR independence of the trees.
Hence our inference will be a conditional one (see Remark \ref{rem:QandR}).
Observe that Dixon's and new overall test statistics yield significant $p$-values,
implying some sort of deviation from CSR independence.
In order to see the type of deviation, we apply the cell-specific tests.
Both versions of the cell-specific tests for each cell are significant,
implying significant deviation from CSR independence.
The cell-specific test statistics are positive for the diagonal cells
$(1,1)$ and $(2,2)$ (and negative for the off-diagonal cells $(1,2)$ and $(2,1)$),
implying segregation for both species.
This is in agreement with what the NNCT suggests
and the findings of (\cite{dixon:1994}).
However, Dixon's cell $(1,1)$ statistics are much larger than cell $(2,2)$ statistics,
which may be interpreted as clustering of Douglas-firs is stronger than
the clustering of ponderosa pines.
Our simulation study indicates that
this might be an artifact of the relative abundance of the tree species.
On the other hand,
new cell $(1,1)$ and cell $(2,2)$ statistics are very similar,
hence the segregation of both tree species are at about the same degree.

\subsection{Swamp Tree Data}
\label{sec:swamp-data}
\begin{figure}[ht]
\centering
\rotatebox{-90}{ \resizebox{3.25 in}{!}{
\includegraphics{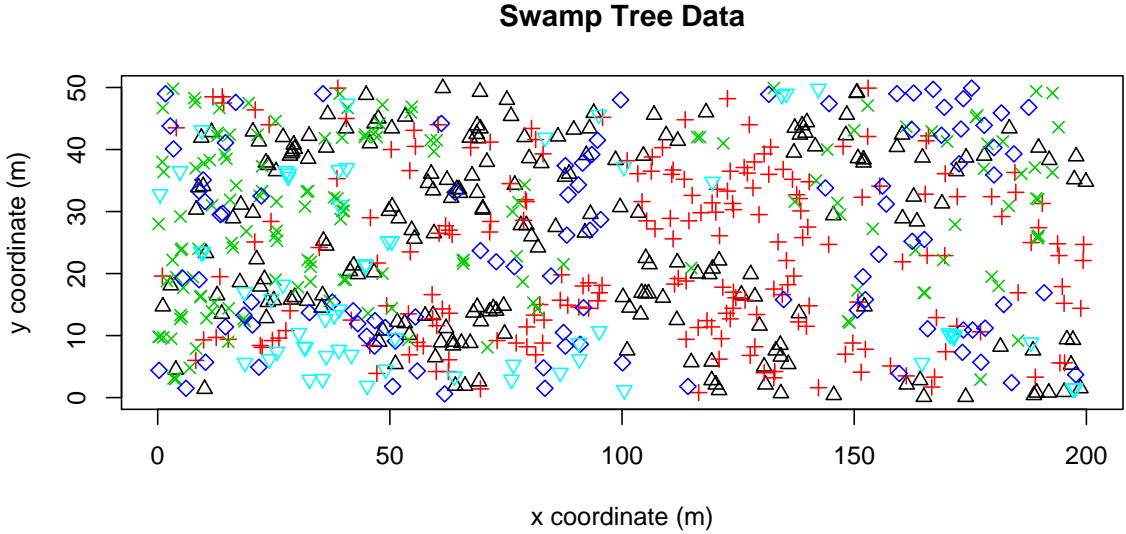} }}
 \caption{
\label{fig:SwampTrees}
The scatter plot of the locations of water tupelos (triangles $\triangle$),
black gum trees (pluses $+$), Carolina ashes (crosses $\times$),
bald cypress trees (diamonds $\diamond$), and other trees (inverse triangles $\triangledown$).}
\end{figure}

\cite{good:1982} considered the spatial interaction between tree species
along the Savannah River, South Carolina, U.S.A.
From this data, \cite{dixon:NNCTEco2002} used a single 50m $\times$ 200m rectangular plot
to illustrate his NNCT-tests.
All live or dead trees with 4.5 cm or more dbh (diameter at breast height)
were recorded together with their species.
Hence it is an example of a realization of a marked multi-variate point pattern.
The plot contains 13 different tree species,
four of which comprises over 90 \% of the 734 tree stems.
The remaining tree stems were categorized as ``other trees".
The plot consists of 215 water tupelos (\emph{Nyssa aquatica}),
205 black gums (\emph{Nyssa sylvatica}), 156 Carolina ashes (\emph{Fraxinus caroliniana}),
98 bald cypresses (\emph{Taxodium distichum}),
and 60 stems from 8 additional species (i.e., other species).
A $5 \times 5$ NNCT-analysis is conducted for this data set.
If segregation among the less frequent species is important,
a more detailed $12 \times 12$ NNCT-analysis should be performed.
The locations of these trees in the study region are plotted in Figure \ref{fig:SwampTrees}
and the corresponding $5 \times 5$ NNCT together with percentages
based on row and grand sums are provided in Table \ref{tab:NNCT-swamp}.
For example, for black gum as the base species and Carolina ash as the NN species,
the cell count is 26 which is 13 \% of the 205 black gums (which is 28 \% of all trees).
Observe that the percentages and Figure \ref{fig:SwampTrees}
are suggestive of segregation for all tree species,
especially for Carolina ashes, water tupelos, black gums, and the ``other" trees
since the observed percentages of species with themselves as the NN are much larger
than the row percentages.

\begin{table}[ht]
\centering
\begin{tabular}{cc|ccccc|c}
\multicolumn{2}{c}{}& \multicolumn{5}{c}{NN}& \\
\multicolumn{2}{c}{}& W.T. & B.G. & C.A. &  B.C. & O.T.   &   sum  \\
\hline
& W.T. &    112 (52 \%) &   40 (19 \%) &  29 (13 \%)  &  20 (11 \%)  &  14 (9 \%)  &  215 (29 \%) \\
& B.G. &    38  (19 \%) &  117 (57 \%) &  26 (13 \%)  &  16 (8 \%) &   8 (4 \%)  &  205 (28 \%) \\
& C.A. &    23  (15 \%) &   23 (15 \%) &  82 (53 \%) &  22 (14 \%) &   6 (4 \%)  &  156 (21 \%) \\
\raisebox{2.5ex}[0pt]{base}
& B.C. &    19  (19 \%) &   29 (30 \%) &  29 (30 \%) &  14 (14 \%) &   7 (7 \%)  &  98  (13 \%) \\
& O.T. &    7   (12 \%) &   8  (13 \%) &  5  (8 \%)  &  7  (12 \%) &  33 (55 \%) &  60  (8 \%)  \\
\hline
&sum   &   199  (27 \%) &  217 (30 \%) & 171 (23 \%) &  79 (11 \%) &  68 (9 \%)  &  734 (100 \%)\\
\end{tabular}
\caption{ \label{tab:NNCT-swamp}
The NNCT for swamp tree data (left)
and the corresponding percentages (right),
where the cell percentages are with respect to the row sums
and marginal percentages are with respect to the total size.
W.T. = water tupelos, B.G. = black gums, C.A. = Carolina ashes, B.C.  = bald cypresses, and O.T. =
other tree species.}
\end{table}

\begin{table}[ht]
\centering
\begin{tabular}{|c|c|c|c|c|c|c|}
\multicolumn{7}{c}{Dixon's overall test} \\
\hline
\multicolumn{7}{c}{$C_D=275.64$ ($\pasy<.0001$, $\pmc<.0001$, $\prand<.0001$)} \\
\hline
\hline
\multicolumn{7}{c}{Dixon's cell-specific tests} \\
\hline
& & W.T. & B.G. & C.A. &  B.C. & O.T.\\
\hline
 & &    6.39     & -3.11 & -2.87    & -1.82 & -0.94    \\
     & $\pasy$  &    ($<.0001$)& (.0019)&  (.0041)   & (.0682) & (.3484)  \\
     & $\pmc$   &    ($<.0001$)& (.0014)&  (.0043)   & (.0702) & (.3489)  \\
\raisebox{3.75ex}[0pt]{W.T.}
     & $\prand$ &    ($<.0001$)& (.0015)&  (.0048)   & (.0670) & (.3286)  \\
\hline
 & &   -3.44     &  8.05& -3.09   & -2.43 & -2.34   \\
     & $\pasy$  &    (.0006)& ($<.0001$)& (.0020)   & (.0150) & (.0194)   \\
     & $\pmc$   &    (.0004)& ($<.0001$)& (.0014)   & (.0179) & (.0192)   \\
\raisebox{3.75ex}[0pt]{B.G.}
     & $\prand$ &    (.0005)& ($<.0001$)& (.0012)   & (.0172) & (.0198)   \\
\hline
 & &   -4.05     & -3.73   &  8.08 &  0.28 & -2.04  \\
     & $\pasy$  &    (.0001) & (.0002)   & ($<.0001$)& (.7820) & (.0410)   \\
     & $\pmc$   &    ($<.0001$)& (.0004)   & ($<.0001$)& (.7810) & (.0430)   \\
\raisebox{3.75ex}[0pt]{C.A.}
     & $\prand$ &    ($<.0001$)& (.0001)   & ($<.0001$)& (.7580) & (.0409)   \\
\hline
 & &   -2.18    & -0.36   &  2.04    &  0.25 &  -0.38   \\
     & $\pasy$  &    (.0295)   & (.7180)   & (.0418)   & (.8011) & (.7008) \\
     & $\pmc$   &    (.0292)   & (.7129)   & (.0410)   & (.7601) & (.6739) \\
\raisebox{3.75ex}[0pt]{B.C.}
     & $\prand$ &    (.0293)   & (.6861)   & (.0419)   & (.7910) & (.6555) \\
\hline
 & &   -3.02    & -2.54   & -2.47 &  -0.39 &  10.77   \\
     & $\pasy$  &    (.0025)   & (.0112)   &(.0135)   & (.6952) &  ($<.0001$)\\
     & $\pmc$   &    (.0028)   & (.0112)   &(.0115)   & (.6582) &  ($<.0001$)\\
\raisebox{3.75ex}[0pt]{O.T.}
     & $\prand$ &    (.0021)   & (.0121)   &(.0158)   & (.6490) &  ($<.0001$)\\
\hline

\end{tabular}
\caption{ \label{tab:pval-swamp-dixon}
Test statistics and $p$-values for Dixon's overall and cell-specific tests
and the corresponding $p$-values (in parentheses).
W.T. = water tupelos, B.G. = black gums, C.A. = Carolina ashes, B.C.  =
bald cypress and O.T. = other tree species.
$\pasy$, $\pmc$, and $\prand$ stand for the $p$-values based on the asymptotic
approximation, Monte Carlo simulation, and randomization of the tests, respectively.}
\end{table}

\begin{table}[ht]
\centering
\begin{tabular}{|c|c|c|c|c|c|c|}
\multicolumn{7}{c}{New overall test} \\
\hline
\multicolumn{7}{c}{$C_N=263.10$ ($\pasy<.0001$, $\pmc<.0001$, $\prand<.0001$)} \\
\hline
\hline
\multicolumn{7}{c}{New cell-specific tests} \\
\hline

& & W.T. & B.G. & C.A. &  B.C. & O.T. \\
\hline
 & &     7.55  & -4.08 & -4.06    & -0.74    & -1.74  \\
     & $\pasy$  &     ($<.0001$)& ($<.0001$)& (.0001)  & (.4584) & (.0819) \\
     & $\pmc$   &     ($<.0001$)& ($<.0001$)& ($<.0001$)  & (.4564) & (.0860) \\
\raisebox{3.75ex}[0pt]{W.T.}
     & $\prand$ &     ($<.0001$)& ($<.0001$)& ($<.0001$)  & (.4645) & (.0824) \\
\hline
 & &   -3.04  &  8.16  & -4.25 & -1.45  & -3.27  \\
     & $\pasy$  &     (.0023)& ($<.0001$)& ($<.0001$)  & (.1479) & (.0011) \\
     & $\pmc$   &     (.0028)& ($<.0001$)& ($<.0001$)  & (.1550) & (.0013) \\
\raisebox{3.75ex}[0pt]{B.G.}
     & $\prand$ &     (.0018)& ($<.0001$)& ($<.0001$)  & (.1493) & (.0008) \\
\hline
 &  &  -3.71 & -4.52  &  7.96 &  1.36  & -2.77  \\
     & $\pasy$  &    (.0002)& ($<.0001$)   &  ($<.0001$)&  (.1745) & (.0056) \\
     & $\pmc$   &    (.0001)& (.0001)   &  ($<.0001$)&  (.1776) & (.0064) \\
\raisebox{3.75ex}[0pt]{C.A.}
     & $\prand$ &    (.0001)& ($<.0001$)   &  ($<.0001$)&  (.1806) & (.0064) \\
\hline
 & &    -1.78   & 0.00   &  1.61    &  0.89  &  -0.82  \\
     & $\pasy$  &    (.0754)   & (.9977)   &  (.1081)   &   (.3725) &  (.4097) \\
     & $\pmc$   &    (.0702)   & (.9952)   &  (.1098)   &   (.3771) &  (.4105) \\
\raisebox{3.75ex}[0pt]{B.C.}
     & $\prand$ &    (.0723)   & (.9958)   &  (.1114)   &   (.3796) &  (.4071) \\
\hline
 & &   -2.72   & -2.90    & -2.94   &  0.21  &  10.71  \\
     & $\pasy$  &    (.0066)   &  (.0037)   & (.0033)   &  (.8335) &  ($<.0001$) \\
     & $\pmc$   &    (.0060)   &  (.0031)   & (.0027)   &  (.8375) &  ($<.0001$) \\
\raisebox{3.75ex}[0pt]{O.T.}
     & $\prand$ &    (.0070)   &  (.0036)   & (.0026)   &  (.8354) &  ($<.0001$) \\
\hline

\end{tabular}
\caption{ \label{tab:pval-swamp-new}
Test statistics and $p$-values for the new overall and cell-specific tests and the
corresponding $p$-values (in parentheses).
The labeling of the species and $p$-values are as in Table \ref{tab:pval-swamp-dixon}.}
\end{table}

The locations of the tree species can be viewed a priori resulting from different processes,
so the more appropriate null hypothesis is the CSR independence pattern.
Hence our inference will be a conditional one (see Remark \ref{rem:QandR}).
We calculate $Q=472$ and $R=454$ for this data set.
We present Dixon's overall test of segregation
and cell-specific test statistics and the associated $p$-values
in Table \ref{tab:pval-swamp-dixon},
where $\pasy$ stands for the $p$-value based on the asymptotic approximation,
$\pmc$ is the $p$-value based on $10000$ Monte Carlo replication of the CSR independence
pattern in the same plot and $\prand$ is based on Monte Carlo
randomization of the labels on the given locations of the trees 10000 times.
Notice that $\pasy$, $\pmc$, and $\prand$ are very similar for each test.
We present the new overall test of segregation
and cell-specific test statistics and the associated $p$-values
in Table \ref{tab:pval-swamp-new},
where $p$-values are calculated as in Table \ref{tab:pval-swamp-dixon}.
Again, all three $p$-values in Table \ref{tab:pval-swamp-new} are similar for each test.

Dixon's and the new overall test of segregation are both significant implying
significant deviation from the CSR independence pattern
for at least for one pair of the tree species.
Then to determine which pairs exhibit segregation or association,
we perform the cell-specific tests.
Dixon's and the new cell-specific tests
agree for all cells (i.e., pairs) in term of significance at .05 level
except for (B.G.,B.C.), (BC,W.T), and (B.C.,C.A.) pairs.
The statistics are all negative for the off-diagonal
cells, except for (B.C.,C.A.) and (C.A., B.C.) pairs.
Based on the Monte Carlo simulation analysis,
the new test is more reliable to attach significance to these situations.
The spatial interaction is significant between each pair
which does not contain bald cypresses.
That is, the new cell-specific test statistics are positive for the diagonal cells
$(i,i)$ for $i=1,2,\ldots,5$ and are significant for $i=1,2,3,5$
at .05 level (which also holds for Dixon's tests);
and are negative for the off-diagonal cells $(i,j)$ with $i,j \in \{1,2,3,5\}$
and $i \ne j$ and significant for most of them.
Hence each tree species except bald cypresses exhibits significant segregation
from each other.
These findings are mostly in agreement with the results of (\cite{dixon:NNCTEco2002}).
Hence except for bald cypresses, each tree species seem
to result from a (perhaps) different first order inhomogeneous Poisson process.

\begin{figure}[t]
\centering
\rotatebox{-90}{ \resizebox{2 in}{!}{\includegraphics{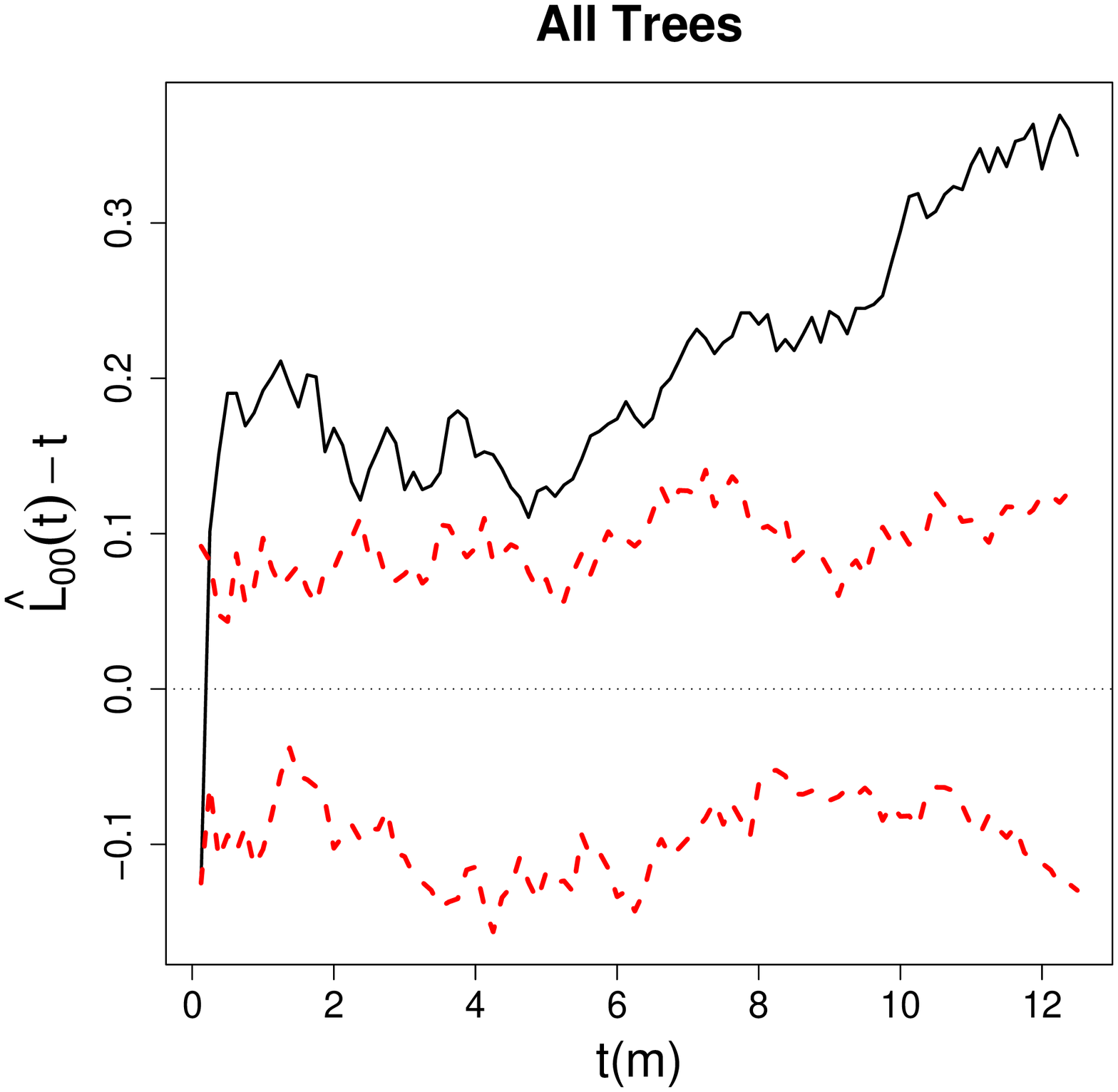} }}
\rotatebox{-90}{ \resizebox{2 in}{!}{\includegraphics{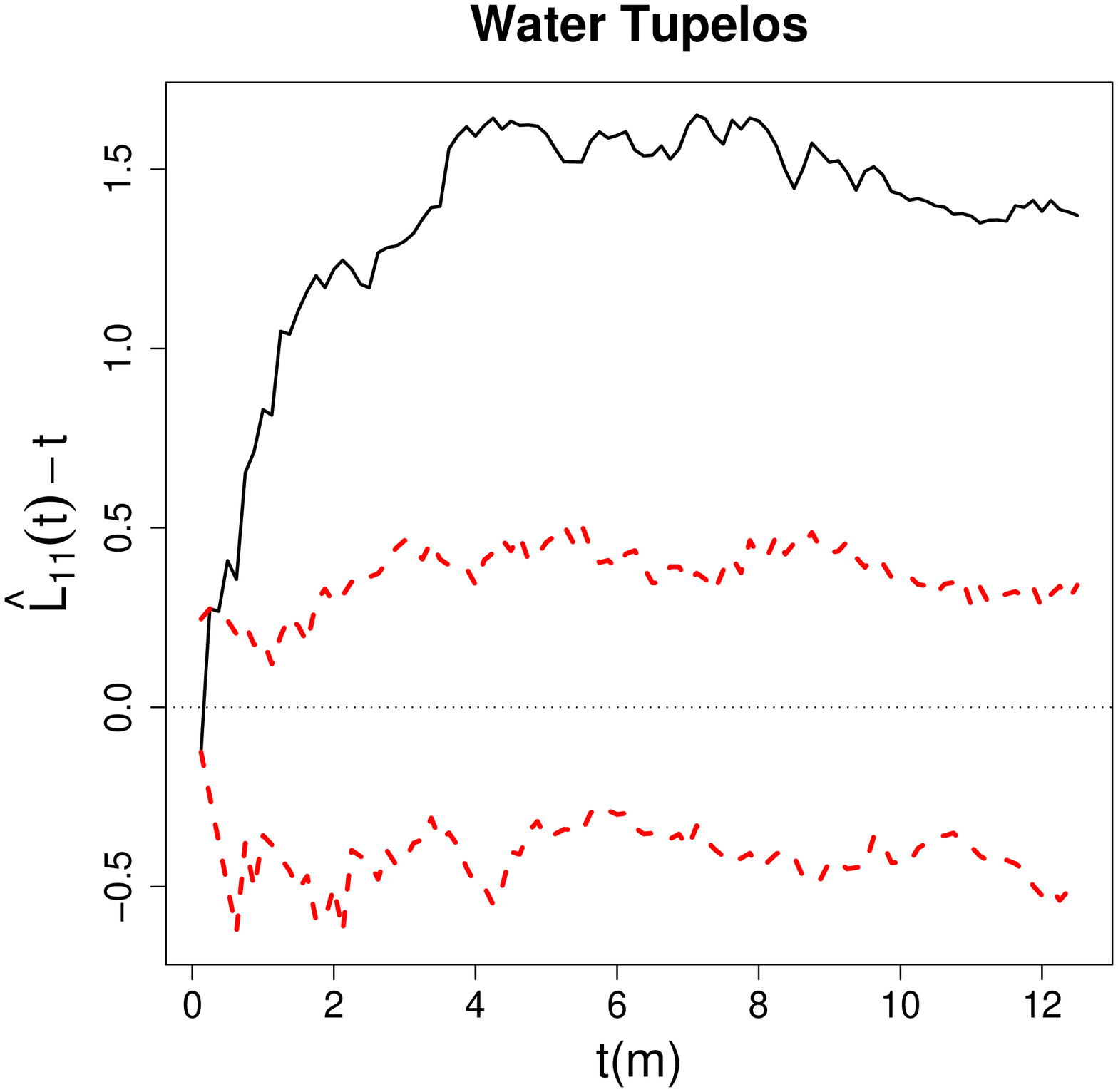} }}
\rotatebox{-90}{ \resizebox{2 in}{!}{\includegraphics{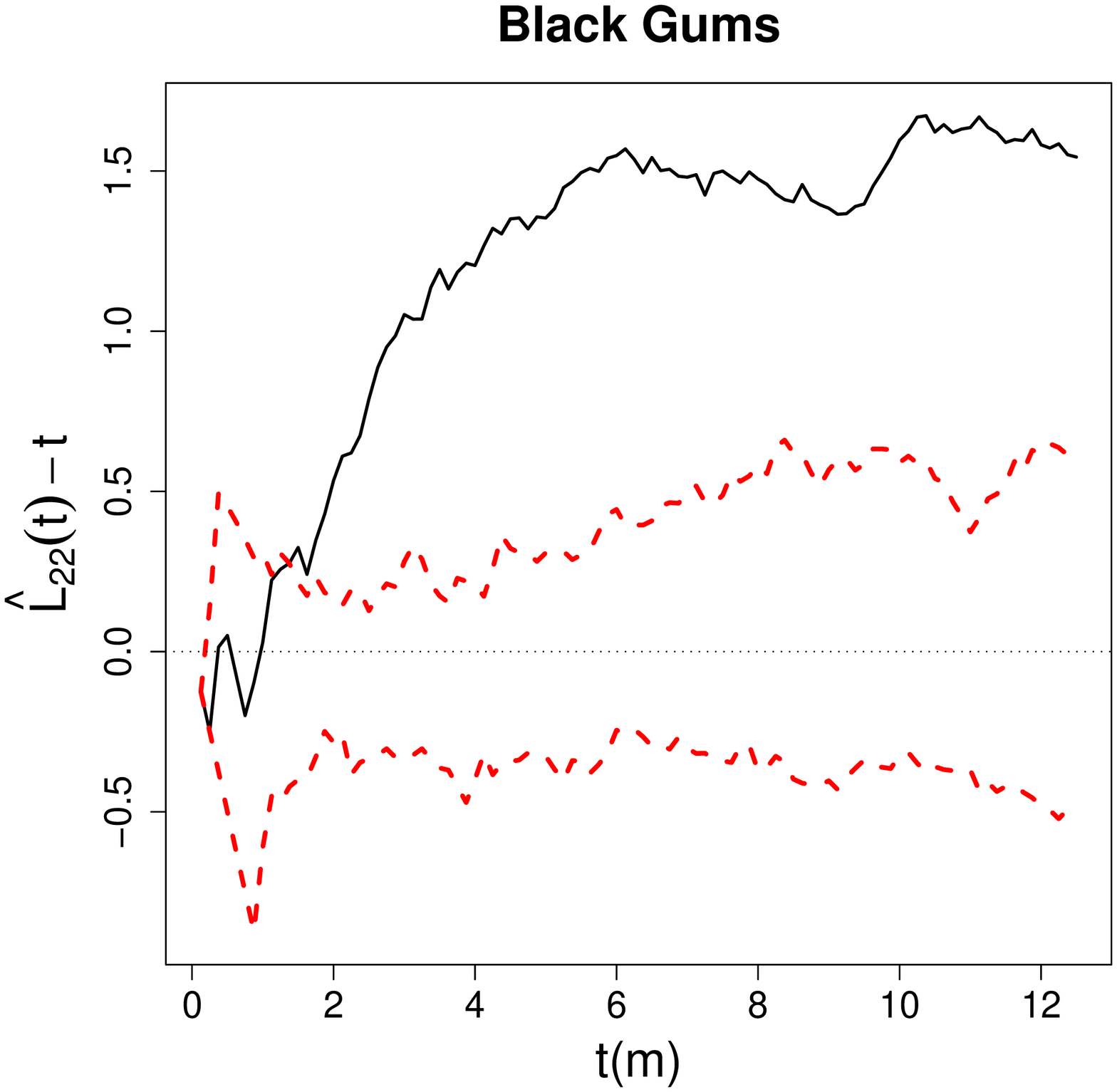} }}
\rotatebox{-90}{ \resizebox{2 in}{!}{\includegraphics{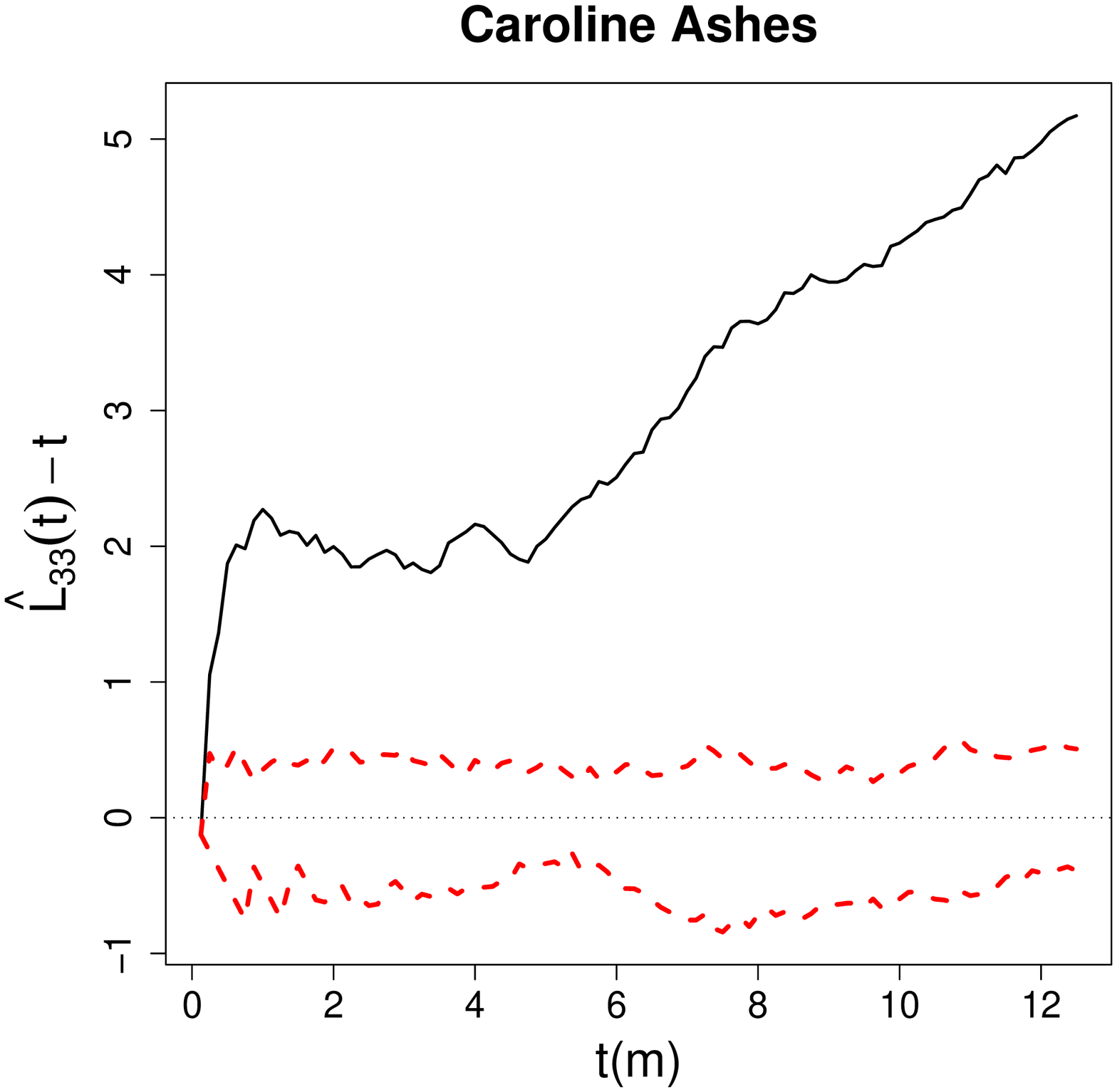} }}
\rotatebox{-90}{ \resizebox{2 in}{!}{\includegraphics{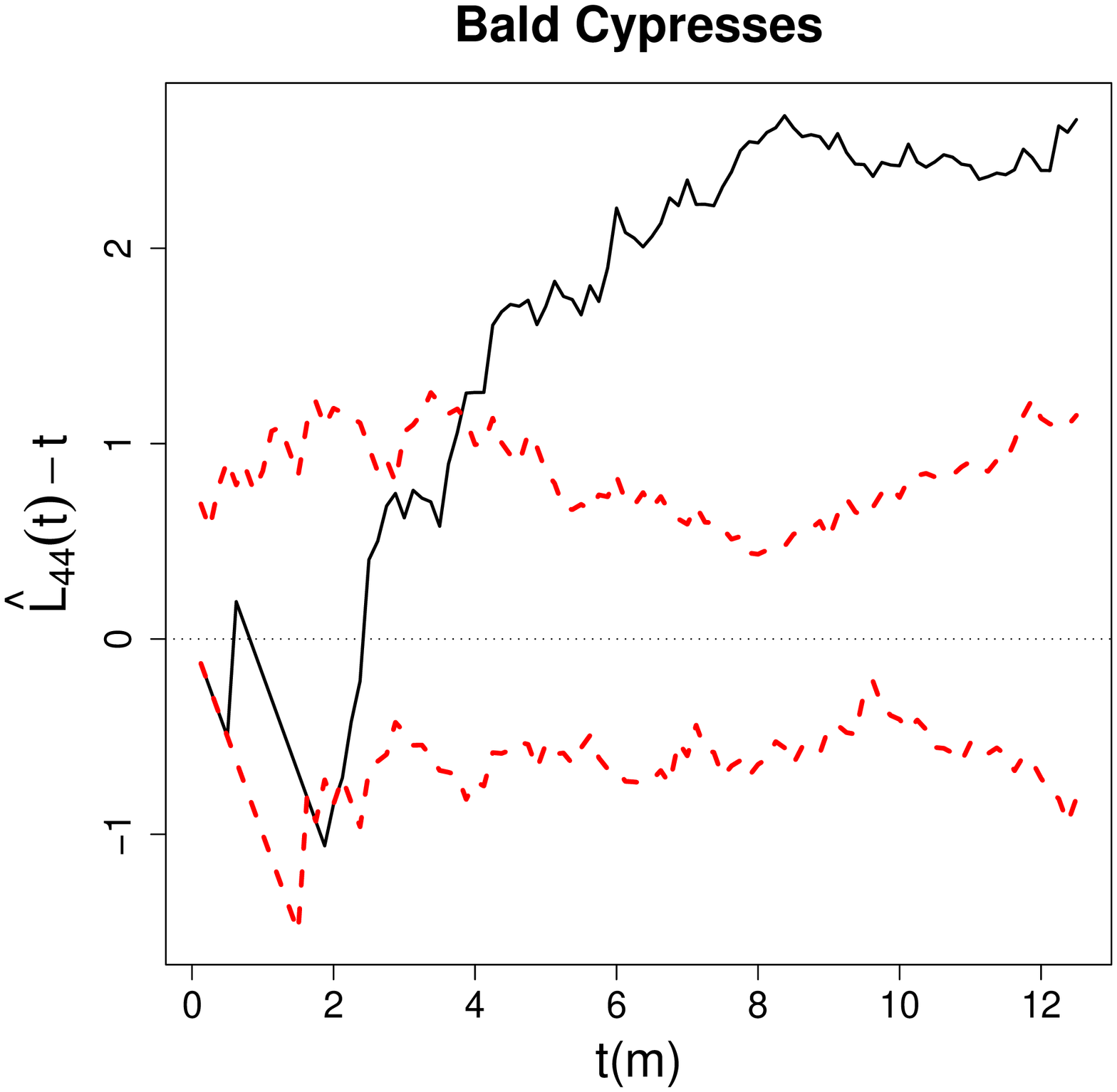} }}
\rotatebox{-90}{ \resizebox{2 in}{!}{\includegraphics{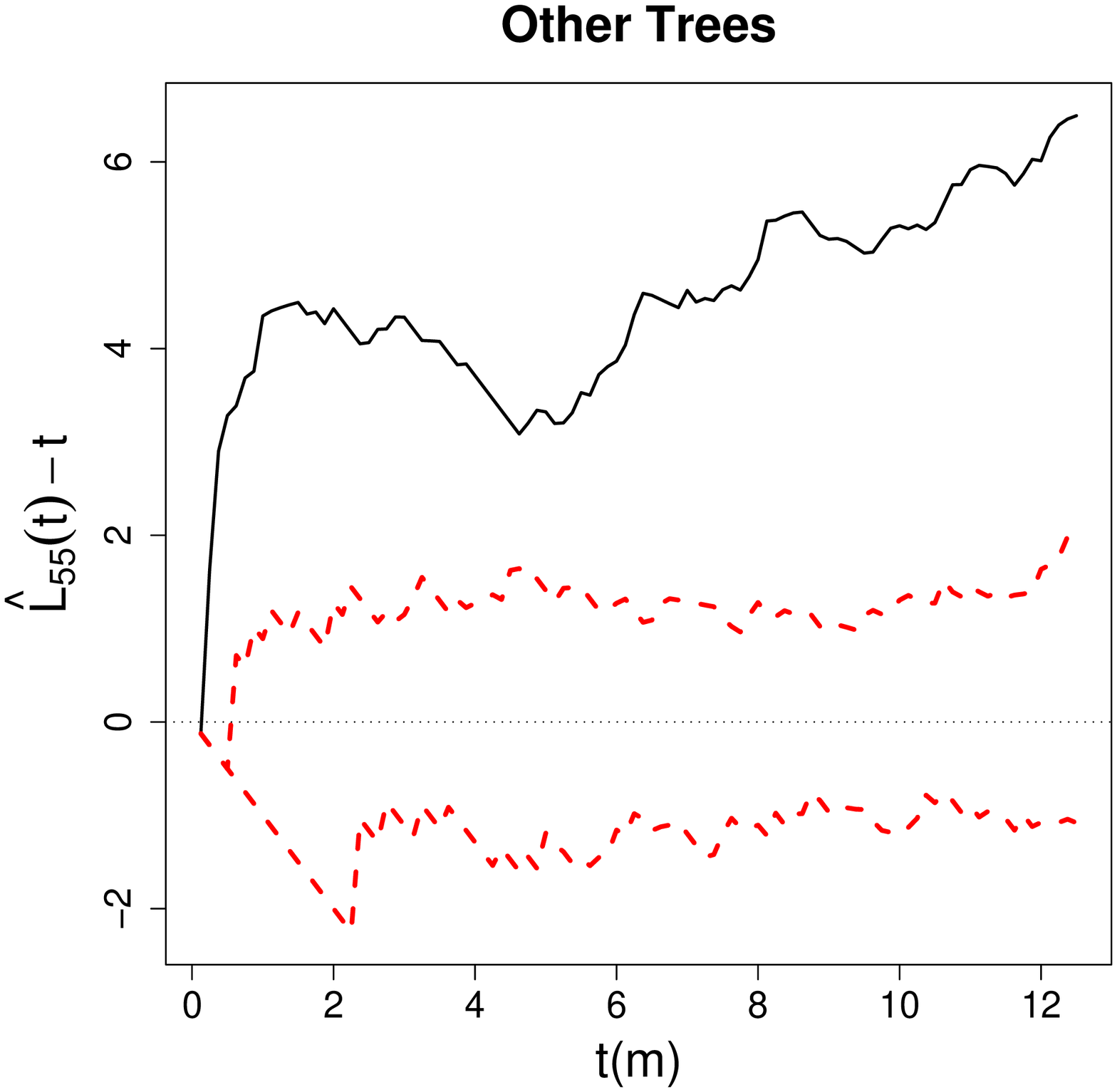} }}
\caption{
\label{fig:swamp-Liihat}
Second-order analysis of swamp tree data.
Functions plotted are Ripley's univariate $L$-functions
$\widehat{L}_{ii}(t)-t$ for $i=0,1,\ldots,5$,
where $i=0$ stands for all data combined, $i=1$ for water tupelos, $i=2$ for black gums,
$i=3$ for Carolina ashes, $i=4$ for bald cypresses, and $i=5$ for other trees.
Wide dashed lines around 0 (which is the theoretical value) are the upper and lower
(pointwise) 95 \% confidence bounds for the
$L$-functions based on Monte Carlo simulation under the CSR independence pattern.}
\end{figure}

\begin{figure}[t]
\centering
\rotatebox{-90}{ \resizebox{2 in}{!}{\includegraphics{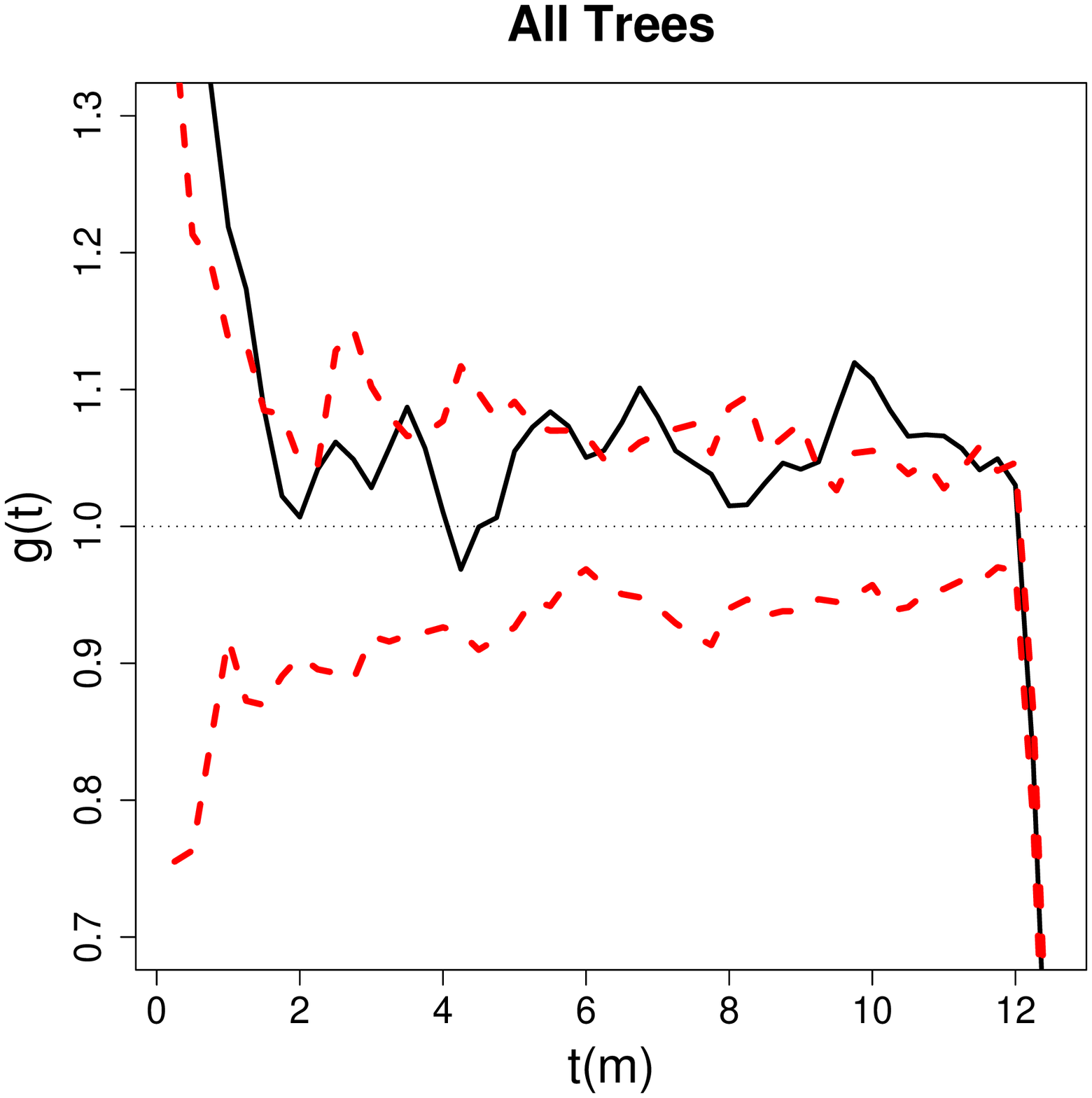} }}
\rotatebox{-90}{ \resizebox{2 in}{!}{\includegraphics{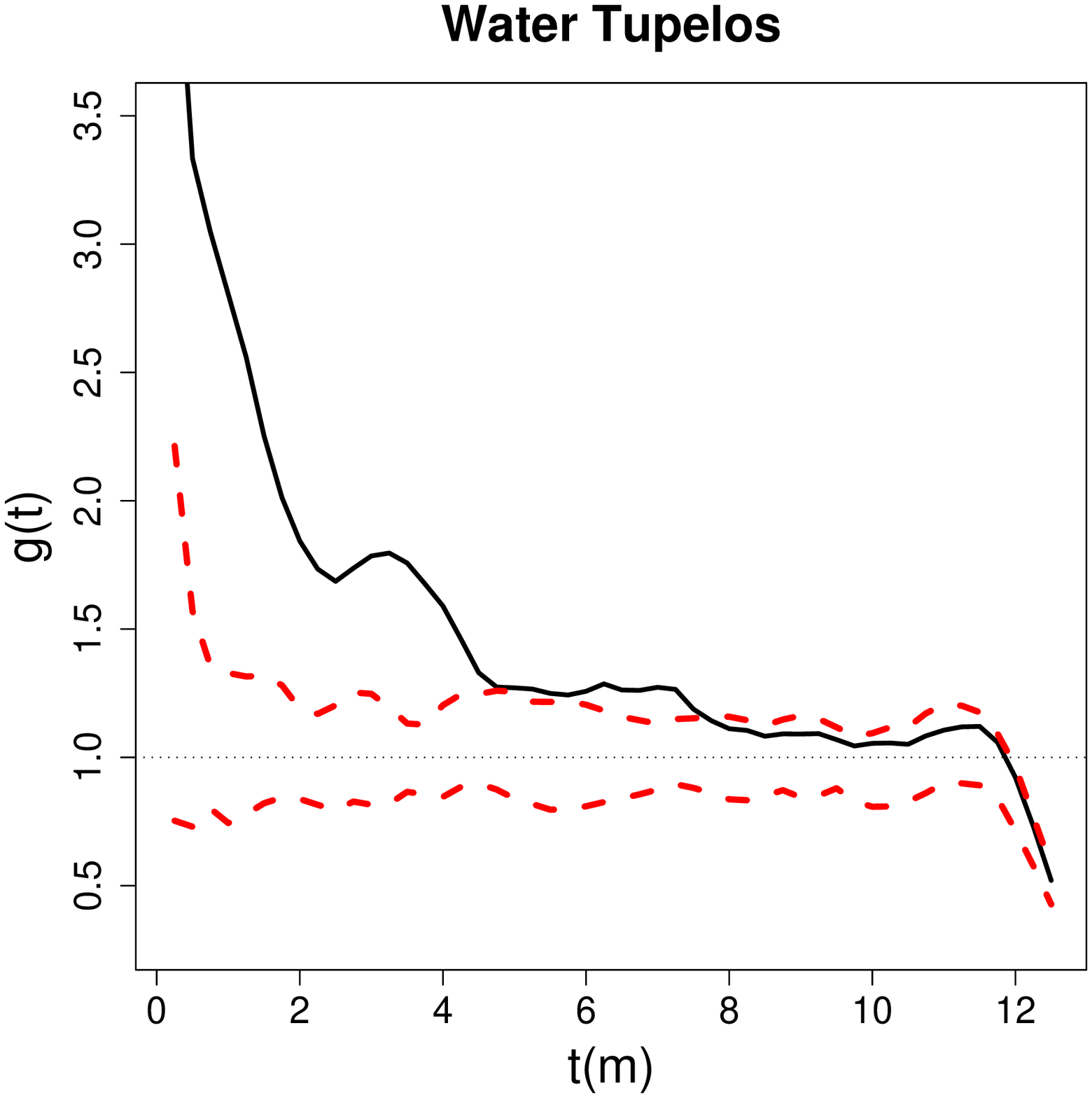} }}
\rotatebox{-90}{ \resizebox{2 in}{!}{\includegraphics{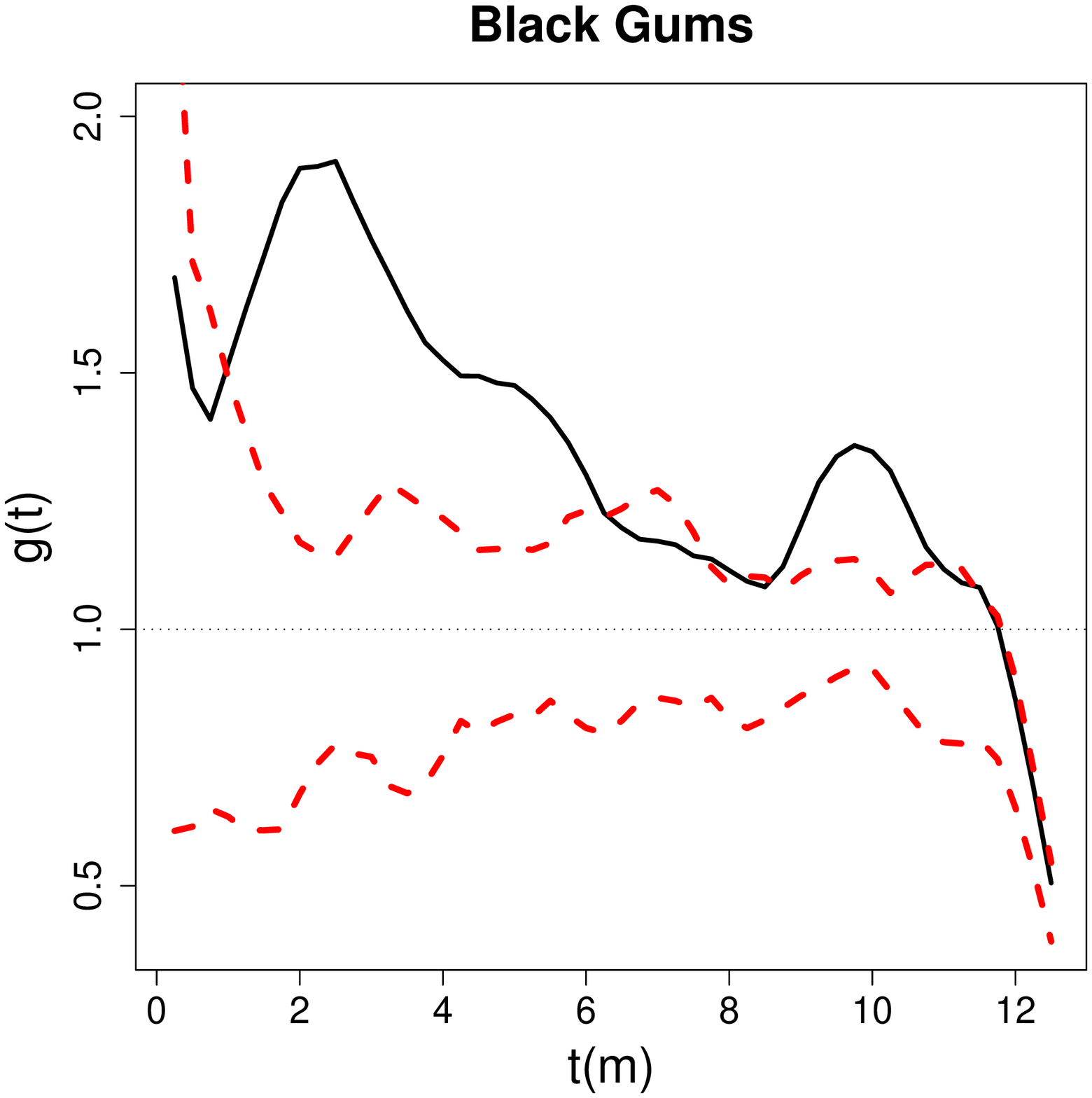} }}
\rotatebox{-90}{ \resizebox{2 in}{!}{\includegraphics{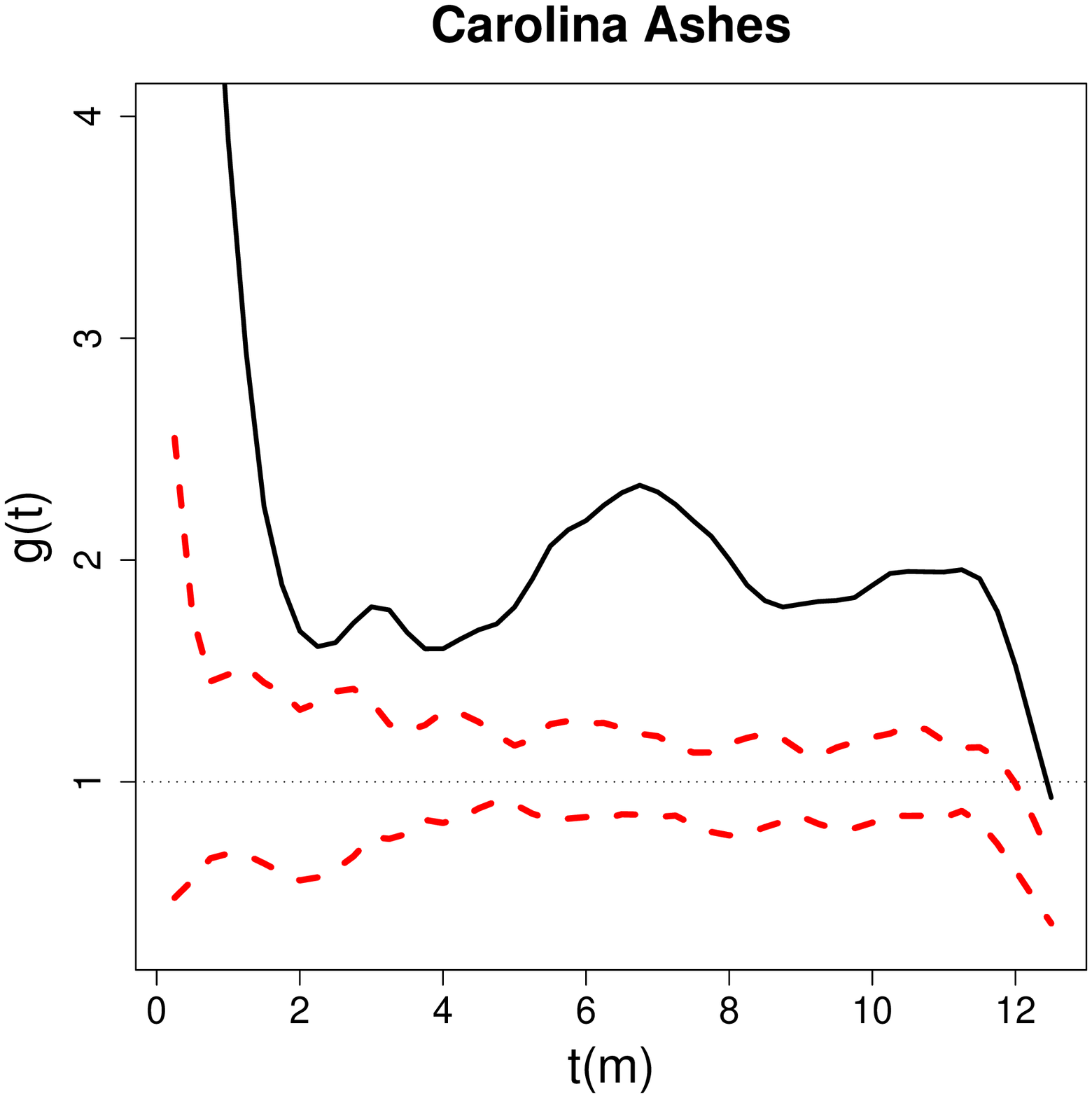} }}
\rotatebox{-90}{ \resizebox{2 in}{!}{\includegraphics{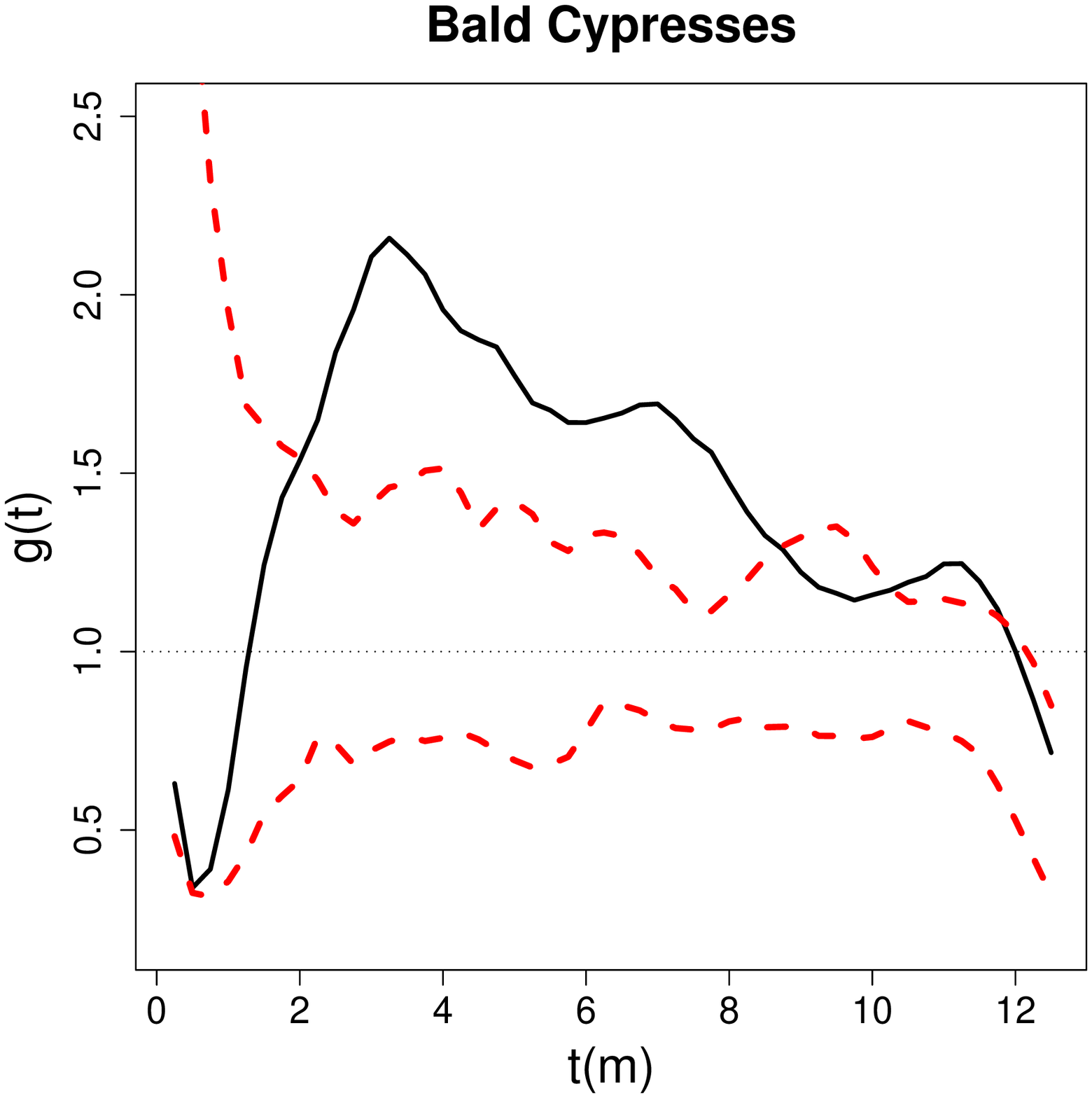} }}
\rotatebox{-90}{ \resizebox{2 in}{!}{\includegraphics{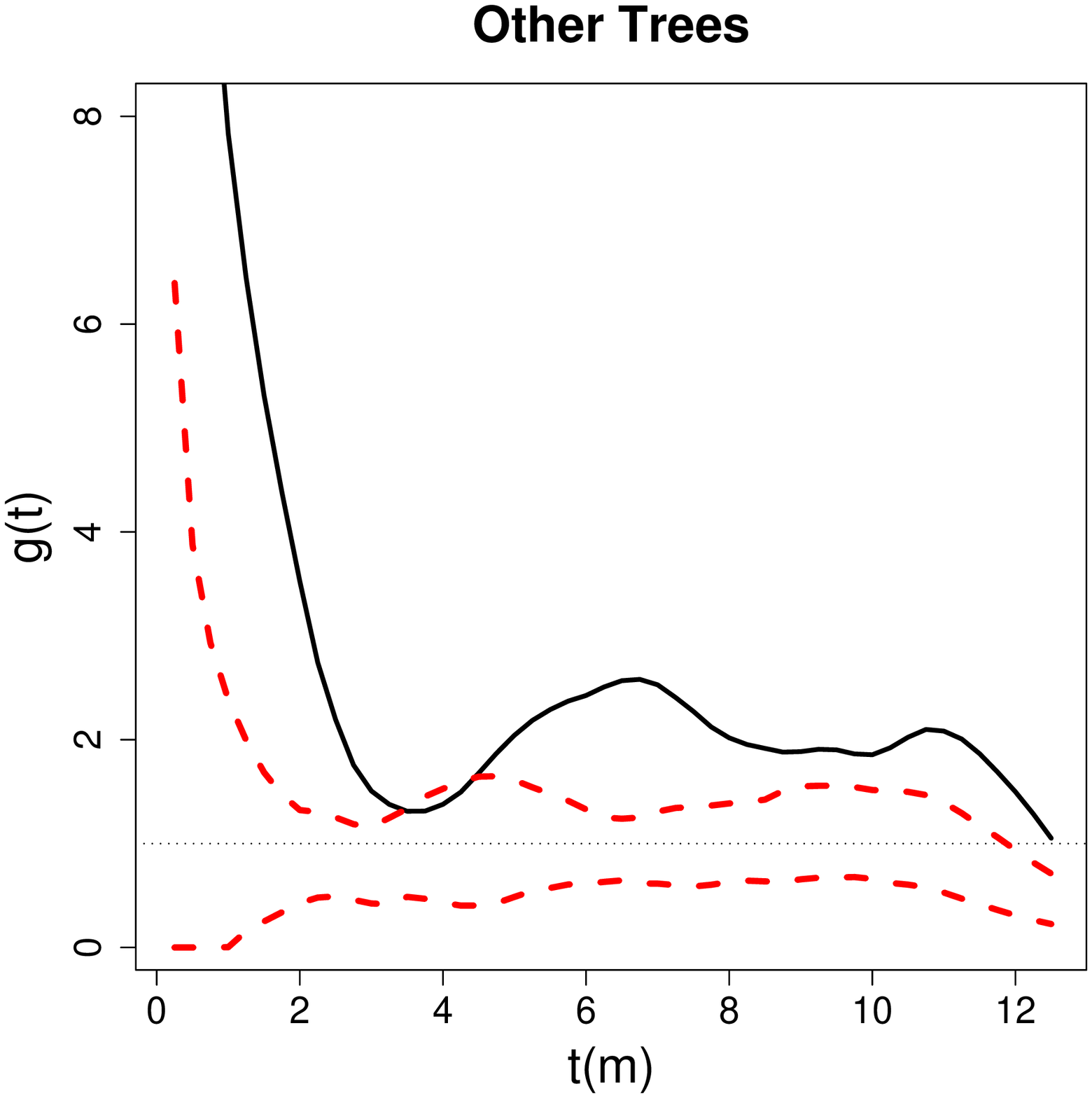} }}
\caption{
\label{fig:swamp-PCFii}
Pair correlation functions for all trees combined and for each species in the swamp tree data.
Wide dashed lines around 1 (which is the theoretical value)
are the upper and lower (pointwise) 95 \% confidence bounds for the
$L$-functions based on Monte Carlo simulation under the CSR independence pattern.}
\end{figure}

Based on the NNCT-tests above, we conclude that
all tree species but bald cypresses exhibit
significant deviation from the CSR independence pattern.
Considering Figure \ref{fig:SwampTrees}, the corresponding NNCT in Table \ref{tab:NNCT-swamp},
and the cell-specific test results in Tables \ref{tab:pval-swamp-dixon} and \ref{tab:pval-swamp-new},
this deviation is toward the segregation of the tree species.
However, these results pertain to small scale interaction at about the average NN distances.
We might also be interested in the causes of the segregation
and the type and level of interaction between the tree species
at different scales (i.e., distances between the trees).
To answer such questions, we also present the second-order analysis of the swamp tree data.
We calculate Ripley's (univariate) $L$-function which is the modified version of $K$ function
as $\widehat{L}_{ii}(t)=\sqrt{\left( \widehat{K}_{ii}(t)/\pi \right)}$
where $t$ is the distance from a randomly chosen event (i.e., location of a tree),
$\widehat{K}_{ii}(t)$ is an estimator of
\begin{equation}
\label{eqn:Kii}
K(t)=\lambda^{-1}\E[\text{\# of extra events within distance $t$ of a randomly chosen event}]
\end{equation}
with $\lambda$ being the density (number per unit area) of events
and is calculated as
\begin{equation}
\label{eqn:Kiihat}
\widehat{K}_{ii}(t)=\widehat{\lambda}^{-1}\sum_i\sum_{j \not= i}w(l_i,l_j)\I(d_{ij}<t)/N
\end{equation}
where $\widehat{\lambda}=N/A$ is an estimate of density ($N$ is the observed number of points
and $A$ is the area of the study region),
$d_{ij}$ is the distance between points $i$ and $j$,
$\I(\cdot)$ is the indicator function,
$w(l_i,l_j)$ is the proportion of the circumference of the circle
centered at $l_i$ with radius $d_{ij}$ that falls in the study area,
which corrects for the boundary effects.
Under CSR independence, $L(t)-t=0$ holds.
If the univariate pattern exhibits aggregation,
then $L(t)-t$ tends to be positive,
if it exhibits regularity then $L(t)-t$ tends to be negative.
The estimator $\widehat{K}(t)$ is approximately unbiased for $K(t)$ at each fixed $t$.
Bias depends on the geometry of the study area and increases with $t$.
For a rectangular region
it is recommended to use $t$ values up to 1/4 of the smaller side
length of the rectangle.
See (\cite{diggle:2003}) for more detail.
So we take the values $t \in [0,12.5]$ in our analysis,
since the smaller side of the rectangular region of swamp tree data is 50 m.
In Figure \ref{fig:swamp-Liihat},
we present the plots of $\widehat{L}_{ii}(t)-t$ functions for each species
as well as the plot of all trees combined.
We also present the upper and lower (pointwise)
95 \% confidence bounds for each $L_{ii}(t)-t$.
Observe that for all trees combined there is significant aggregation of trees
(the $L_{00}(t)-t$ curve is above the upper confidence bound) at all scales (i.e., distances).
Water tupelos exhibit significant aggregation for the range of the plotted distances;
black gums exhibit significant aggregation for distances $t>1$ m;
Carolina ashes exhibit significant aggregation for the range of plotted distances;
bald cypresses exhibit no deviation from CSR independence for $t \lesssim 5$ m,
then they exhibit significant spatial aggregation for $t>4$ m;
other trees exhibit significant aggregation for the range of plotted distances.
Hence, segregation of the species might be due to different
levels and types of aggregation of the species in the study region.

\begin{figure}[t]
\centering
\rotatebox{-90}{ \resizebox{2 in}{!}{\includegraphics{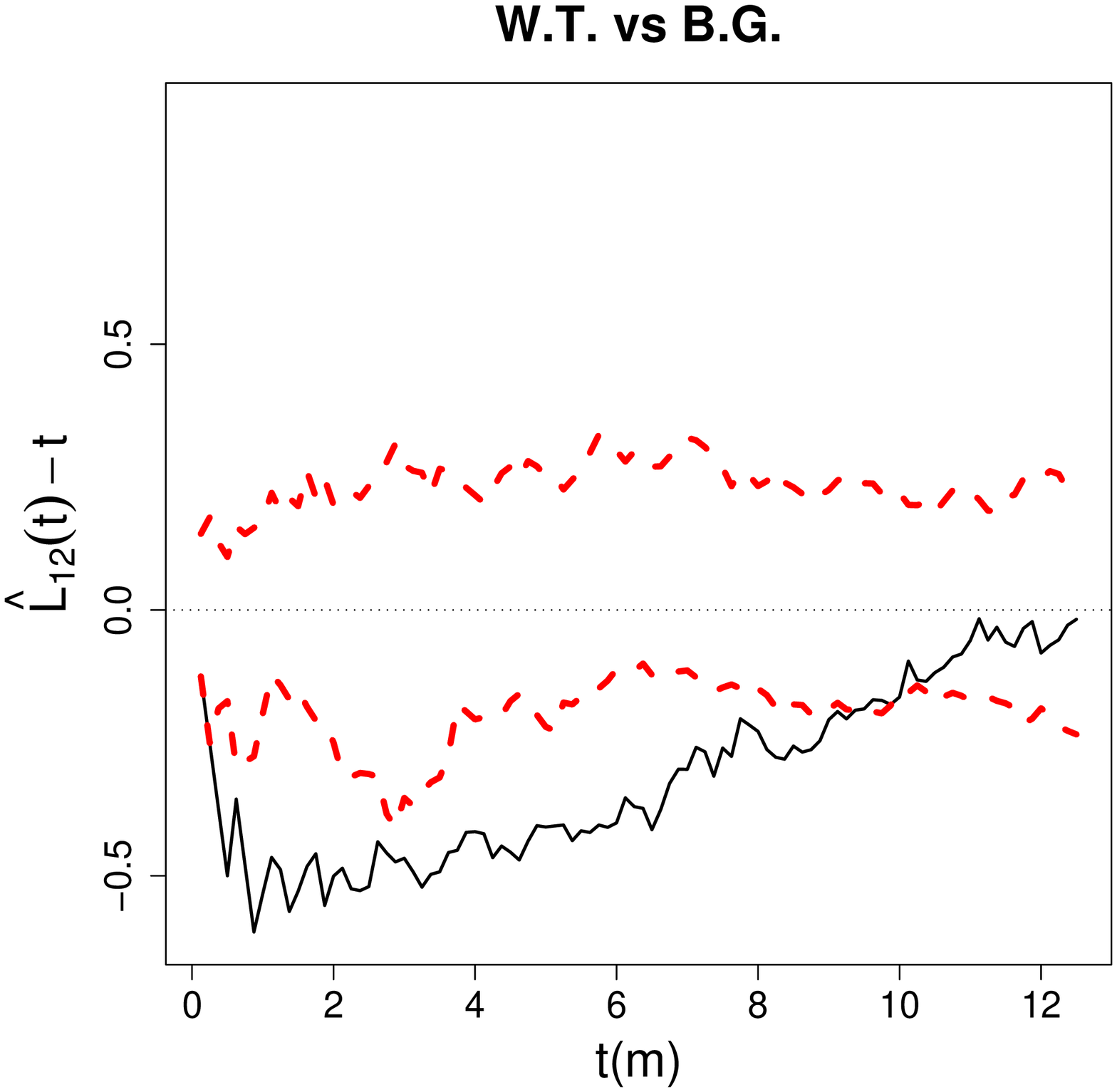} }}
\rotatebox{-90}{ \resizebox{2 in}{!}{\includegraphics{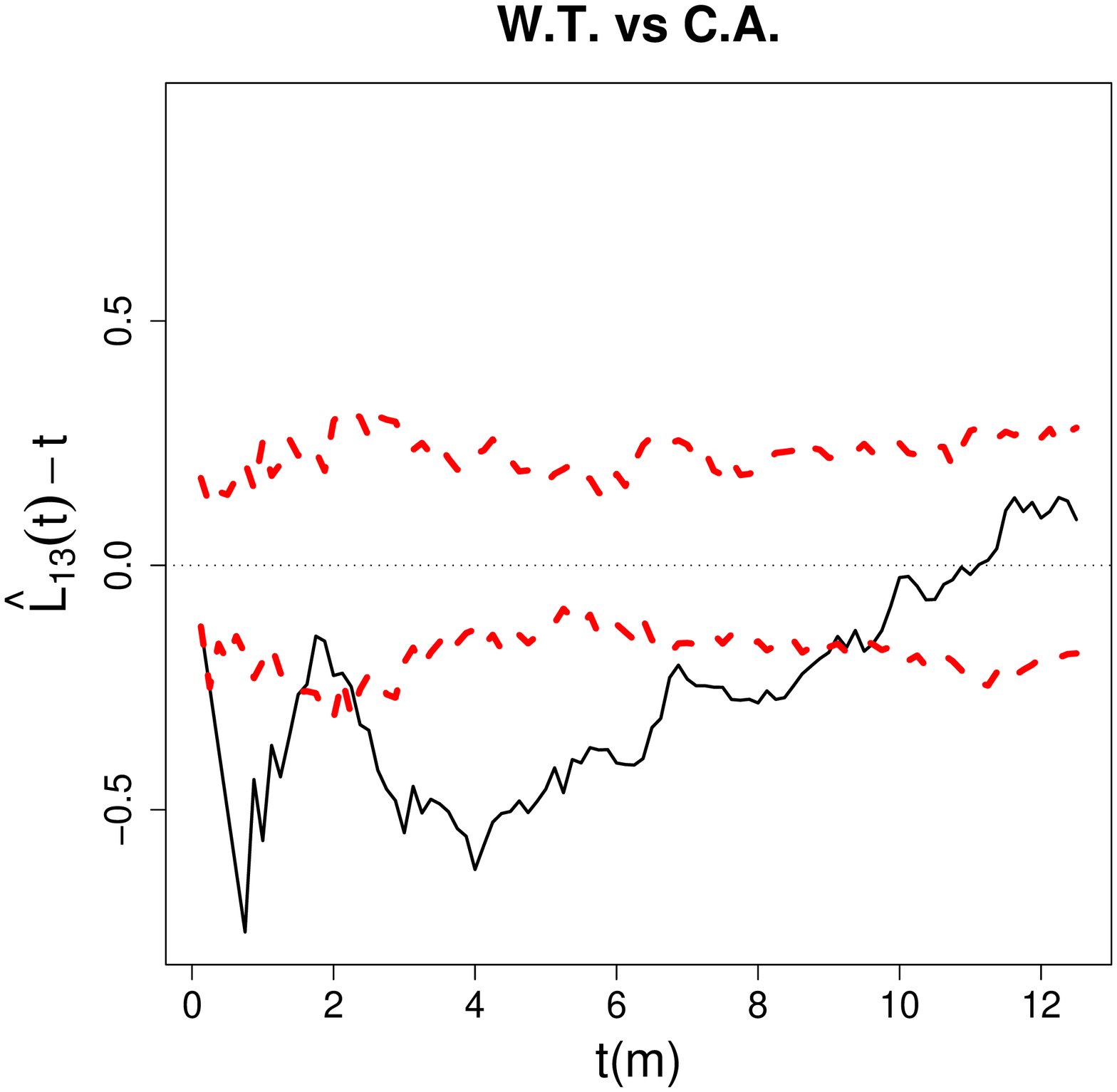} }}
\rotatebox{-90}{ \resizebox{2 in}{!}{\includegraphics{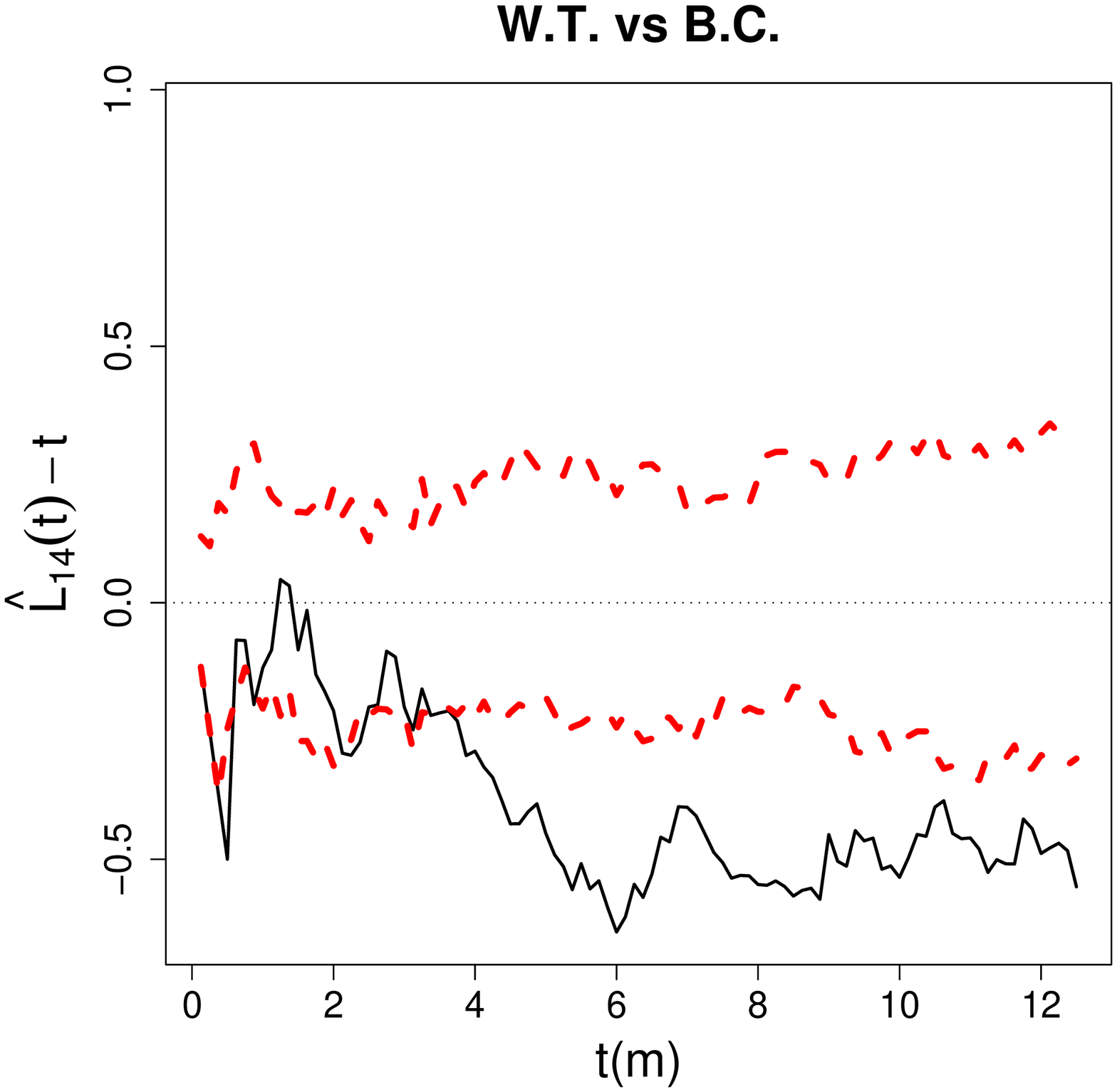} }}
\rotatebox{-90}{ \resizebox{2 in}{!}{\includegraphics{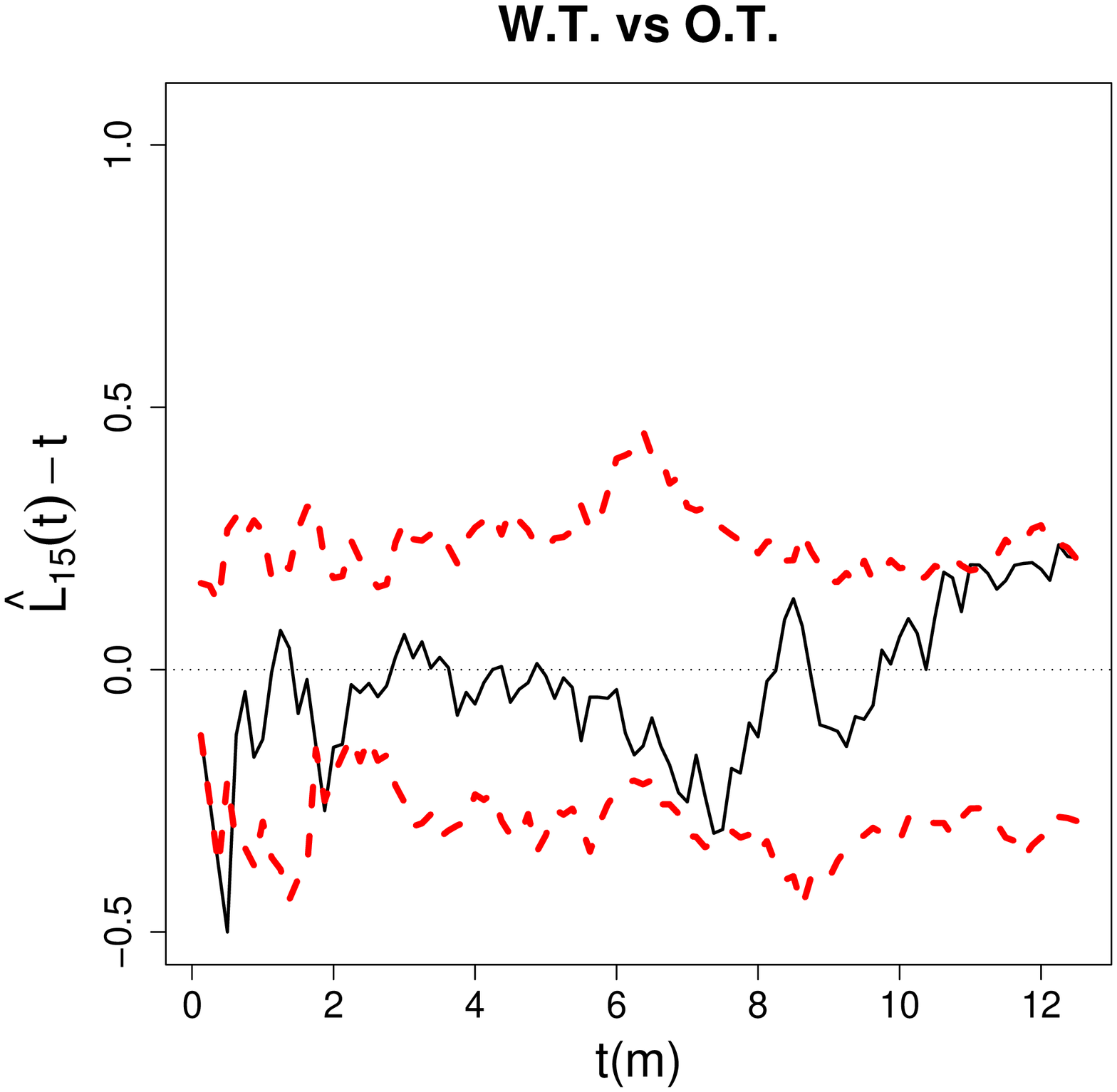} }}
\rotatebox{-90}{ \resizebox{2 in}{!}{\includegraphics{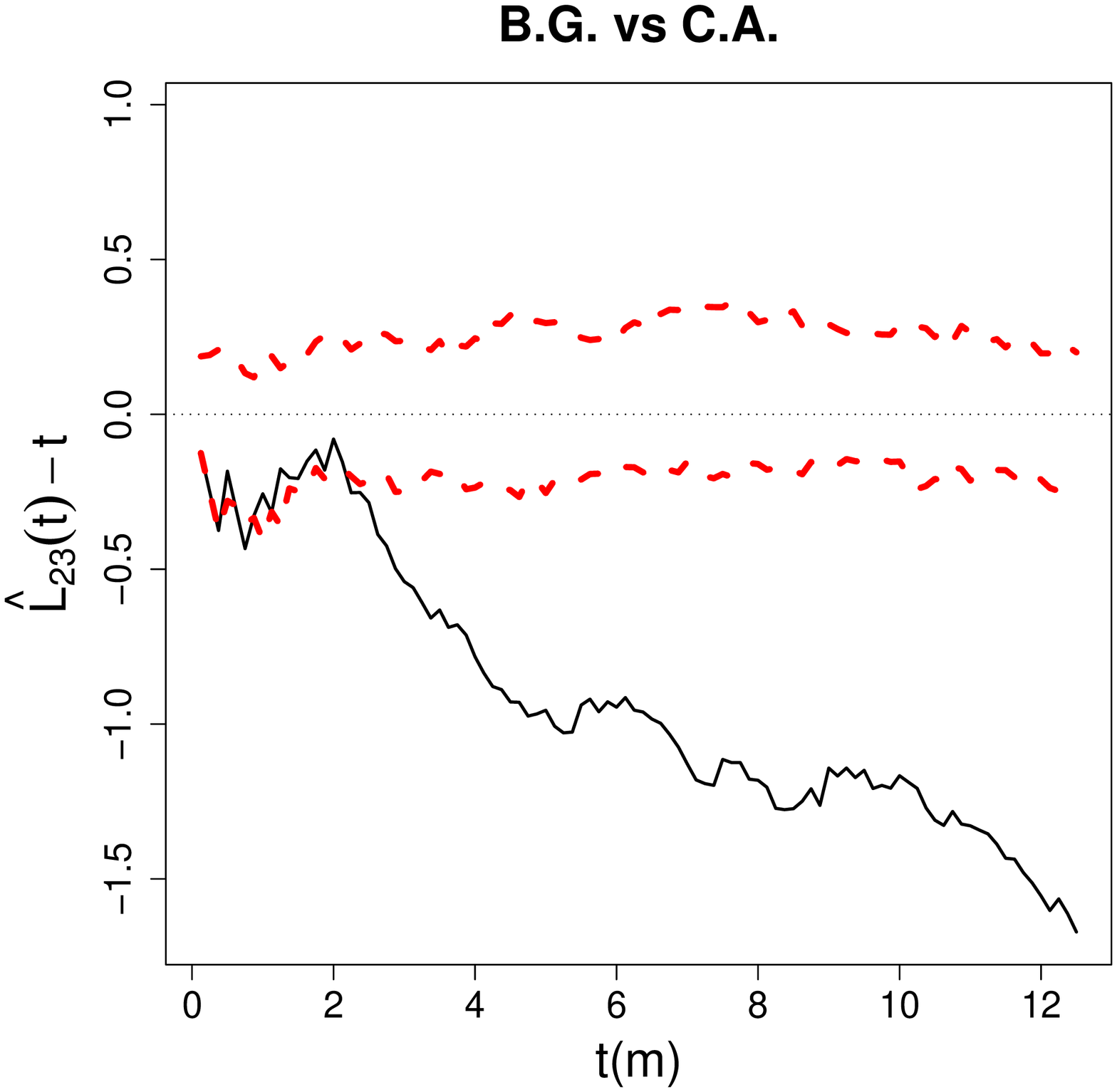} }}
\rotatebox{-90}{ \resizebox{2 in}{!}{\includegraphics{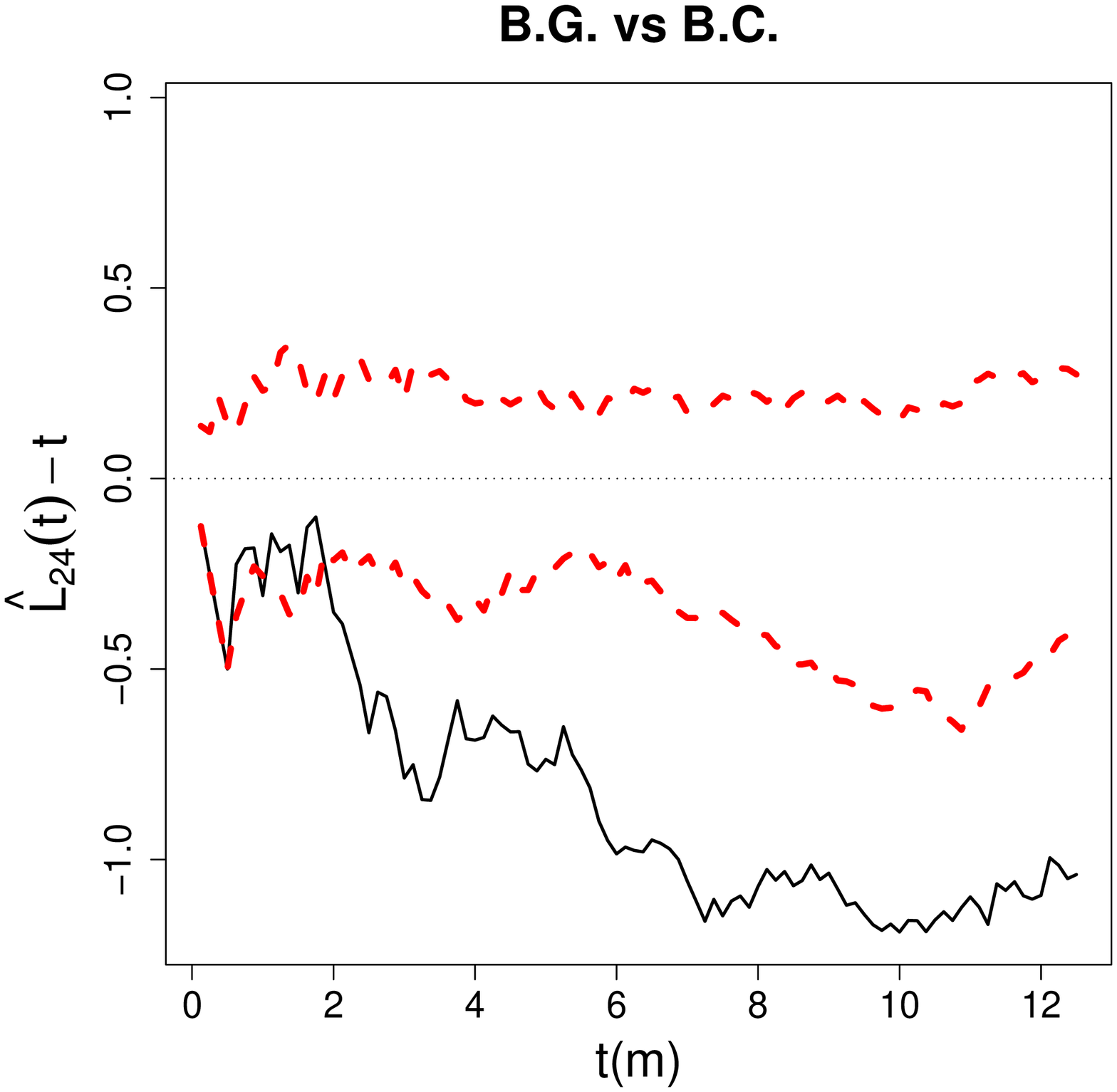} }}
\rotatebox{-90}{ \resizebox{2 in}{!}{\includegraphics{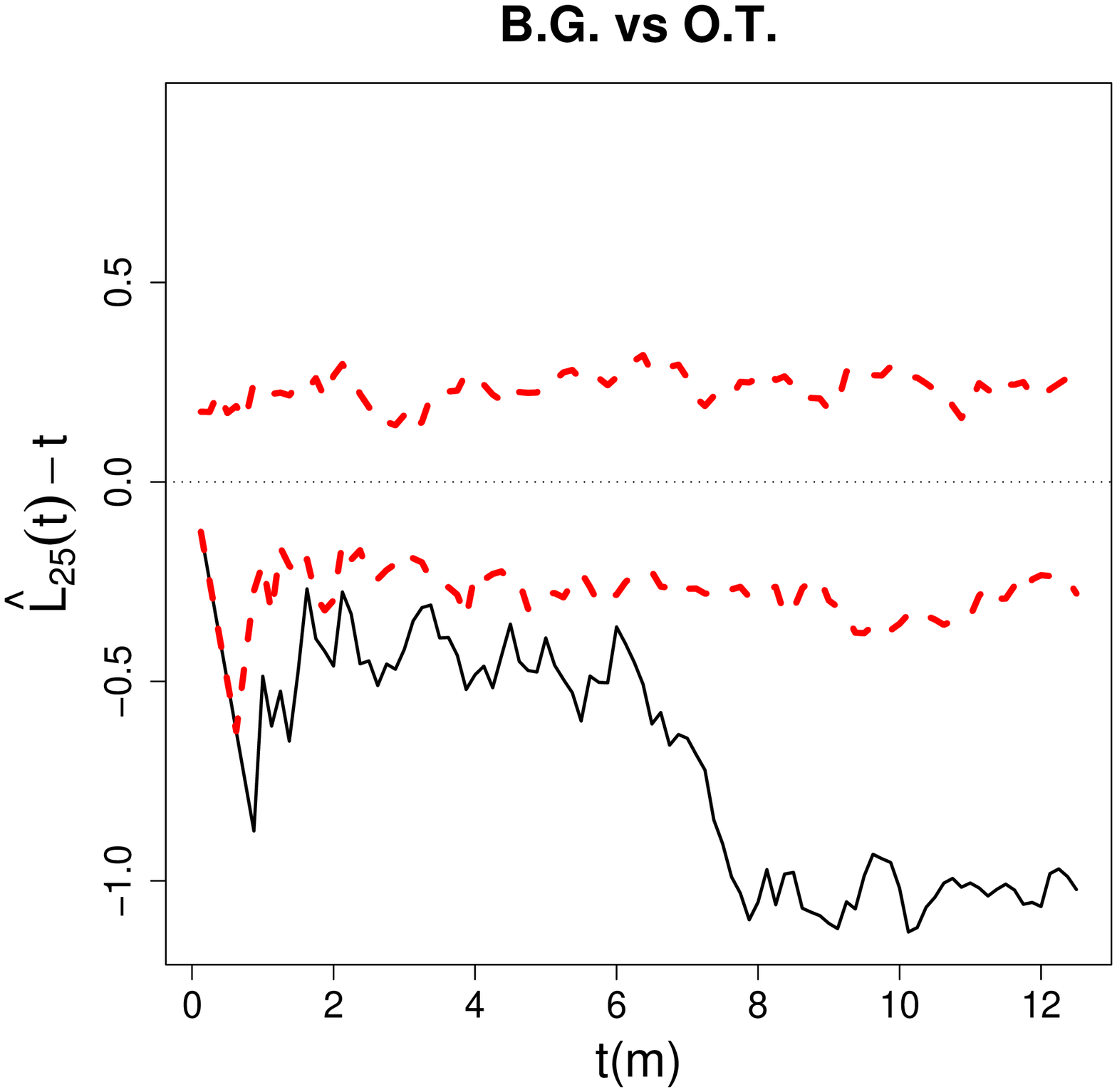} }}
\rotatebox{-90}{ \resizebox{2 in}{!}{\includegraphics{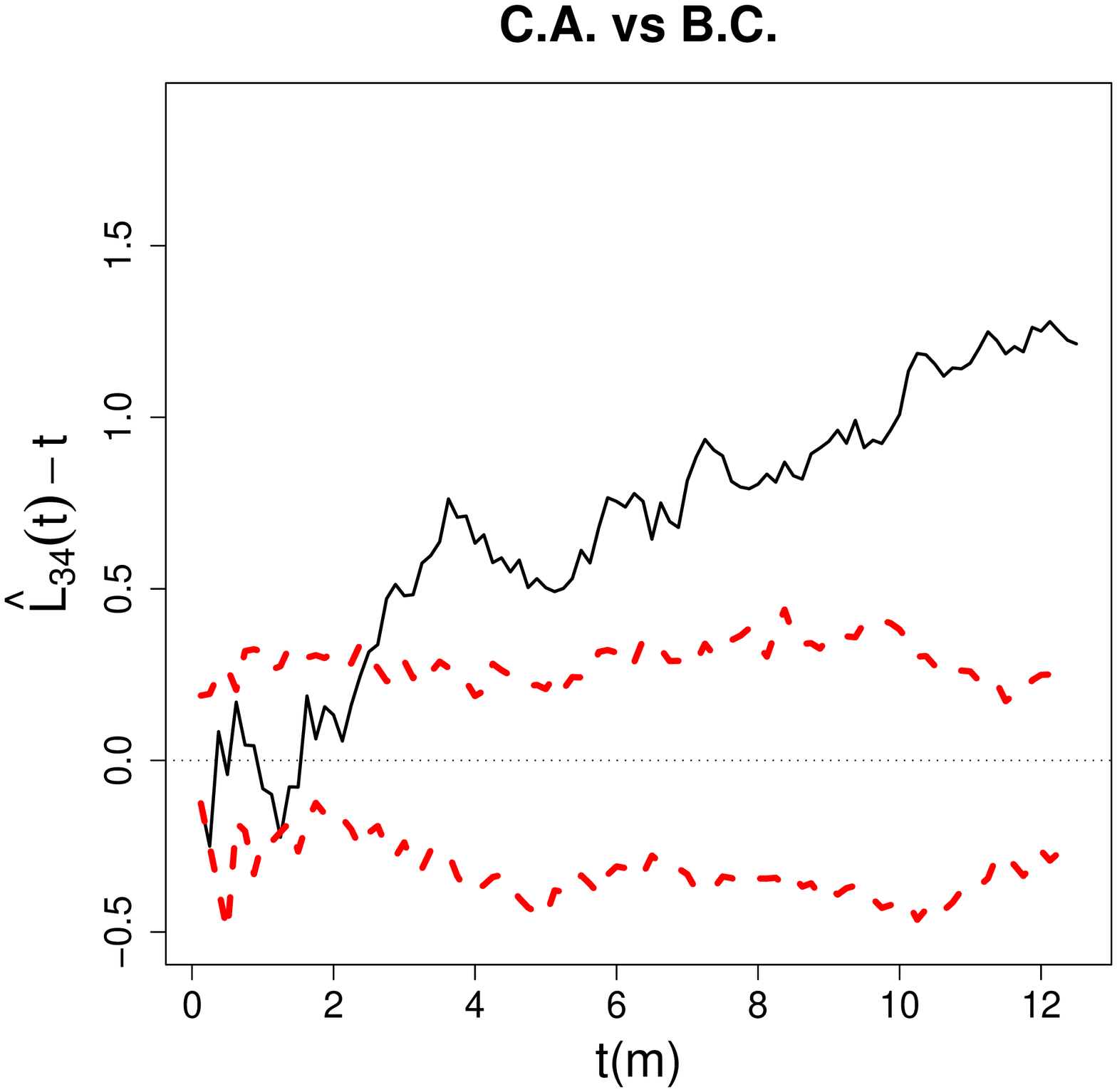} }}
\rotatebox{-90}{ \resizebox{2 in}{!}{\includegraphics{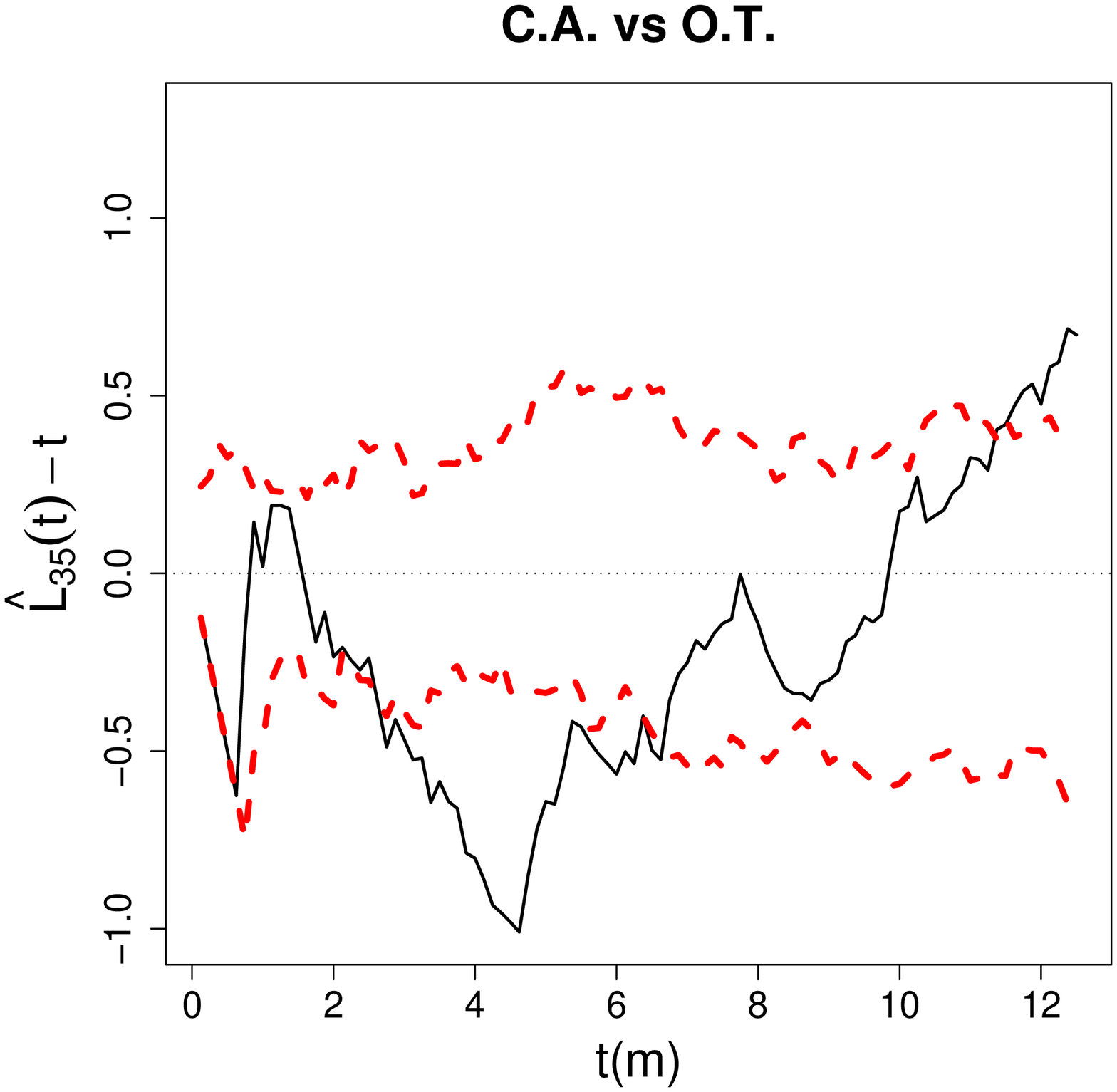} }}
\rotatebox{-90}{ \resizebox{2 in}{!}{\includegraphics{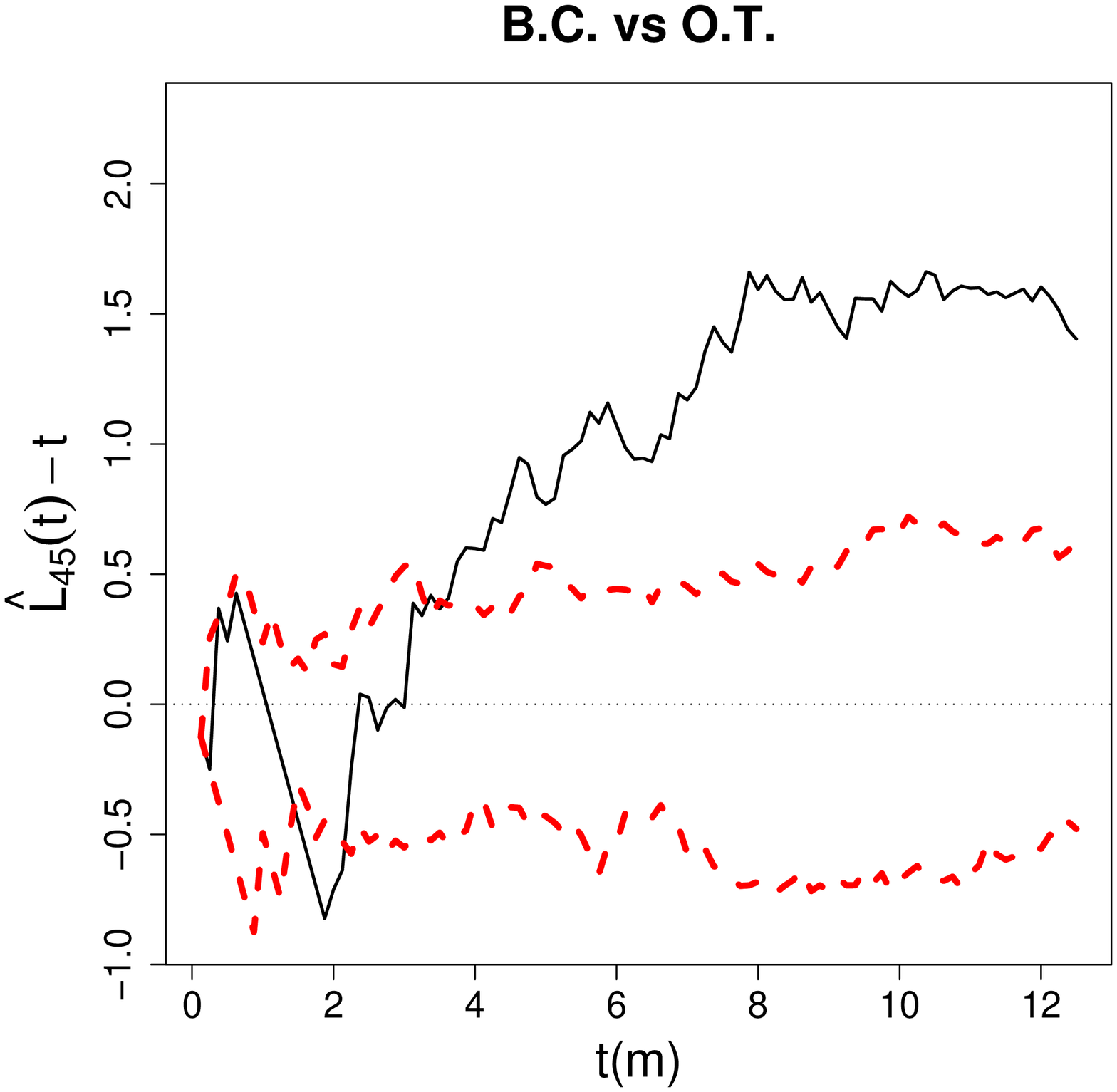} }}
\caption{
\label{fig:swamp-Lijhat}
Second-order analysis of swamp tree data.
Functions plotted are Ripley's bivariate $L$-functions
$\widehat{L}_{ij}(t)-t$ for $i,j=1,2,\ldots,5$ and $i \not= j$
where $i=0$ stands for all data combined, $i=1$ for water tupelos, $i=2$ for black gums,
$i=3$ for Carolina ashes, $i=4$ for bald cypresses, and $i=5$ for other trees.
Wide dashed lines around 0 (which is the theoretical value)
are the upper and lower (pointwise) 95 \% confidence bounds for the
$L$-functions based on Monte Carlo simulations under the CSR independence pattern.
W.T. = water tupelos, B.G.
= black gum, C.A. = Carolina ashes, B.C.  = bald cypresses, and O.T. =
other tree species.}
\end{figure}

We also calculate Ripley's bivariate $L$-function
as $\widehat{L}_{ij}(t)=\sqrt{\left( \widehat{K}_{ij}(t)/\pi \right)}$
where $\widehat{K}_{ij}(t)$ is an estimator of
$$K_{ij}(t)=\lambda_j^{-1}\E[\text{\# of extra type $j$ events within distance $t$ of a randomly chosen type $i$ event}]$$
with $\lambda_j$ being the density of type $j$ events.
Then $\widehat{K}_{ij}(t)$ is calculated as
\begin{equation}
\label{eqn:Kijhat}
\widehat{K}_{ij}(t)=\left( \widehat{\lambda}_i\widehat{\lambda}_j A \right)^{-1}
\sum_i\sum_{j}w(i_k,j_l)\I(d_{i_k,j_l}<t),
\end{equation}
where $d_{i_k,j_l}$ is the distance between $k^{th}$ type $i$ and $l^{th}$ type $j$ points,
$w(i_k,j_l)$ is the proportion of the circumference of the circle
centered at $k^{th}$ type $i$ point with radius $d_{i_k,j_l}$ that falls in the study area,
which is used for edge correction.
Notice that by construction, $\widehat{L}_{ij}(t)$ is symmetric in $i$ and $j$,
that is, $\widehat{L}_{ij}(t)=\widehat{L}_{ji}(t)$ for all $i,j$.
Under CSR independence, $L_{ij}(t)-t=0$ holds.
If the bivariate pattern is segregation,
then $L_{ij}(t)-t$ tends to be negative,
if it is association then $L_{ij}(t)-t$ tends to be positive.
See (\cite{diggle:2003}) for more detail.
In Figure \ref{fig:swamp-Lijhat},
we present the bivariate plots of $\widehat{L}_{ij}(t)-t$ functions
together with the upper and lower (pointwise) 95 \% confidence bounds
for each pair of species (due to the symmetry of $\widehat{L}_{ij}(t)$
there are only 10 different pairs).
Observe that for distances up to $t \approx 10$ m,
water tupelos and black gums exhibit significant segregation
($\widehat{L}_{12}(t)-t$ is below the lower confidence bound)
and for the rest of the plotted distances their interaction is not significantly
different from the CSR independence pattern;
water tupelos and Carolina ashes are significantly segregated up to about $t \approx 10$ m;
water tupelos and bald cypresses do not have significant deviation
from the CSR independence pattern for distances up to 4 m,
for larger distances they exhibit significant segregation;
water tupelos and the other trees do not deviate from CSR independence
for the range of the plotted distances.
Black gums and Carolina ashes are significantly segregated for $t>2$ m;
black gums and bald cypresses are significantly segregated for  $t>2$ m;
black gum and other trees are significantly segregated for all the distances plotted.
Carolina ashes and bald cypresses are significantly associated for distances
larger than 3 m;
and Carolina ashes and the other trees exhibit significant segregation
for $3< t < 7$ m and for $t>11$ m they exhibit significant association.
On the other hand, bald cypresses and other trees are significantly associated
for distance larger than 4 m.

\begin{figure}[t]
\centering
\rotatebox{-90}{ \resizebox{2 in}{!}{\includegraphics{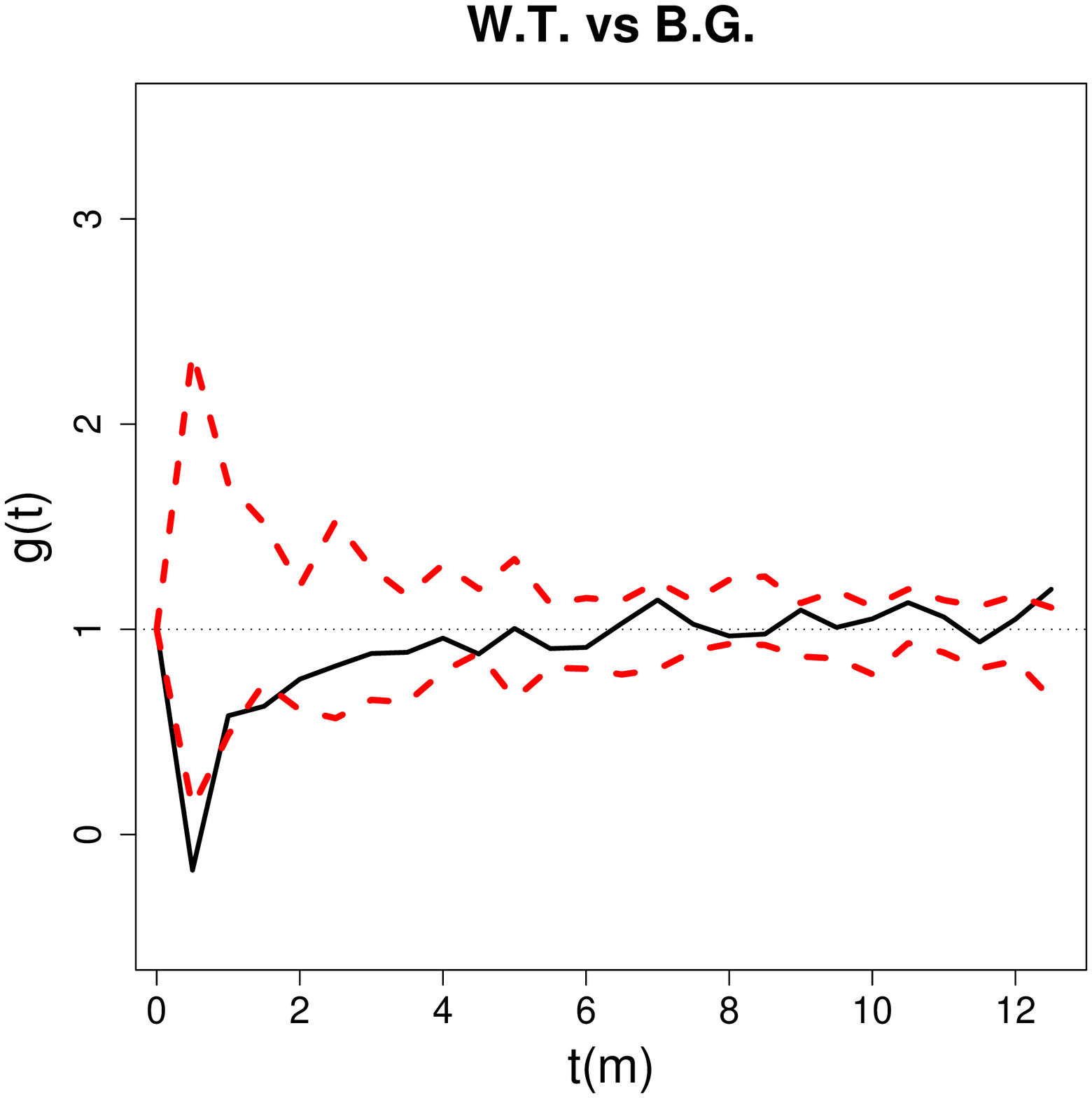} }}
\rotatebox{-90}{ \resizebox{2 in}{!}{\includegraphics{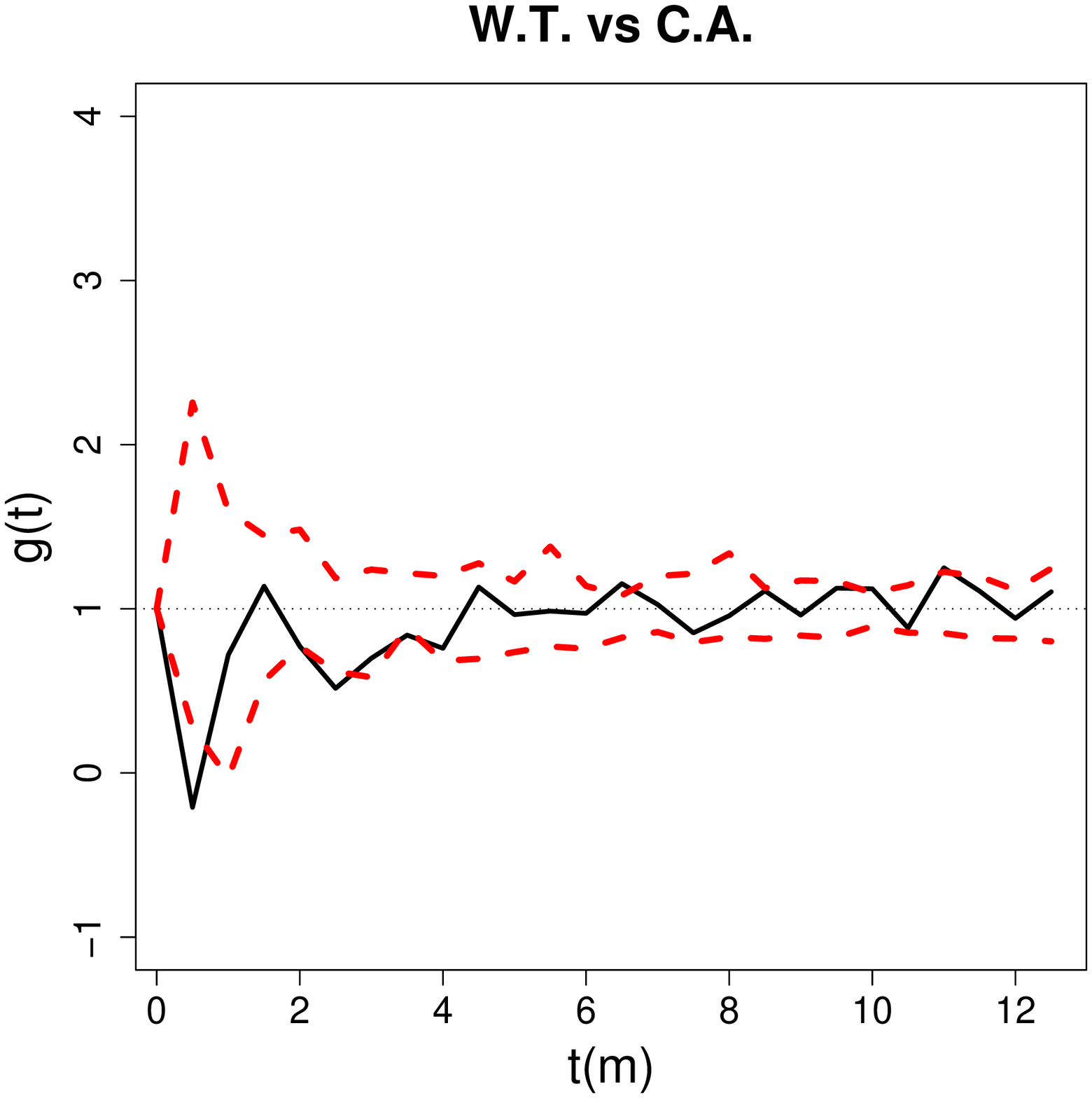} }}
\rotatebox{-90}{ \resizebox{2 in}{!}{\includegraphics{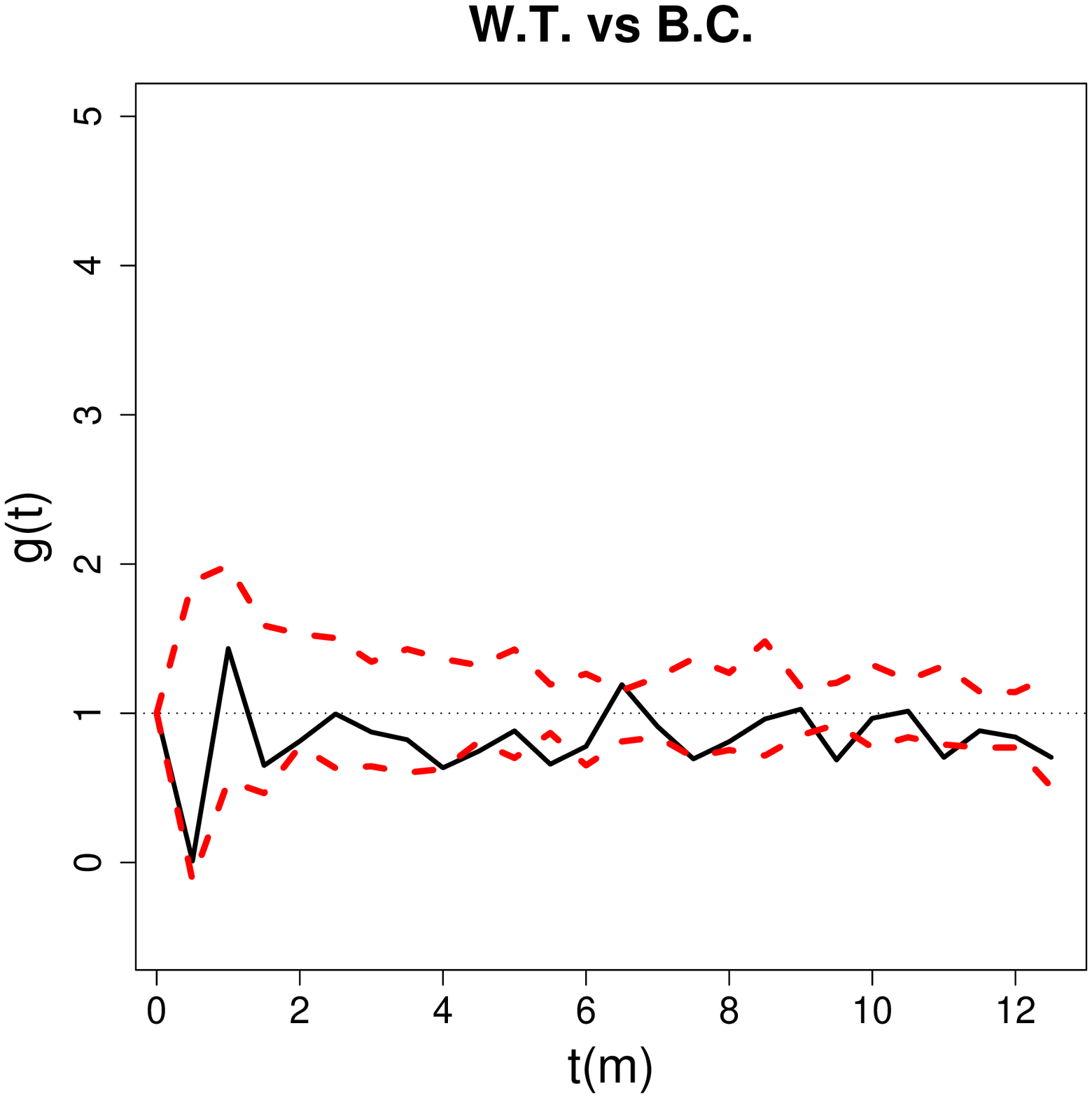} }}
\rotatebox{-90}{ \resizebox{2 in}{!}{\includegraphics{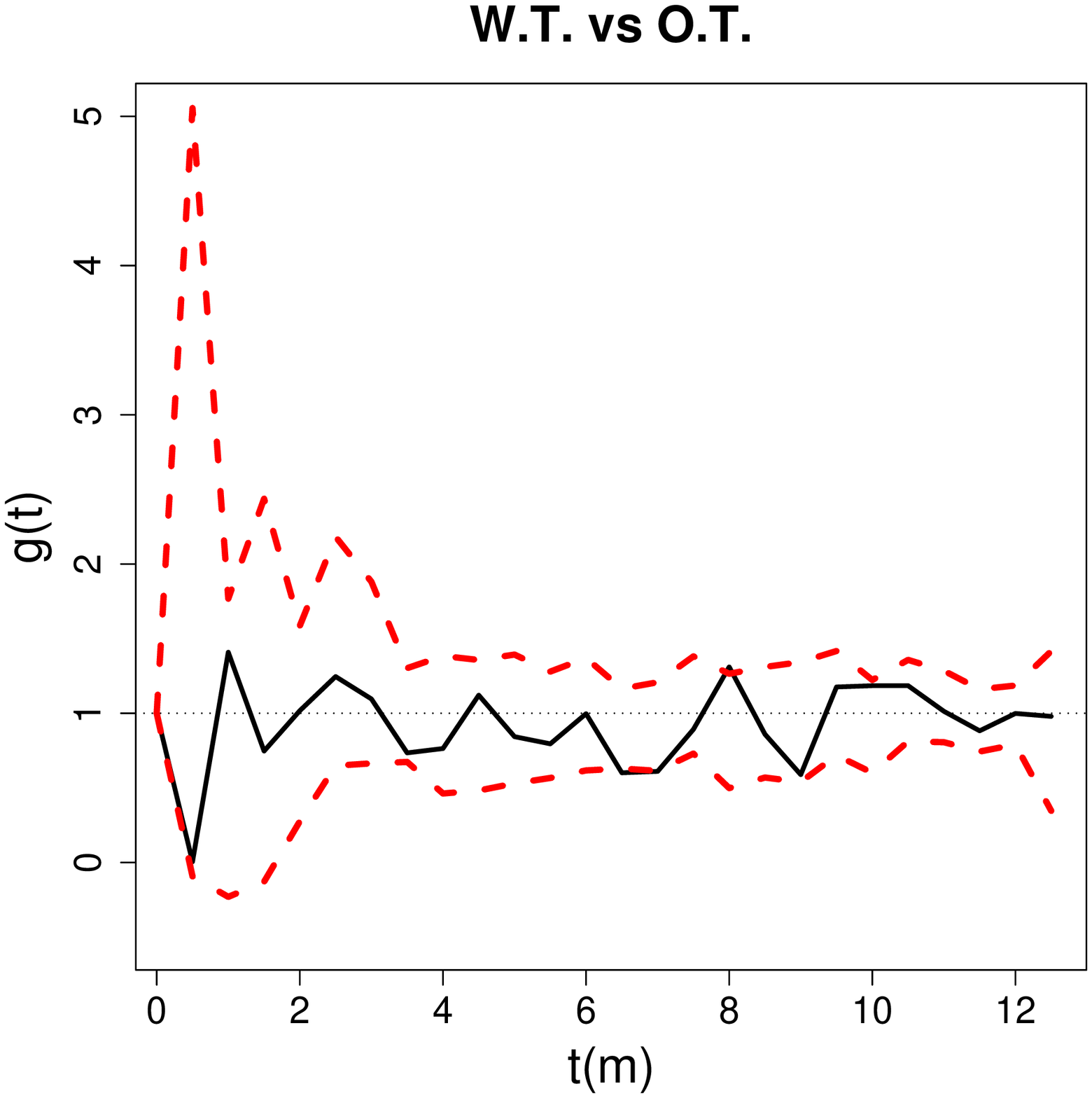} }}
\rotatebox{-90}{ \resizebox{2 in}{!}{\includegraphics{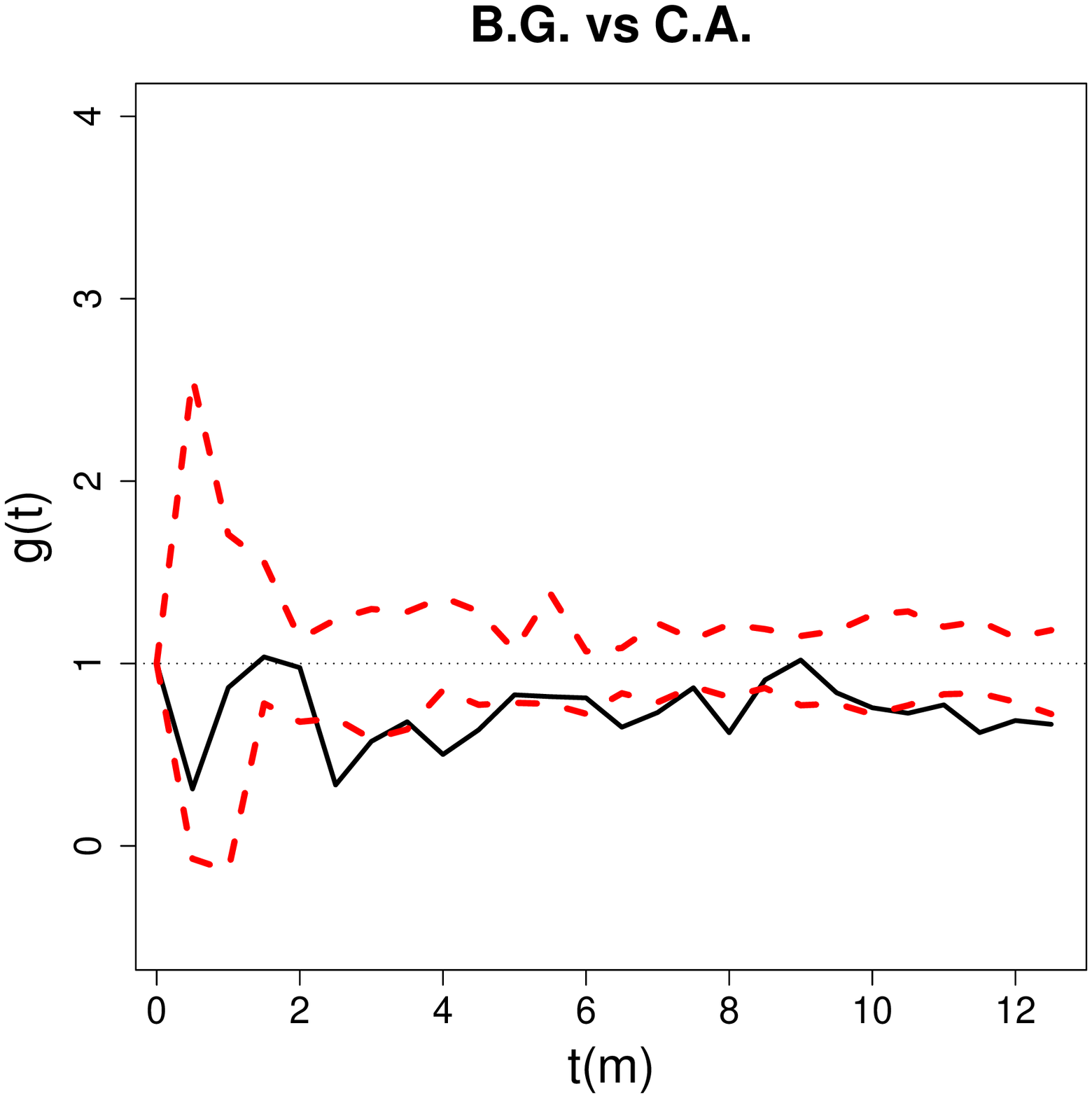} }}
\rotatebox{-90}{ \resizebox{2 in}{!}{\includegraphics{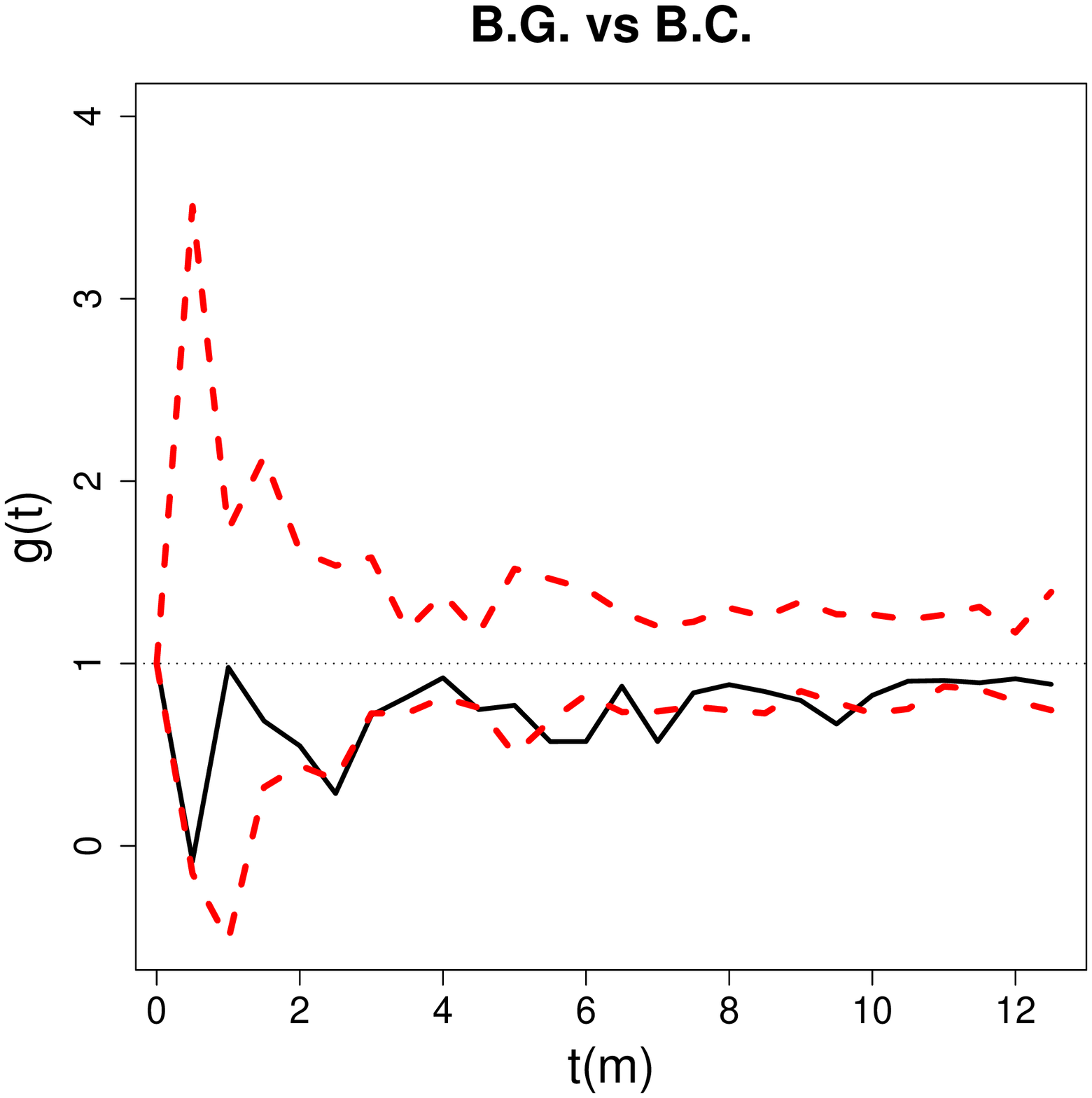} }}
\rotatebox{-90}{ \resizebox{2 in}{!}{\includegraphics{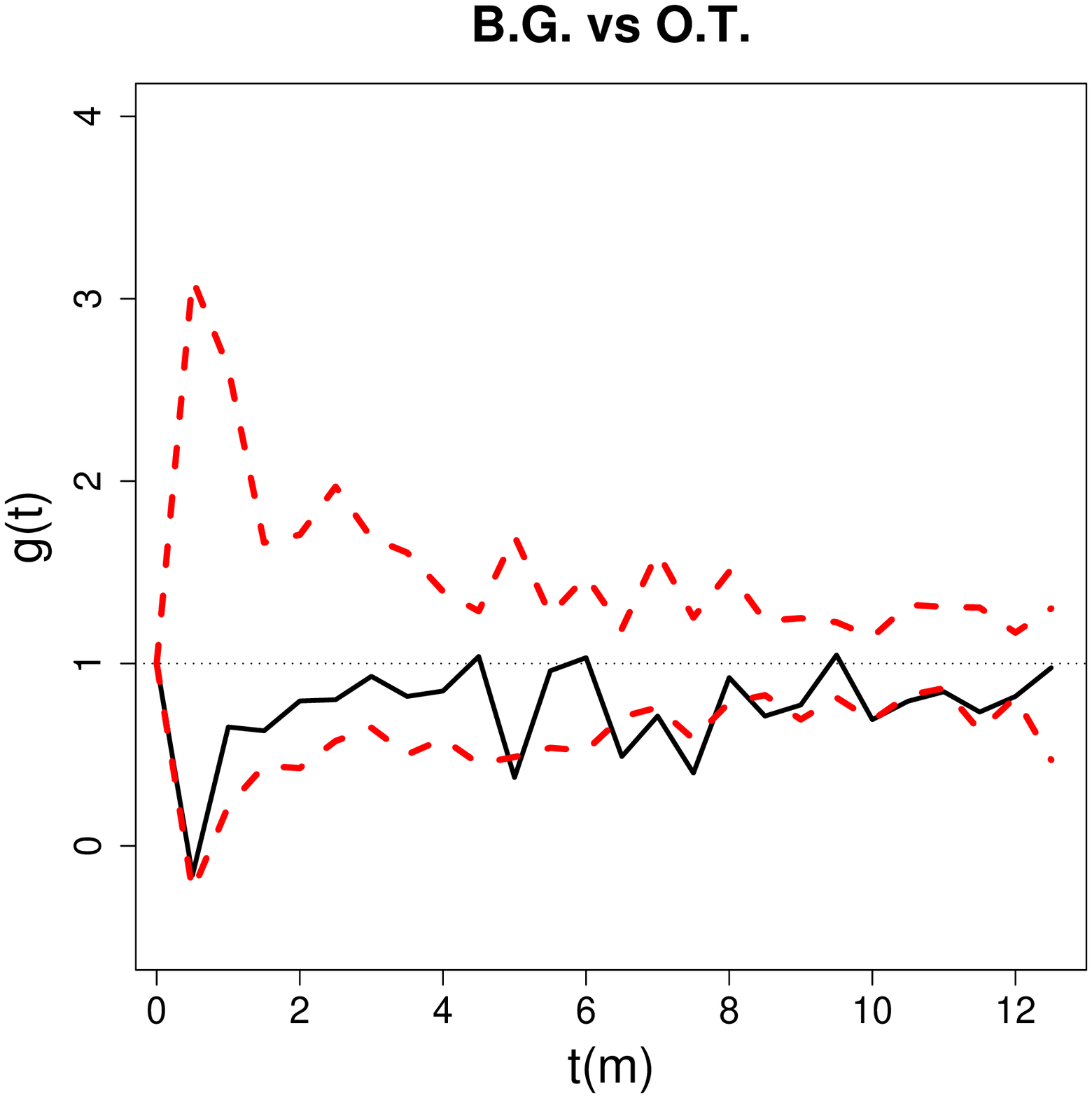} }}
\rotatebox{-90}{ \resizebox{2 in}{!}{\includegraphics{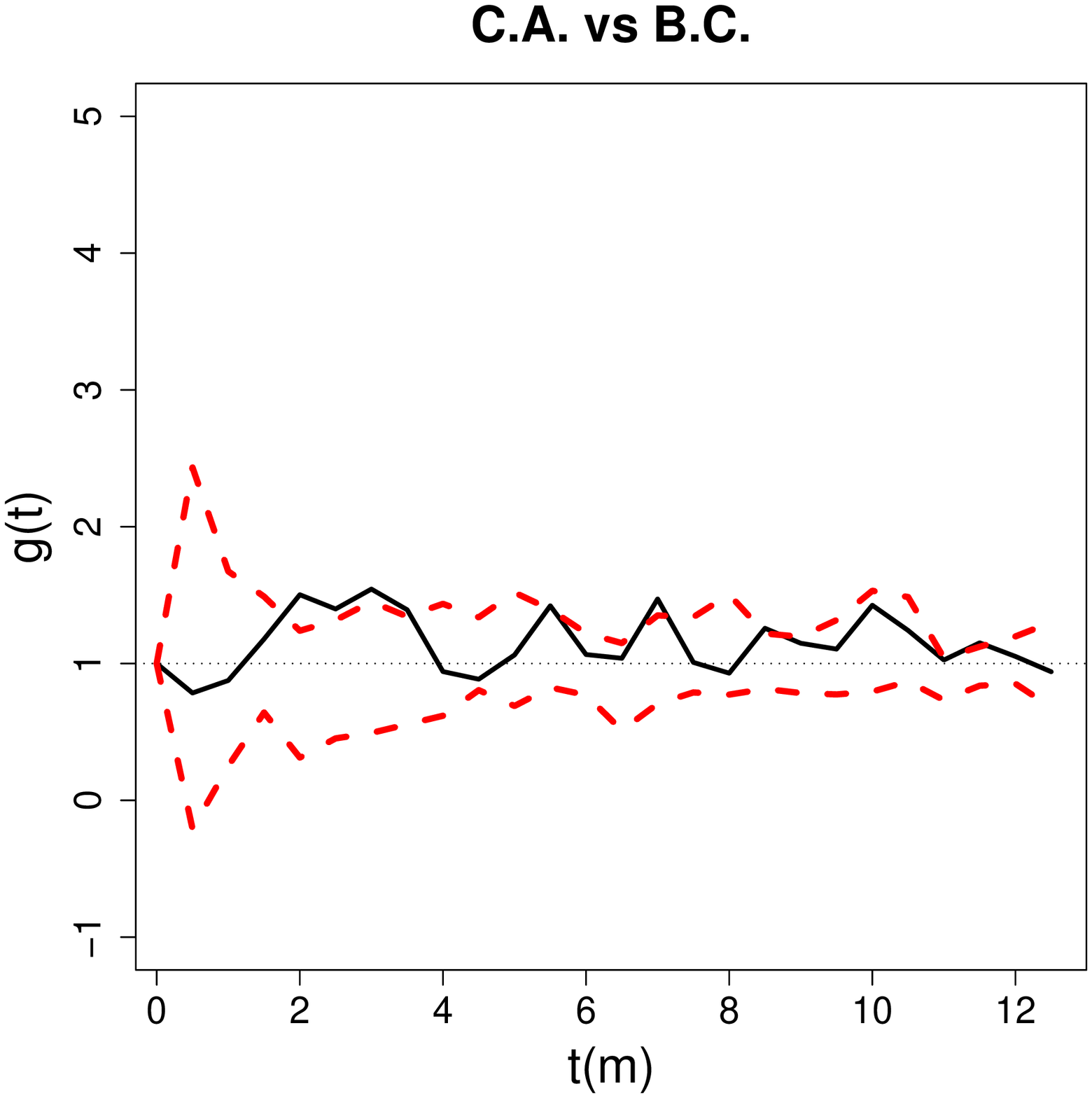} }}
\rotatebox{-90}{ \resizebox{2 in}{!}{\includegraphics{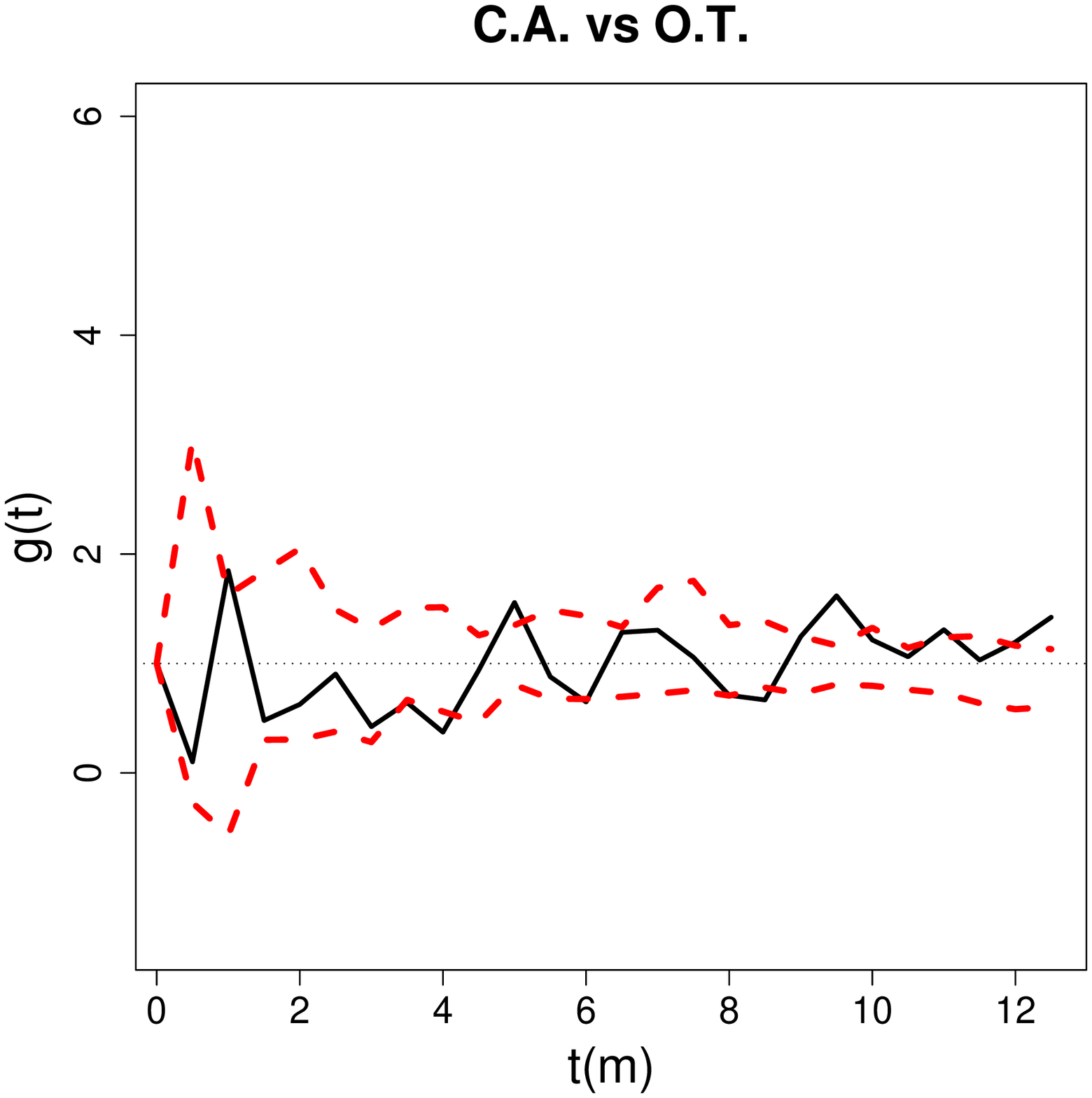} }}
\rotatebox{-90}{ \resizebox{2 in}{!}{\includegraphics{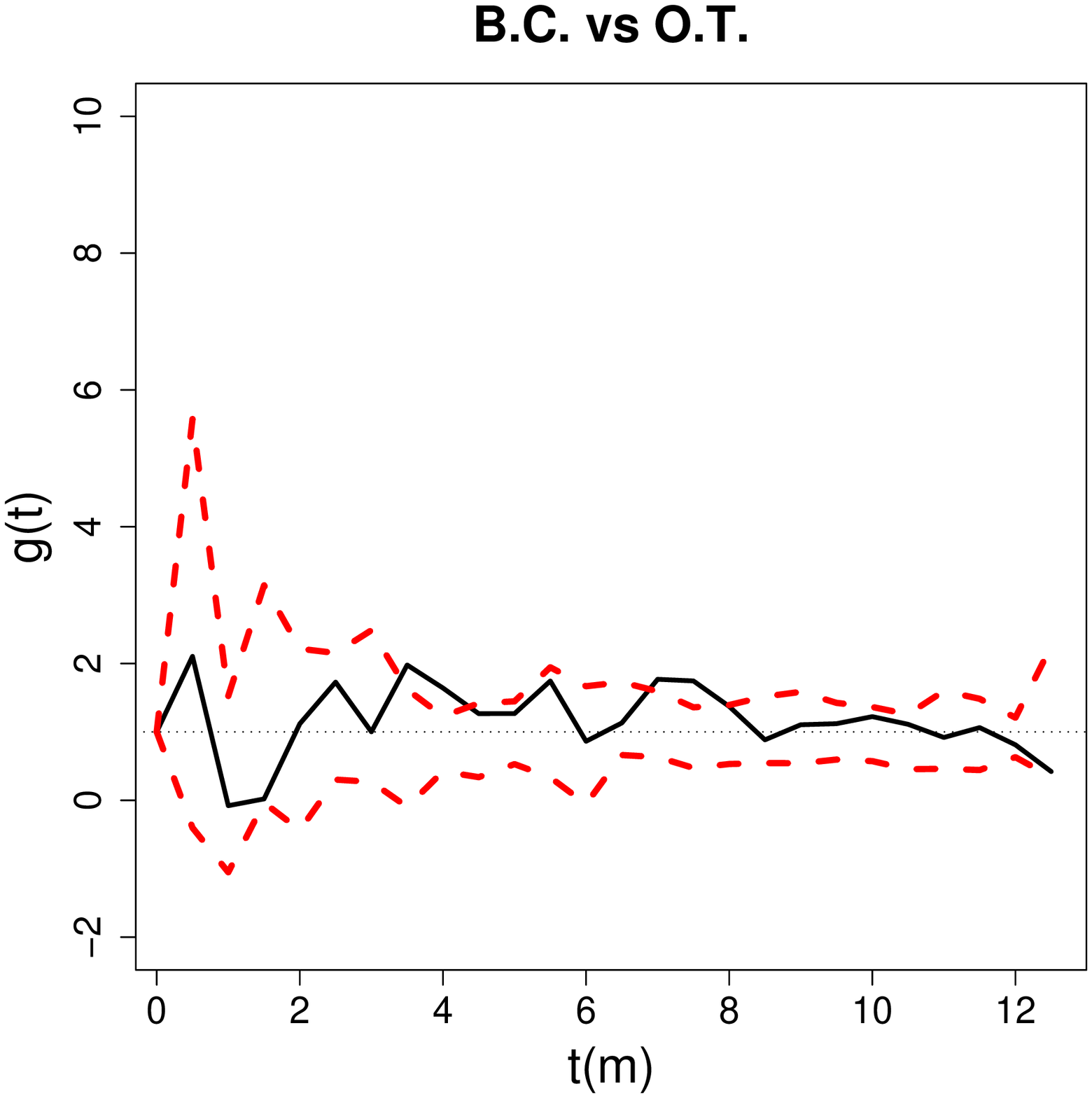} }}
\caption{
\label{fig:swamp-PCFij}
Pair correlation functions for each pair of species in the swamp tree data.
Wide dashed lines around 1 (which is the theoretical value)
are the upper and lower (pointwise) 95 \% confidence bounds for the
$L$-functions based on Monte Carlo simulations under the CSR independence pattern.
W.T. = water tupelos, B.G. = black gums, C.A. = Carolina ashes,
B.C.  = bald cypresses, and O.T. = other tree species.}
\end{figure}

But Ripley's $K$-function is cumulative,
so interpreting the spatial interaction at larger distances
is problematic (\cite{wiegand:2007}).
The (accumulative) pair correlation function $g(t)$
is better for this purpose (\cite{stoyan:1994}).
The pair correlation function of a (univariate)
stationary point process is defined as
$$g(t) = \frac{K'(t)}{2\,\pi\,t}$$
where $K'(t)$ is the derivative of $K(t)$.
For a univariate stationary Poisson process, $g(t)=1$;
values of $g(t) < 1$ suggest inhibition (or regularity) between points;
and values of $g(t) > 1$ suggest clustering (or aggregation).
The pair correlation functions for all trees
and each species for the swamp tree data are plotted
in Figure \ref{fig:swamp-PCFii}.
Observe that all trees are aggregated around
distance values of 0-1,3,4,5,7,9-10 m;
water tupelos are aggregated for distance values of 0-4 and 5-7 m;
black gums are aggregated for distance values of 1-6 and 8-11 m;
Carolina ashes are aggregated for all the range of the plotted distances;
bald cypresses are aggregated for distance values of 2-8 and around 11 m;
and other trees are aggregated for all distance values except 3-5 m.
Comparing Figures \ref{fig:swamp-Liihat} and \ref{fig:swamp-PCFii},
we see that Ripley's $L$ and pair correlation functions
detect the same patterns but with different distance values.
That is, Ripley's $L$ implies that the particular pattern is significant
for a wider range of distance values compared to $g(t)$,
since Ripley's $L$ is cumulative, so the values of $L$ at
small scales confound the values of $L$ at larger scales (\cite{loosmore:2006}).
Hence the results based on pair correlation function $g(t)$
are more reliable.

The same definition of the pair correlation function
can be applied to Ripley's bivariate $K$ or $L$-functions as well.
The benchmark value of $K_{ij}(t) = \pi \, t^2$ corresponds to $g(t) = 1$;
$g(t) < 1$ suggests segregation of the classes;
and $g(t) > 1$ suggests association of the classes.
The bivariate pair correlation functions for the
species in swamp tree data are plotted in Figure \ref{fig:swamp-PCFij}.
Observe that
water tupelos and black gums are segregated for distance values of 0-1 m;
water tupelos and Carolina ashes are segregated for values of 0-1 and 2.5 m
and are associated for values about 6 m;
water tupelos and bald cypresses are segregated for 0-1, 5.5, 9.5, and 11 m
and are associated for 6.5 m;
water tupelos and other trees are segregated for 0-0.5 and 7 m
and are associated for 8 m;
black gums and Carolina ashes are segregated for 2-2.5, 3.5-4.5, 6-8.5, and 9.5-12 m;
black gums and bald cypresses are segregated for 3.5, 5.5-6.5,7, and 9.5 m;
black gums and other trees are segregated for 5 and 6-7.5 m;
Carolina ashes and bald cypresses are associated for 1.5-3, 5.5., and 7 m;
Carolina ashes and other trees are associated for 5 and 9-10 m;
and bald cypresses and other trees are segregated for 4 m
and are associated for 3-4 and 6.5-7.5 m.

However the pair correlation function estimates might have critical behavior
for small $t$ if $g(t)>0$ since the estimator variance and hence
the bias are considerably large.
This problem gets worse especially in cluster processes (\cite{stoyan:1996}).
See for example Figures \ref{fig:swamp-PCFii} and \ref{fig:swamp-PCFij}
where the confidence bands for smaller $t$ values are much wider compared
to those for larger $t$ values.
So pair correlation function analysis is more reliable for larger distances
and it is safer to use $g(t)$ for distances larger than the average NN distance in the data set.
Comparing Figure \ref{fig:swamp-Liihat} with Figure \ref{fig:swamp-PCFii}
and Figure \ref{fig:swamp-Lijhat} with Figure \ref{fig:swamp-PCFij}
we see that Ripley's $L$ and pair correlation functions
usually detect the same large-scale pattern but at different ranges of distance values.
Ripley's $L$ suggests that the particular pattern is significant
for a wider range of distance values compared to $g(t)$,
since 
values of $L$ at small scales confound the values of $L$
at larger scales where $g(t)$ is more reliable to use (\cite{loosmore:2006}).

While second order analysis (using Ripley's $K$ and $L$-functions or pair correlation function)
provides information on the univariate and bivariate patterns
at all scales (i.e., for all distances),
NNCT-tests summarize the spatial interaction for the smaller scales
(for distances about the average NN distance in the data set).
In particular, for the swamp tree data average NN distance ($\pm$ standard deviation)
is about 1.8 ($\pm$ 1.04) meters and notice that Ripley's $L$-function and
NNCT-tests yield similar results for distances about 2 meters.

\section{Discussion and Conclusions}
\label{sec:disc-conc}
In this article we introduce new overall and cell-specific tests
of segregation based on nearest neighbor contingency tables (NNCTs).
Such tests are referred to as NNCT-tests.
We also consider Dixon's NNCT-tests,
discuss the differences in these (new and Dixon's) NNCT-tests,
present the asymptotic properties of them,
compare the tests using extensive Monte Carlo simulations
under RL and CSR independence and under various segregation and association alternatives
for two and three classes.
We also illustrate the tests on two examples and compare them with Ripley's
$L$-function (\cite{ripley:2004}).

NNCT-tests (i.e., overall and cell-specific tests of segregation)
are used in testing randomness in the nearest neighbor (NN)
structure between two or more classes.
The overall test is used for testing any deviation
from randomness in all the NNCT cells combined;
cell-specific test for cell $(i,j)$ is used for testing any deviation from randomness
in cell $(i,j)$, i.e., NN structure in which base class is $i$ and NN class is $j$.
This statistic tests the segregation or lack of it if $i=j$;
the association or lack of it between classes $i$ and $j$ if $i \neq j$.
The randomness in the NN structure is implied by the RL or CSR independence patterns.
We demonstrate that under the CSR independence pattern,
NNCT-tests are conditional on $Q$ and $R$,
while under the RL pattern, these tests are unconditional.
In the two-class case, cell-specific tests are essentially different only for two cells,
since cell $(1,1)$ and $(1,2)$ yield the same test statistic in absolute value for Dixon's
cell-specific test, likewise for cells $(2,1)$ and $(2,2)$.
Similarly, cell $(1,1)$ and $(2,1)$ yield the same test statistic in absolute value for the new
cell-specific test, likewise for cells $(1,2)$ and $(2,2)$.

Based on our Monte Carlo simulations,
we conclude that the asymptotic approximation for the cell-specific-tests
is appropriate only when the corresponding cell count in the NNCT is larger than 10;
and for the overall test when all cell counts are larger than 4.
When at least one cell count is less than 5,
we recommend the Monte Carlo randomization version of the overall tests;
and when a cell count is less than 10,
we recommend the Monte Carlo randomization of the cell-specific tests.
When each cell count is larger than 5,
the new versions of the segregation tests have empirical significance levels
closer to the nominal level.
Type I error rates (empirical significance levels)
of the new cell-specific tests are more robust to the differences in sample sizes
(i.e., differences in relative abundance).
When some cell count(s) are less than 5 for overall test and less than 10
for the cell-specific tests,
we compare the power of the tests using Monte Carlo critical values.
For large samples, the power comparisons can be made using both the asymptotic
or Monte Carlo critical values.
For the segregation alternatives,
we conclude that the new cell-specific and overall tests have higher power estimates
compared to those of Dixon's tests.
For the association alternatives,
we observe that the best performer NNCT-test
depends on the cell and level of association.
When testing against association,
the new cell-specific tests have higher power estimates for the upper triangular
cells in the NNCT and Dixon's cell-specific tests have higher power estimates
for the lower triangular cells in the NNCT.
We recommend the new cell-specific and overall tests for the segregation alternatives.
For the association alternatives, we recommend both versions of the overall and
cell-specific tests.

The CSR independence pattern assumes that the study region
is unbounded for the analyzed pattern,
which is not the case in practice.
Edge effects are a constant problem in the analysis of empirical
(i.e., bounded) data sets and much effort has gone into the
development of edge corrections methods (\cite{yamada:2003}).
So the edge (or boundary) effects might confound the test results
if the null pattern is the CSR independence.
Two correction methods for the edge effects on NNCT-tests,
namely buffer zone correction and toroidal correction,
are investigated in (\cite{ceyhan:cell-class-edge-correct})
where it is shown that the empirical sizes of the
NNCT-tests are mildly affected by the toroidal edge correction.
However, the (outer) buffer zone edge correction method
seems to have slightly stronger influence on the tests compared to toroidal correction.
But for these tests, buffer zone correction
does not change the sizes significantly for most sample size combinations.
This is in agreement with the findings of \cite{barot:1999}
who say NN methods only require a small buffer area around the study region.
A large buffer area does not help much since one only
needs to be able to see far enough away from an event to find its NN.
Once the buffer area extends past the likely NN distances
(i.e., about the average NN distances),
it is not adding much helpful information for NNCTs.
Hence we recommend inner or outer buffer zone correction for NNCT-tests
with the width of the buffer area being about the average NN distance.
We do not recommend larger buffer areas,
since they are wasteful with little additional gain.
On the other hand, we recommend the use of toroidal edge correction
with points within the average NN distance in the additional copies
around the study region.
For larger distances, the gain might not be worth the effort.

NNCT-tests summarize the pattern in the data set for small scales,
more specifically, they provide information on the pattern
around the average NN distance between all points.
On the other hand, pair correlation function $g(t)$
and Ripley's classical $K$ or $L$-functions and other variants provide
information on the pattern at various scales.
However, the classical $L$-function is not appropriate for the null pattern of RL
when locations of the points have spatial inhomogeneity.
For such cases, Diggle's $D$-function (\cite{diggle:2003} p. 131)
is more appropriate in testing the bivariate spatial clustering at various scales.

Ripley's classical $K(t)$ or $L(t)$ functions can be used when the null pattern
can be assumed to be CSR independence, that is when the null pattern
assumes first-order homogeneity for each class.
When the null pattern is the RL of points from an inhomogeneous
Poisson process they are not appropriate (\cite{kulldorff:2006});
Cuzick-Edward's $k$-NN tests are designed for testing bivariate spatial interaction
and mostly used for spatial clustering of cases in epidemiology;
Diggle's $D$-function is a modified version of Ripley's $K$-function (\cite{diggle:2003})
and adjusts for any inhomogeneity in the locations of, e.g., cases and controls.
Furthermore, there are variants of $K(t)$ that explicitly correct for inhomogeneity
(see \cite{baddeley:2000b}).
Ripley's $K-$, Diggle's $D$-functions and pair correlation functions are designed to analyze
univariate or bivariate spatial interaction at various scales (i.e., inter-point distances).
Our example illustrates that for distances around the average NN distance,
NNCT-tests and Ripley's bivariate $L$-function yield similar results.

The NNCT-tests and Ripley's $L$-function provide
similar information in the two-class case at small scales.
For $q$-class case with $q>2$ classes,
overall tests provide information on the (small-scale)
while the Ripley's $L$-function requires performing all bivariate
spatial interaction analysis.
The cell-specific tests can serve as pairwise post hoc analysis
only when the overall test is significant.
Furthermore, the cell-specific tests are testing
the spatial clustering of one class or bivariate interaction between two classes
as part of the multivariate interaction between all the classes.
On the other hand, Ripley's univariate $K$- or $L$-functions
are restricted to one class and bivariate $K$- or $L$-functions
are restricted to two classes they pertain to,
ignoring the potentially important multivariate interaction
between all classes in the study area.
However, there are forms of the $J$-function
which is derived from the well-known $G$ and $F$ functions (\cite{lieshout:1999})
and deal with this multi-type setting (i.e., consider the pattern of type $i$ in
the context of the pattern of all other types).
\cite{lieshout:1999} define two basic types of $J$-functions.
First is a type-$i$-to-type-$j$ function
which considers the points of type $i$ in the context of the points of type $j$.
The second one is the type-$i$-to-any-type function
which considers the points of type $i$ in the context of points of
all types including type $i$.
Other forms can be derived from them by re-defining the types.
For example, if we want to consider the points of type $i$ in the context of
points of all other types,
then we collapse all the other types $j$
(i.e., all $j$ which are not equal to $i$) into a single
type $i'$ and then use the type-$i$-to-type-$i'$ function.
Several authors have written about the bivariate $K$-function,
which is of the type-$i$-to-type-$j$ form (\cite{diggle:1991}, \cite{haase:1995},
and \cite{diggle:2003}).
Type-$i$-to-type-$j$ $K$-function can easily be modified to
type-$i$-to-any-type $K$-function.
Thus essentially there is only one family of multi-type $K$-functions in literature.
But type-$i$-to-type-$j$ $K$-function is comparable with a NNCT analysis
based on a $2\times2$ NNCT restricted to the classes $i$ and $j$.
Similarly, type-$i$-to-type-$i'$ $K$-function
is comparable with the NNCT analysis based on a
$2 \times 2$ NNCT with classes $i$ and the rest of the classes labeled as $i'$.
Since pairwise analysis of $q$ classes with $2 \times 2$ NNCTs
might yield conflicting results compared to $q \times q$ NNCT analysis (\cite{dixon:NNCTEco2002}),
Ripley's $L$-function and NNCT-tests might also yield conflicting results at small distances.
Hence Ripley's $L$-function and NNCT-tests may provide similar but not identical information
about the spatial pattern and the latter might provide
small-scale interaction that is not detected by the former.
Since the pair correlation functions are derivatives of Ripley's $K$-function,
most of the above discussion holds for them also,
except $g(t)$ is reliable only for large scale interaction analysis.
Hence NNCT-tests and pair correlation function are not comparable but
provide complimentary information about the pattern in question.

Cell-specific tests for diagonal cells in a NNCT and
Ripley's univariate $K$- or $L$-functions (and hence pair correlation functions)
are symmetric, as they measure the spatial clustering of one class only.
On the other hand, Ripley's bivariate $K$- or $L$-functions
and pair correlation functions are symmetric  in the two classes they pertain to.
But cell-specific tests for two classes (i.e., off-diagonal cells in the NNCT)
are not symmetric.
Hence, at small scales, the cell-specific test for an off-diagonal cell,
provides the type and different levels of spatial interaction
for the corresponding two classes,
while Ripley's $L$-function and pair correlation function
provide only the type of spatial interaction,
but can not distinguish the class-specific level of interaction
for each of the two classes in question.

For a data set for which CSR independence is the reasonable null pattern,
we recommend the overall segregation test
if the question of interest is the spatial interaction at small scales
(i.e., about the mean NN distance).
If it yields a significant result, then to determine which pairs
of classes have significant spatial interaction,
the cell-specific tests can be performed.
One can also perform Ripley's $K$ or $L$-function
and only consider distances up to around the average NN distance
and compare the results with those of NNCT analysis.
If the spatial interaction at higher scales is of interest,
pair correlation function is recommended (\cite{loosmore:2006}),
due to the cumulative nature of Ripley's $K$- or $L$-functions for larger distances.
On the other hand, if the RL pattern is the reasonable null pattern for the data,
we recommend the NNCT-tests if the small-scale interaction is of interest
and Diggle's $D$-function if the spatial interaction at higher scales is also of interest.

\section*{Acknowledgments}
I would like to thank an anonymous associate editor and two referees,
whose constructive comments and suggestions greatly improved the presentation
and flow of the paper.
Most of the Monte Carlo simulations presented in this article
were executed on the Hattusas cluster of
Ko\c{c} University High Performance Computing Laboratory.


\begin{thebibliography}{}

\bibitem[Armstrong and Irvine, 1989]{armstrong:1989}
Armstrong, J.~E. and Irvine, A.~K. (1989).
\newblock Flowering, sex ratios, pollen-ovule ratios, fruit set, and
  reproductive effort of a dioecious tree, \emph{{M}yristica {I}nsipida}
  (\emph{Myristicacea}), in two different rain forest communities.
\newblock {\em American Journal of Botany}, 76:75--85.

\bibitem[Baddeley et~al., 2000]{baddeley:2000b}
Baddeley, A., M{\o}ller, J., and Waagepetersen, R. (2000).
\newblock Non- and semi-parametric estimation of interaction in inhomogeneous
  point patterns.
\newblock {\em Statistica Neerlandica}, 54(3):329–--350.

\bibitem[Baddeley and Turner, 2005]{baddeley:2005}
Baddeley, A. and Turner, R. (2005).
\newblock spatstat: An {R} package for analyzing spatial point patterns.
\newblock {\em Journal of Statistical Software}, 12(6):1--42.

\bibitem[Barot et~al., 1999]{barot:1999}
Barot, S., Gignoux, J., and Menaut, J.~C. (1999).
\newblock Demography of a savanna palm tree: predictions from comprehensive
  spatial pattern analyses.
\newblock {\em Ecology}, 80:1987--2005.

\bibitem[Ceyhan, 2007]{ceyhan:cell-class-edge-correct}
Ceyhan, E. (2007).
\newblock Edge correction for cell- and class-specific tests of segregation
  based on nearest neighbor contingency tables.
\newblock In {\em Proceedings of the International Conference on Environment:
  Survival and Sustainability, Near East University.}

\bibitem[Coomes et~al., 1999]{coomes:1999}
Coomes, D.~A., Rees, M., and Turnbull, L. (1999).
\newblock Identifying aggregation and association in fully mapped spatial data.
\newblock {\em Ecology}, 80(2):554--565.

\bibitem[Cressie, 1993]{cressie:1993}
Cressie, N. A.~C. (1993).
\newblock {\em Statistics for Spatial Data}.
\newblock Wiley, New York.

\bibitem[Cuzick and Edwards, 1990]{cuzick:1990}
Cuzick, J. and Edwards, R. (1990).
\newblock Spatial clustering for inhomogeneous populations (with discussion).
\newblock {\em Journal of the Royal Statistical Society, Series B}, 52:73--104.

\bibitem[Diggle, 2003]{diggle:2003}
Diggle, P.~J. (2003).
\newblock {\em Statistical Analysis of Spatial Point Patterns}.
\newblock Hodder Arnold Publishers, London.

\bibitem[Diggle and Chetwynd, 1991]{diggle:1991}
Diggle, P.~J. and Chetwynd, A.~G. (1991).
\newblock Second-order analysis of spatial clustering for inhomogeneous
  populations.
\newblock {\em Biometrics}, 47:1155--1163.

\bibitem[Dixon, 1994]{dixon:1994}
Dixon, P.~M. (1994).
\newblock Testing spatial segregation using a nearest-neighbor contingency
  table.
\newblock {\em Ecology}, 75(7):1940--1948.

\bibitem[Dixon, 2002]{dixon:NNCTEco2002}
Dixon, P.~M. (2002).
\newblock Nearest-neighbor contingency table analysis of spatial segregation
  for several species.
\newblock {\em Ecoscience}, 9(2):142--151.

\bibitem[Good and Whipple, 1982]{good:1982}
Good, B.~J. and Whipple, S.~A. (1982).
\newblock Tree spatial patterns: {S}outh {C}arolina bottomland and swamp
  forests.
\newblock {\em Bulletin of the Torrey Botanical Club}, 109:529--536.

\bibitem[Goreaud and P\'{e}lissier, 2003]{goreaud:2003}
Goreaud, F. and P\'{e}lissier, R. (2003).
\newblock Avoiding misinterpretation of biotic interactions with the intertype
  ${K}_{12}$-function: population independence vs. random labelling hypotheses.
\newblock {\em Journal of Vegetation Science}, 14(5):681–--692.

\bibitem[Haase, 1995]{haase:1995}
Haase, P. (1995).
\newblock Spatial pattern analysis in ecology based on {R}ipley's
  {$K$}-function: {I}ntroduction and methods of edge correction.
\newblock {\em The Journal of Vegetation Science}, 6:575--582.

\bibitem[Hamill and Wright, 1986]{hamill:1986}
Hamill, D.~M. and Wright, S.~J. (1986).
\newblock Testing the dispersion of juveniles relative to adults: A new
  analytical method.
\newblock {\em Ecology}, 67(2):952--957.

\bibitem[Herler and Patzner, 2005]{herler:2005}
Herler, J. and Patzner, R.~A. (2005).
\newblock Spatial segregation of two common {G}obius species ({T}eleostei:
  {G}obiidae) in the {N}orthern {A}driatic {S}ea.
\newblock {\em Marine Ecology}, 26(2):121--129.

\bibitem[Herrera, 1988]{herrera:1988}
Herrera, C.~M. (1988).
\newblock Plant size, spacing patterns, and host-plant selection in
  \emph{{O}syris quadripartita}, a hemiparasitic dioecious shrub.
\newblock {\em Journal of Ecology}, 76:995--1006.

\bibitem[Kulldorff, 2006]{kulldorff:2006}
Kulldorff, M. (2006).
\newblock Tests for spatial randomness adjusted for an inhomogeneity: A general
  framework.
\newblock {\em Journal of the American Statistical Association},
  101(475):1289--1305.

\bibitem[Lahiri, 1996]{lahiri:1996}
Lahiri, S.~N. (1996).
\newblock On consistency of estimators based on spatial data under infill
  asymptotics.
\newblock {\em Sankhya: The Indian Journal of Statistics, Series A},
  58(3):403--417.

\bibitem[Loosmore and Ford, 2006]{loosmore:2006}
Loosmore, N. and Ford, E. (2006).
\newblock Statistical inference using the $g$ or $k$ point pattern spatial
  statistics.
\newblock {\em Ecology}, 87:1925--1931.

\bibitem[Mat'ern, 1986]{matern:1986}
Mat'ern, B. (1986).
\newblock Spatial variation.
\newblock In {\em Lecture Notes in Statistics, edited by D. Brillinger et al.},
  volume~36. Springer-Verlag, Berlin.

\bibitem[Meagher and Burdick, 1980]{meagher:1980}
Meagher, T.~R. and Burdick, D.~S. (1980).
\newblock The use of nearest neighbor frequency analysis in studies of
  association.
\newblock {\em Ecology}, 61(5):1253--1255.

\bibitem[Moran, 1948]{moran:1948}
Moran, P. A.~P. (1948).
\newblock The interpretation of statistical maps.
\newblock {\em Journal of the Royal Statistical Society, Series B},
  10:243--251.

\bibitem[Nanami et~al., 1999]{nanami:1999}
Nanami, S.~H., Kawaguchi, H., and Yamakura, T. (1999).
\newblock Dioecy-induced spatial patterns of two codominant tree species,
  \emph{{P}odocarpus nagi} and \emph{{N}eolitsea aciculata}.
\newblock {\em Journal of Ecology}, 87(4):678--687.

\bibitem[Pielou, 1961]{pielou:1961}
Pielou, E.~C. (1961).
\newblock Segregation and symmetry in two-species populations as studied by
  nearest-neighbor relationships.
\newblock {\em Journal of Ecology}, 49(2):255--269.

\bibitem[Ripley, 2004]{ripley:2004}
Ripley, B.~D. (2004).
\newblock {\em Spatial Statistics}.
\newblock Wiley-{I}nterscience, New York.

\bibitem[Searle, 2006]{searle:2006}
Searle, S.~R. (2006).
\newblock {\em Matrix Algebra Useful for Statistics}.
\newblock Wiley-{I}ntersciences.

\bibitem[Stoyan and Stoyan, 1994]{stoyan:1994}
Stoyan, D. and Stoyan, H. (1994).
\newblock {\em Fractals, random shapes and point fields: methods of geometrical
  statistics.}
\newblock John Wiley and Sons, New York.

\bibitem[Stoyan and Stoyan, 1996]{stoyan:1996}
Stoyan, D. and Stoyan, H. (1996).
\newblock Estimating pair correlation functions of planar cluster processes.
\newblock {\em Biometrical Journal}, 38(3):259--271.

\bibitem[van Lieshout and Baddeley, 1999]{lieshout:1999}
van Lieshout, M.~N.~M. and Baddeley, A.~J. (1999).
\newblock Indices of dependence between types in multivariate point patterns.
\newblock {\em Scandinavian Journal of Statistics}, 26:511--532.

\bibitem[Waagepetersen, 2007]{waagepetersen:2007}
Waagepetersen, R.~P. (2007).
\newblock An estimating function approach to inference for inhomogeneous
  {N}eyman-{S}cott processes.
\newblock {\em Biometrics}, 63(1):252--258.

\bibitem[Waller and Gotway, 2004]{waller:2004}
Waller, L.~A. and Gotway, C.~A. (2004).
\newblock {\em Applied Spatial Statistics for Public Health Data}.
\newblock Wiley-Interscience, NJ.

\bibitem[Whipple, 1980]{whipple:1980}
Whipple, S.~A. (1980).
\newblock Population dispersion patterns of trees in a {S}outhern {L}ouisiana
  hardwood forest.
\newblock {\em Bulletin of the Torrey Botanical Club}, 107:71--76.

\bibitem[Wiegand et~al., 2007]{wiegand:2007}
Wiegand, T., Gunatilleke, S., and Gunatilleke, N. (2007).
\newblock Species associations in a heterogeneous {S}ri {L}ankan dipterocarp
  forest.
\newblock {\em The {A}merican {N}aturalist}, 170(4):77--95.

\bibitem[Yamada and Rogersen, 2003]{yamada:2003}
Yamada, I. and Rogersen, P.~A. (2003).
\newblock An empirical comparison of edge effect correction methods applied to
  {$K$}-function analysis.
\newblock {\em Geographical Analysis}, 35(2):97--109.

\end{thebibliography}



\end{document}